\documentclass[12pt]{article}
\usepackage{a4}
\usepackage{graphicx}
\usepackage{epsf}
\usepackage{amssymb}
\usepackage{cite}
\newcommand{\be}{\begin{equation}}
\newcommand{\ee}{\end{equation}}
\newcommand{\bea}{\begin{eqnarray}}
\newcommand{\eea}{\end{eqnarray}}

\begin{document}
\def\C{{\mathbb{C}}}
\def\R{{\mathbb{R}}}
\def\s{{\mathbb{S}}}
\def\T{{\mathbb{T}}}
\def\Z{{\mathbb{Z}}}
\def\W{{\mathbb{W}}}
\def\Bbb{\mathbb}
\def\BZ{\Bbb Z} \def\BR{\Bbb R}
\def\BW{\Bbb W}
\def\BM{\Bbb M}
\def\BC{\Bbb C} \def\BP{\Bbb P}
\def\CP{\BC\BP}
\begin{titlepage}
\title{On the Thermodynamic Geometry and Critical Phenomena of AdS Black Holes}
\author{}
\date{
Anurag Sahay, Tapobrata Sarkar, Gautam Sengupta
\thanks{\noindent E-mail:~ ashaya, tapo, sengupta @iitk.ac.in}
\vskip0.4cm
{\sl Department of Physics, \\
Indian Institute of Technology,\\
Kanpur 208016, \\
India}}
\maketitle
\abstract{
\noindent
In this paper, we study various aspects of the equilibrium thermodynamic state space geometry of AdS black holes. We first examine the 
Reissner-Nordstrom-AdS (RN-AdS) and the Kerr-AdS black holes. In this context, the state space scalar curvature of these
black holes is analysed in various regions of their thermodynamic parameter space. This provides important new insights into the structure and significance
of the scalar curvature. We further investigate critical phenomena, and the behaviour of the scalar curvature near criticality, for KN-AdS black holes in 
two mixed ensembles, introduced and elucidated in our earlier work arXiv:1002.2538 [hep-th]. The critical exponents are identical to those in the 
RN-AdS and Kerr-AdS cases in the canonical ensemble. This suggests an universality in the scaling behaviour near critical points of AdS black holes. 
Our results further highlight qualitative differences in the thermodynamic state space geometry for 
electric charge and angular momentum fluctuations of these. }
\end{titlepage}

\section{Introduction}

It has been widely recognized over the past few decades that thermodynamic properties of black holes (see, e.g \cite{td1},\cite{td2},\cite{td3}) and references
therein) provide an important analytical tool for understanding
several issues involving quantum theories of gravity. The topic has been intensely discussed in the recent past, and several key features have
been uncovered, especially in the context of black holes arising in String Theory. However, a knowledge of the exact statistical description 
of black hole microstates is still elusive, although thermodynamic studies of black holes do indicate extremely interesting phase structures and critical 
phenomena in these systems.

In particular, black holes in asymptotically anti de-Sitter space have been at the center of extensive research since the discovery of Maldacena's AdS/CFT correspondence\cite{maldacena}. The gauge/gravity duality is expected to play a major role in our efforts to understand the underlying microscopic statistical interactions in black 
holes via the gauge theory living on the boundary. For example, this duality has led to the celebrated correspondence between
Hawking-Page phase transitions in asymptotically AdS black holes and confinement-deconfinement phase transitions in the boundary quantum field theory.

In a separate context, a geometrical perspective of equilibrium thermodynamics has been developed over the last few decades. 
As is well known, an extrinsic geometrical perspective of thermodynamics
has been elucidated following early treatments by Tisza \cite{tisza} and Callen \cite{callen}. However an intrinsic geometrical
structure inherent in thermodynamics turned out to be much more elusive.  Following the early work of Weinhold \cite{weinhold}, 
in the energy representation, it was possible to define a Riemannian metric on the space of the equilibrium thermodynamic states
of a system. However it was not possible to interpret this physically. In a later development Ruppeiner \cite{rupp} provided an 
intrinsic Riemannian geometrical framework in the entropy representation for the equilibrium thermodynamic state space of a system. The line elemnt in this case could be related to the probability measure of thermodynamic fluctuations connecting two
equilibrium states and has been refered to as the thermodynamic geometry of a system. The first analysis
of black hole thermodynamics in the framework of thermodynamic geometry was presented in \cite{ferrara} for extremal black holes in supergravity theories. In the last few years the thermodynamic geometry of both extremal and non extremal black holes have been
a subject of investigations and have provided interesting insights into the thermodynamics, phase transition and critical phenomena
of black holes. However there are still numerous unresolved issues which merit further detailed analysis of diverse black holes in this framework.

 To this end, in \cite{paper}, we studied the thermodynamic geometry of Kerr-Newman-AdS (KN-AdS) black holes in the grand canonical and two novel ``mixed'' ensembles, 
wherein one of the two thermodynamic charges, (i.e the angular momentum or the electric charge), was held fixed while the other charge was allowed to be 
exchanged with the surroundings of black hole. For the mixed ensembles, we were able to establish new phase structures exhibiting liquid-gas
like first order phase transitions culminating in a critical point. The scalar curvature, $R$, of the state space Riemannian geometries corresponding to these 
ensembles show a divergence at the critical point analogous to that for conventional critical phenomena. Interestingly, we showed that the scalar curvature 
also carries information about first order phase transitions in these black hole systems. This is encoded in the multivalued branch structure of the scalar curvature 
in the phase coexistence regions. We examined this issue in detail, first in the context of a conventional Van der Waals model and followed it up with an 
extensive analysis of the mixed ensemble black holes, finding that the two systems closely resemble each other in the behaviour of their scalar curvature 
near first and second order phase transitions. 

In the present work we complement our earlier investigation by undertaking a detailed investigation of the scaling behaviour of the thermodynamic functions and 
the corresponding scalar curvatures near the critical points of these black holes. For the two mixed ensembles, we calculate the critical exponents by developing 
a general perturbative scheme for thermodynamic quantities depending on two independent control parameters, like the temperature and the electric potential. We then 
compare the critical exponents with the known case of Reissner-Nordstrom-AdS (RN-AdS) black holes, as obtained in \cite{johnson1} and \cite{wu}, 
and also check whether these exponents follow from a more general scaling law for the singular part of the free energy near criticality. We also obtain the 
scaling behaviour of $R$ and compare it with its known critical behaviour for conventional thermodynamic systems, as explained in \cite{rupp}.

As regards thermodynamic geometry, it is believed that the sign of the state space scalar curvature $R$ is an indicator of the nature of microscopic interactions 
undelying the thermodynamic system. For example, it was observed in the case of ideal quantum gases that $R$ is positive for repulsive fermi interactions 
and negative for attractive bosonic interactions \cite{jan1}. Similarly, for the case of anyon gases obeying fractional statistics, $R$ was shown to become 
positive for repulsive quantum interactions and negative for attractive quantum interactions, \cite{mirza1}. However, a convincing physical or analytical justification 
for this attribute of $R$ is still forthcoming. The issue of the sign of R becomes more pressing in the case of black holes, where it is well known that $R$ for different black holes usually changes sign on varying the independent thermodynamic charges. Moreover, the problem is compounded by a lack of knowledge of the microscopic 
interactions in the case of black holes. In our previous work, we had addressed the issue of the sign of $R$ by first obtaining a global picture of its sign variations. 
This was achieved by  plotting its zeroes in the parameter space of fluctuating variables for various ensembles. Our approach was mainly kinematic, in the sense 
that the investigations were confined to observing the sign variations in $R$ for the grand canonical and the two mixed ensembles of KN-AdS black holes, with brief comments on the differences in the various cases. Other approach, which we had studied extensively, was to adopt the view that, irrespective of its signature, the \emph{magnitude} of the state space scalar curvature , $|R|$, is a measure of strength of interactions in the black hole, \cite{rupp1}. From the plots of 
$|R|$ vs. $t$ we could make some interesting observations regarding the stability of small black hole branches vis. a vis the large black hole branches, 
which sometimes ran counter to common expectations.

In the present work we address the issue of  the sign variations in $R$ in a more systematic fashion. We start with the assertion that just as in the case of conventional 
thermodynamic systems, the state space scalar curvature for black holes too is indicative of the (yet unknown) nature of microscopic interactions. In order to
substantiate this viewpoint, we carefully study the scalar curvature corresponding to different kinds of thermodynamic fluctuations for black holes in AdS space. 
Starting with the simpler cases of RN-AdS and Kerr-AdS black holes, each of them having a two dimensional state space geometry, we move on to discuss the 
KN-AdS black hole in the grand canonical ensemble, which has a three dimensional state space geometry consisting of fluctuations in the mass, electric charge and the angular momentm  $m,q$ and $j$ respectively. We then 
study the scalar curvature corresponding to the two dimensional state space of KN AdS black holes in the mixed ensembles. One of the main conclusions 
following our investigation of the different scalar curvatures is that the presence of $q$ fluctuations always brings about a change in the sign of $R$, whereas 
the $j$ fluctuations do not effect any such sign change.

This paper is organized as follows. In section $2$ we briefly discuss the thermodynamics of KN-AdS black holes, mainly using it to establish our notations 
and obtain formulae of thermodynamic functions for later use, and then revisit the thermodynamics of RN-AdS black holes using our notation, mainly emphasizing 
on the grand canonical ensemble. Next we study the state space scalar curvature of these black holes in detail. In section $3$ the Kerr-AdS black holes are subjected 
to a similar analysis, ending with a comparison between the scalar curvature of RN-AdS and Kerr-AdS black holes. In sections $4$ and $5$ respectively,
the scalar curvature for KN-AdS black holes in the grand canonical ensemble and the two mixed ensembles alluded to earlier, is studied in details, starting with 
a discussion of their phase behaviour. Further, in section $5$, the perturbative calculations for the critical exponents of these mixed ensemble black holes is undertaken. 
Finally, we summarize our results in section $5$.

\section{Thermodynamics of KN-AdS and RN-AdS Black Holes}

In this section, as a warm up exercise, we will first review certain aspects of the thermodynamics of KN-AdS and RN-AdS black holes, before analyzing the geometry of 
their equilibrium state spaces. The main purpose of this section is to set the notations and conventions used in the rest of the paper. However, we emphasize that our 
formulation will be entirely in terms of thermodynamic variables, in contrast to analyses involving black holes parameters as commonly appear in the literature. 
In particular, subsection (2.2) contains qualitatively new analyses. 

\subsection{KN-AdS Black Holes }

We begin with the Smarr formula for the KN-AdS black holes, which is \cite{calda},
\begin{equation}
M = \left[\frac{S}{4\pi} + \frac{\pi}{4S}\left(4J^2 + Q^4\right) + \frac{Q^2}{2} + \frac{J^2}{l^2}
+ \frac{S}{2\pi l^2}\left(Q^2 + \frac{S}{\pi} + \frac{S^2}{2\pi^2l^2}\right)\right]^{1 \over 2}
\end{equation}

where $M$ is the mass of the black hole, $S$ is the entropy and $J$, $Q$ are the angular momentum and the electric charge respectively. This formula differs from that of asymptotically flat Kerr-Newman
black holes in the last two terms which arise because of  a finite AdS length scale. It can be verified from the Smarr formula that the mass $M$ is a homogenous function 
of degree $1/2$ in $S,J,Q^2$ and $l^2$, (\cite{calda}),

\begin{equation}
M({\lambda}S,{\lambda}J,{\lambda}Q^2,{\lambda}l^2)={\lambda}^{1/2}M(S,J,Q^2,l^2)
\end{equation}
where ${\lambda}$ is a constant. Since we will not be treating the AdS length  $l$ as a thermodynamic variable, i.e we are not considering ensembles in which 
$l$ may fluctuate, we absorb it into the Smarr formula by taking the scaling constant $\lambda$ equal to $1/l^2$. The rescaled thermodynamic variables are
then defined as
\begin{equation}
m = \frac{M}{l},~ s = \frac{S}{l^2},~ q = \frac{Q}{l},~j = \frac{J}{l^2}
\end{equation}
The Smarr formula, after the rescaling, becomes
\begin{equation}
\label{reducedsmarr}
m=\frac{1}{2}\left ({\frac { s^4 + 2s^3\pi + s^2\pi^2\left(2q^2 + 1\right) + \pi^3s\left(2q^2 + 4j^2\right) + \pi^2\left(q^4 + 4j^2\right)}{{\pi }^{3}{s}}}\right )^{\frac{1}{2}}
\end{equation}
We can calculate the conjugate quantities like the temperature and potentials using the first law of thermodynamics, i.e by differentiating the Smarr formula for 
$m$ with respect to the charges $s$, $q$ and $j$.
\begin{equation}
dm = tds + \omega dj + \phi dq,
\end{equation}
where the rescaled angular velocity, temperature and the electric potential are defined as
\begin{equation}
\omega=l\Omega~,~t=lT~,~\phi=\Phi
\end{equation}

Although the expressions are standard,
we reproduce some of them for convenience. The electric potential $\phi$ and the angular velocity $\omega$, as functions of the
charges (i.e $s$, $q$, $j$) are given by
\begin{eqnarray}
\phi = \frac{\pi^{\frac{1}{2}} q \left(s^2 + s\pi + q^2\pi^2\right)} {s^{1 \over 2}\left[s^4 + 2s^3\pi + s^2\pi^2\left(1 + 2q^2\right)  + 2q^2\pi^3s +
4\pi^3j^2\left(\pi + s\right)\right]^{1 \over 2}}\label{knphiom1}\\
\omega = \frac{2\pi^{\frac{3}{2}}j\left(\pi + s\right)} {s^{1 \over 2}\left[s^4 + 2s^3\pi + s^2\pi^2\left(1 + 2q^2\right)  + 2q^2\pi^3s +
4\pi^3j^2\left(\pi + s\right)\right]^{1 \over 2}}
\label{knphiom}
\end{eqnarray}
whereas the temperature of the black hole is given by
\begin{equation}
t = \frac{3s^4 + 4s^3\pi + s^2\pi^2\left(1 + 2q^2\right) - 4\pi^4 j^2 - \pi^4q^4}{4\pi^{3 \over 2}s^{3 \over 2}
\left[s^4 + 2s^3\pi + s^2\pi^2\left(1 + 2q^2\right)  + 2q^2\pi^3s +
4\pi^3j^2\left(\pi + s\right)\right]^{1 \over 2}}
\label{kntemp}
\end{equation}
Having tabulated the basic formulae for KN-AdS black holes in terms of thermodynamic parameters, we now discuss the RN-AdS black holes.

\subsection{RN-AdS Black Holes}
Thermodynamic quantities for the RN-AdS black hole may be calculated by setting $j=0$ in the expressions given in the previous subsection. We begin 
with the expression for the temperature, 
\begin{equation}
t =  {1 \over 4}\left({\frac {3\,{s}^{2}+s\pi -{q}^{2}{\pi }^{2}}{{\pi }^{3/2}{s}^{3/2}}}\right)
\label{rntemp}
\end{equation}
From the expression for $t$ we can obtain the charge at the extremal limit in terms of entropy $s$ by
\begin{equation}
q_{ex}(s)=\frac{1}{\pi}\,\sqrt{s\,(3s+\pi)}
\end{equation}
The electric potential is given by
\begin{equation}
\phi =  \frac{q\sqrt{\pi}}{\sqrt{s}}
\label{rnphi}
\end{equation}
The value of $\phi$ obtained from thermodynamics is the potential measured at infinity with respect to the horizon, \cite{calda}.
Let us also, for future reference, calculate the heat capacities and the susceptibility. The heat capacity at constant charge, denoted by
$c_q$, is given by
\begin{equation}
c_q =t\,\left(\frac{\partial s}{\partial t}\right)_q= {\frac {2s \left( 3\,{s}^{2}+s\pi -{q}^{2}{\pi }^{2} \right) }{3\,{s}^{2}-s\pi +3\,{q}^{2}{\pi }^{2}}}
\end{equation}
and the heat capacity at constant potential, $c_{\phi}$ is given by
\begin{equation}
c_{\phi}=t\,\left(\frac{\partial s}{\partial t}\right)_{\phi}= {\frac { 2s \left( 3\,{s}^{2}+s\pi -{q}^{2}{\pi }^{2} \right) }{3\,{s}^{2}-s\pi +{q}^{2}{\pi }^{2}}}
\label{cphi}
\end{equation}
It can be seen that both the heat capacities vanish at extremality. The ``capacitance'' or the isothermal susceptibility,
\begin{equation}
{\chi}_t=\left(\frac{\partial q}{\partial {\phi}}\right)_{t}={\frac { \left( 3\,{s}^{2}-s\,\pi +3\,{\pi }^{2}{q}^{2} \right) \sqrt {s}}{\sqrt {\pi } \left( 3\,{s}^{2}-s\pi +{q}^{2}{\pi }^{2} \right)
}},
\end{equation}
can be expressed in terms of the heat capacities as
\begin{equation}
{\chi}_t=\frac{c_{\phi}}{c_q}\frac{\sqrt{s}}{\sqrt{\pi}}
\end{equation}

Thermodynamics of RN-AdS black holes has been discussed extensively in \cite{johnson1} and \cite{johnson2}. There it was established that in the canonical 
ensemble these black holes show a liquid-gas like phase coexistence behaviour between a small black hole phase and a large black hole phase, culminating 
in a critical point of second order phase transition.  However, in the grand canonical ensemble, they
undergo a Hawking-Page phase transition to a thermal AdS space-time at low temperatures. We shall briefly describe the phase behaviour in the different ensembles before proceeding to discuss the thermodynamic geometry of these black holes. As we have pointed out earlier, our presentation here is qualitatively different from that in the 
standard literature, and we find it more convenient to study the system using the parameters that will be used in the subsequent analysis of thermodynamic geometry 
for these black holes. 
 
In the canonical ensemble the internal energy is allowed to fluctuate while the electric charge 
is held fixed, so that the black hole remains in thermal equilibrium with the heat reservoir held at a constant temperature $t$. This ensemble is aptly described by 
the Hehlmoltz potential
\begin{equation}
\label{rnhelmo}
f(t,q)=m-t\,s=\frac{1}{4}\,{\frac {3\,{\pi }^{2}{q}^{2}+\pi \,s-{s}^{2}}{\sqrt {s}{\pi }^{3/2}}}
\end{equation}
The variables ($t, q$) in the argument of $f$ indicate the ``control'' parameters for the canonical ensemble, which can be tuned independent of each other.
In the grand canonical ensemble however the charge is unconstrained and the black hole is in a thermal as well as electrical equilibrium with its reservoir held at a constant temperature $t$ and a potential $\phi$. The Gibbs free energy for this ensemble is
 \begin{equation}
 \label{rnGibbs}
 g(t,\phi)=m-ts-\phi q=-\frac{1}{4}\,{\frac {{\pi }^{2}{q}^{2}-\pi \,s+{s}^{2}}{\sqrt {s}{\pi }^{3/2}}}
\end{equation}
where ($t,\phi$) indicate the independent control parameters for this ensemble, even though we have expressed it in terms of $s$ and $q$.
\begin{figure}[t!]
\begin{minipage}[b]{0.5\linewidth}
\centering
\includegraphics[width=3in,height=2.5in]{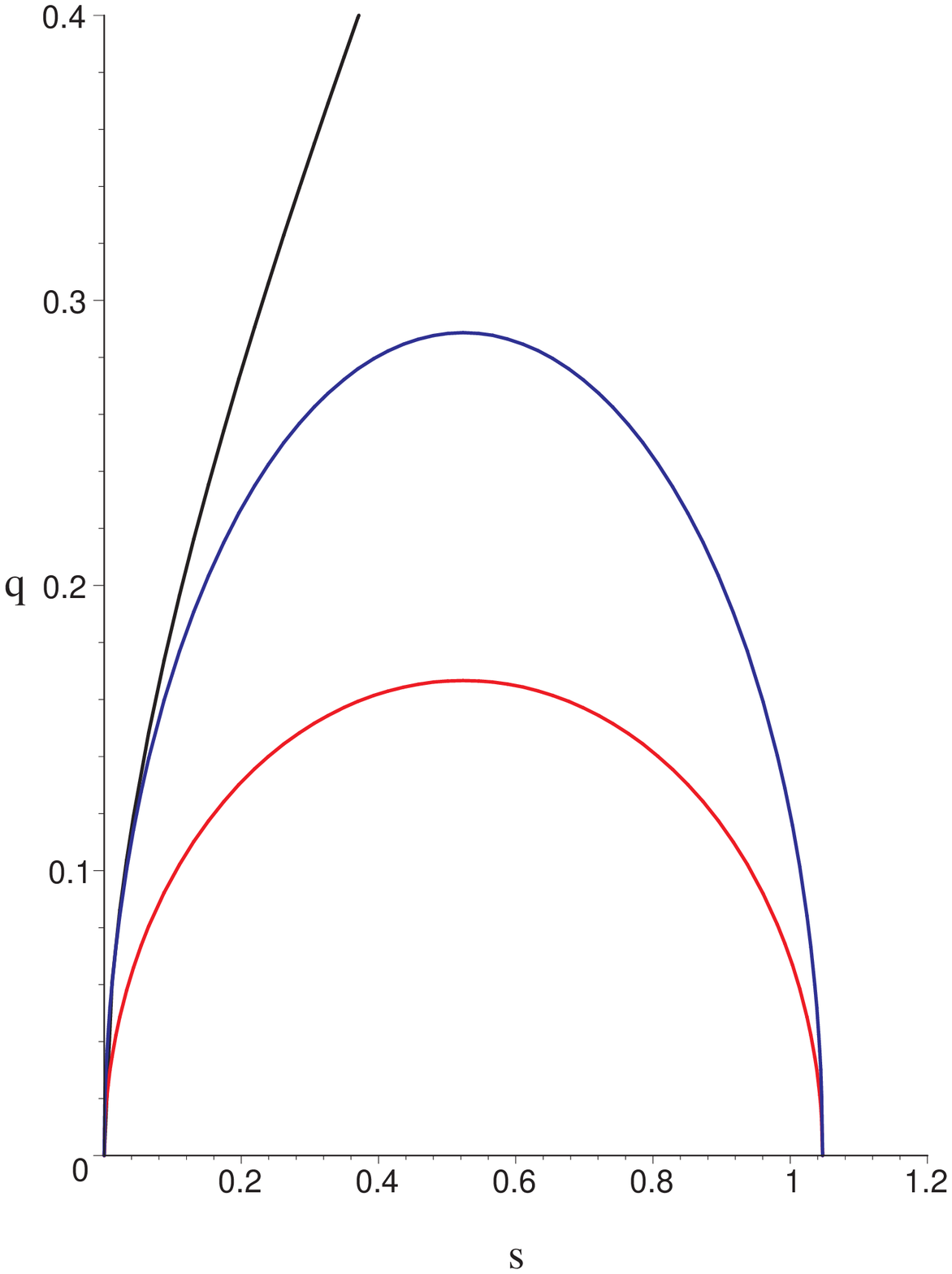}
\caption{Stability curves of RN-AdS black holes. Heat capacities $c_q$ and $c_{\phi}$ diverge along the upper blue and the lower red semi circular 
curves respectively. The extremal curve is black colored.}
\label{rnads1}
\end{minipage}
\hspace{0.6cm}
\begin{minipage}[b]{0.5\linewidth}
\centering
\includegraphics[width=2.7in,height=2.7in]{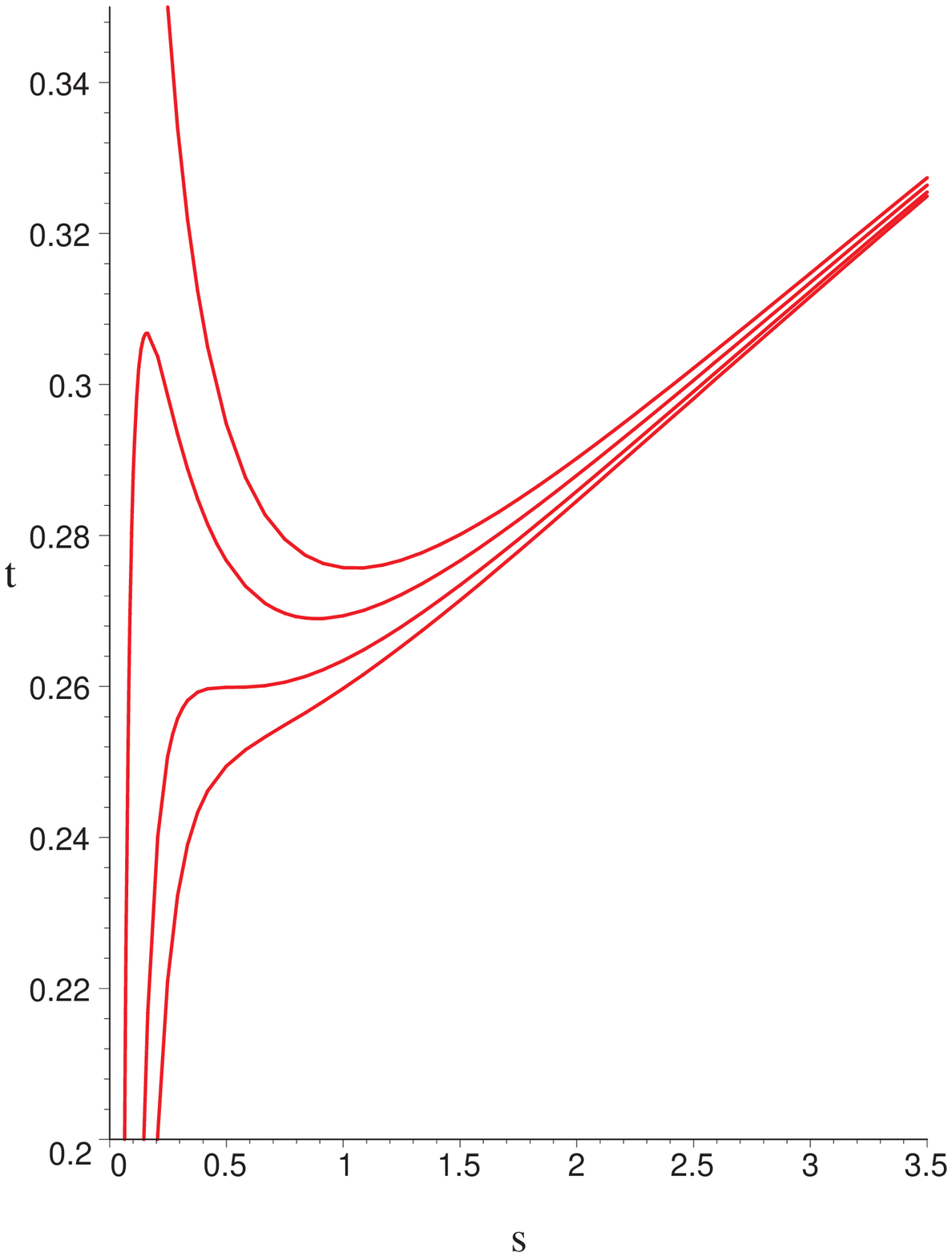}
\caption{Isocharge plots of $t$ vs. $s$ with  $q=0,\,0.12,\,0.167\, (q_c)$, and $0.19$ from the top to bottom. The critical isocharge curve in the middle shows an inflection at $s_c={{\pi}}/{6}$ and $t_c={\sqrt 3}/{{\sqrt 2}\pi}$}
\label{rnads2}
\end{minipage}
\end{figure}

In order to describe the phase behaviour in the two ensembles we first present the relations between $q$ and $s$ obtained by solving for the divergences in the heat capacities. For the divergence in $c_q$ the electric charge may be expressed as an implicit function of $s$ as,
\begin{equation}
q_{1}(s)={\frac {\sqrt {s(\pi -3\,s})}{\sqrt{3}\pi }}~~ ,
\end{equation}
whereas, along the divergence of $c_{\phi}$, we obtain
\begin{equation}
\label{rncphidiv}
q_{2}(s)={\frac {\sqrt {s(\pi -3\,s})}{\pi }}
\end{equation}
The stable and the unstable regions in the canonical ensemble are separated by the locus of the divergences of the heat capacity $c_q$ while those in the grand canonical 
ensemble are separated by the divergences of the heat capacity $c_{\phi}$.
In fig.(\ref{rnads1}) we plot the curves  $q_{1}(s)$ (red) and $ q_{2}(s)$ (blue) in the $q$-$s$ plane. The heat capacities $c_q$ and $c_{\phi}$ are negative inside and positive outside their respective curves. We shall refer to the stability curve corresponding to the canonical ensemble as the $c_q$-spinodal curve and the 
one corresponding to the grand canonical ensemble as the $c_{\phi}$-spinodal curve. The black curve representing $t=0$ separates the naked singularity 
region on the left from the physical region on the right. Expectedly, the canonical ensemble being the more constrained one, has a greater region of stability 
than the grand canonical ensemble.

From fig.(\ref{rnads1}) we can deduce the phase coexistence behaviour in the canonical ensemble as follows. Any constant $q$ line intersects the $c_q$-spinodal
 curve twice for $q<q_{c}(=1/6)$, thereby separating the small black hole branch from the large black hole branch by an unstable branch. This phase 
 coexistence behaviour terminates at the critical charge, $q_c=1/6$, where the constant $q$ line becomes tangent to the $c_q$-spinodal curve, thus effecting a 
 continuous second order phase transition between the small black hole and the large black hole. We shall discuss the critical exponents for the black hole in a canonical 
ensemble in brief in a later section where we present the critical behaviour for other ensembles.

In fig.(\ref{rnads2}) we present isocharge plots in the $t$-$s$ plane for various fixed values of charge. The $q=0$ curve at the top, corresponding to the 
AdS-Schwarzschild black hole, is  characterized by a finite temperature minima or
a turning point, below which there is no black hole solution and above which the unstable and stable branches coexist at all temperatures. We shall term this 
phase behaviour as ``Davies'' phase behaviour and the turning point temperature as the Davies temperature, $t_d$. This phase behaviour will turn out to be 
quite generic, as will be seen in later sections. Addition of a constant charge $q$ causes a low temperature small black hole branch to form, or, in other words, 
it causes a ``deconfinement'' at all temperatures, \cite{johnson1}. On increasing the charge to $q_c$, the $t$-$s$ curve shows an inflection at the critical point.

 We now turn our attention to the black hole in a grand canonical ensemble, which is the one relevant for analysis in the framework
 of  the thermodynamic geometry.  In this ensemble the thermal AdS space 
 with a constant pure gauge potential equal to that of the RN-AdS black hole can serve as a reference background (i.e, the zero of the Gibbs potential) since it 
 is also a solution to the Einstein equations, (\cite{johnson1}, \cite{calda}, \cite{johnson2}). Therefore, while for negative values of $g$ the black holes are globally 
 stable, for positive values of $g$, they become unstable to the formation of thermal AdS via a Hawking-Page transition. The locus of zeroes of the Gibbs 
 potential in the $q$-$s$ plane can be obtained from the relation
 \begin{equation}
 \label{rnzeroGibbs}
 q_3(s)={\frac {\sqrt {s \left( \pi -s \right) }}{\pi }}
 \end{equation}
 It will also be convenient to express the temperature as a function of $s$ and $\phi$ using eqs. (\ref{rntemp}) and (\ref{rnphi}).
 \begin{equation}
 \label{rntgc}
 t=\frac{1}{4}\,{\frac {\pi +3\,s-\pi \,{\phi}^{2}}{\sqrt {s}{\pi }^{3/2}}}
 \end{equation}
 Using this expression for the temperature it may be verified that the extremal black holes exist only for $\phi \geq 1$. Eliminating $q$ from eq.(\ref{rnphi}) and eq.(\ref{rnzeroGibbs}) and then using eq.(\ref{rntgc}), the Hawking-Page phase transition temperature may be expressed in terms of the potential $\phi$ as
 \begin{equation}
 \label{rnthp}
 t_{hp}={\frac {\sqrt {1-{\phi}^{2}}}{\pi }}
 \end{equation}
\begin{figure}[t!]
\begin{minipage}[b]{0.5\linewidth}
\centering
\includegraphics[width=3in,height=2.5in]{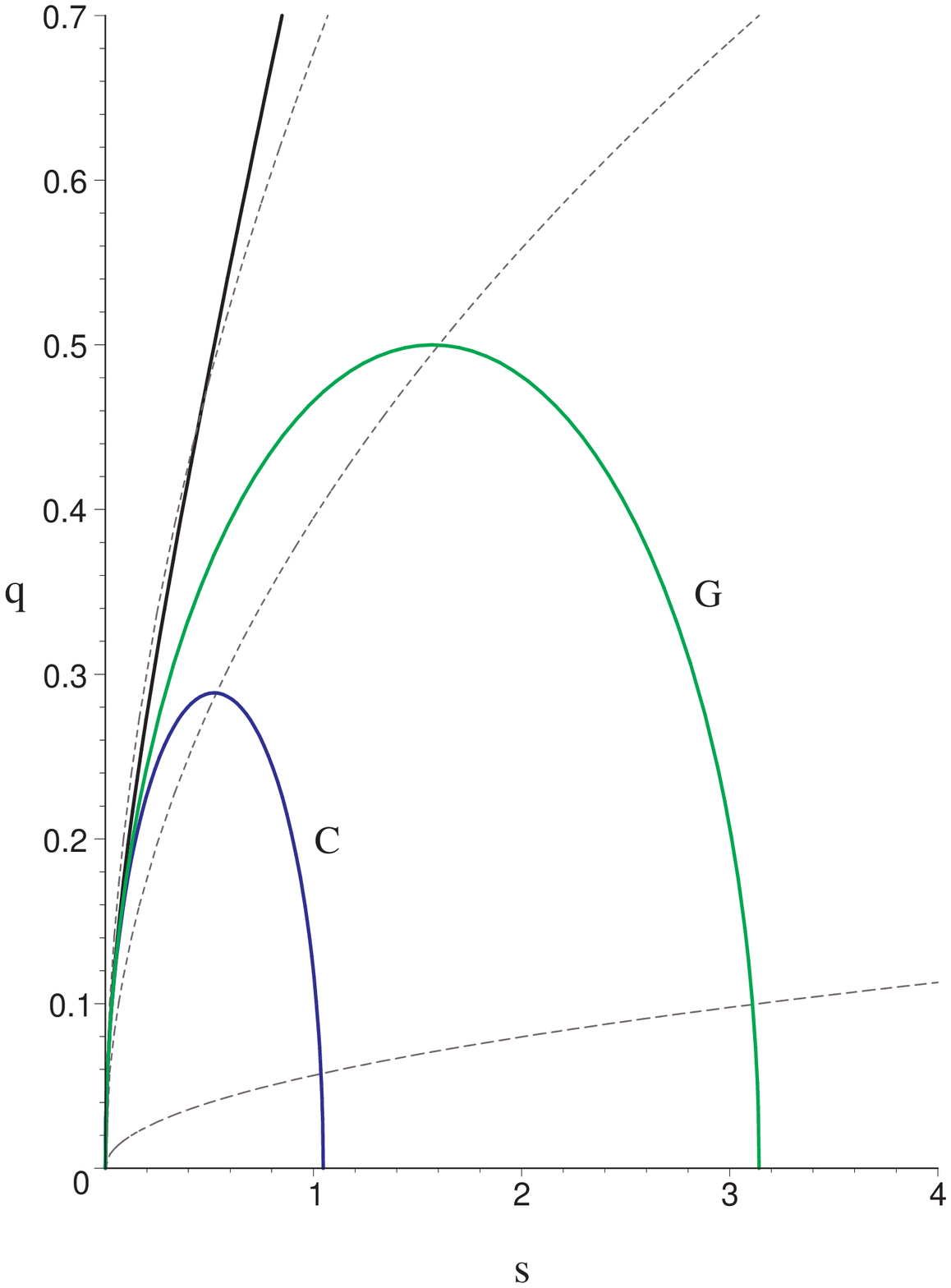}
\caption{RN-AdS plots showing $c_{\phi}$-spinodal curve in blue, zeroes of Gibbs free energy in green and isopotential curves in dotted grey at $\phi = 0.1$, $0.7$, and $1.2$ from bottom to top. Extremal curve is black in color..}
\label{rnads3}
\end{minipage}
\hspace{0.6cm}
\begin{minipage}[b]{0.5\linewidth}
\centering
\includegraphics[width=2.7in,height=2.7in]{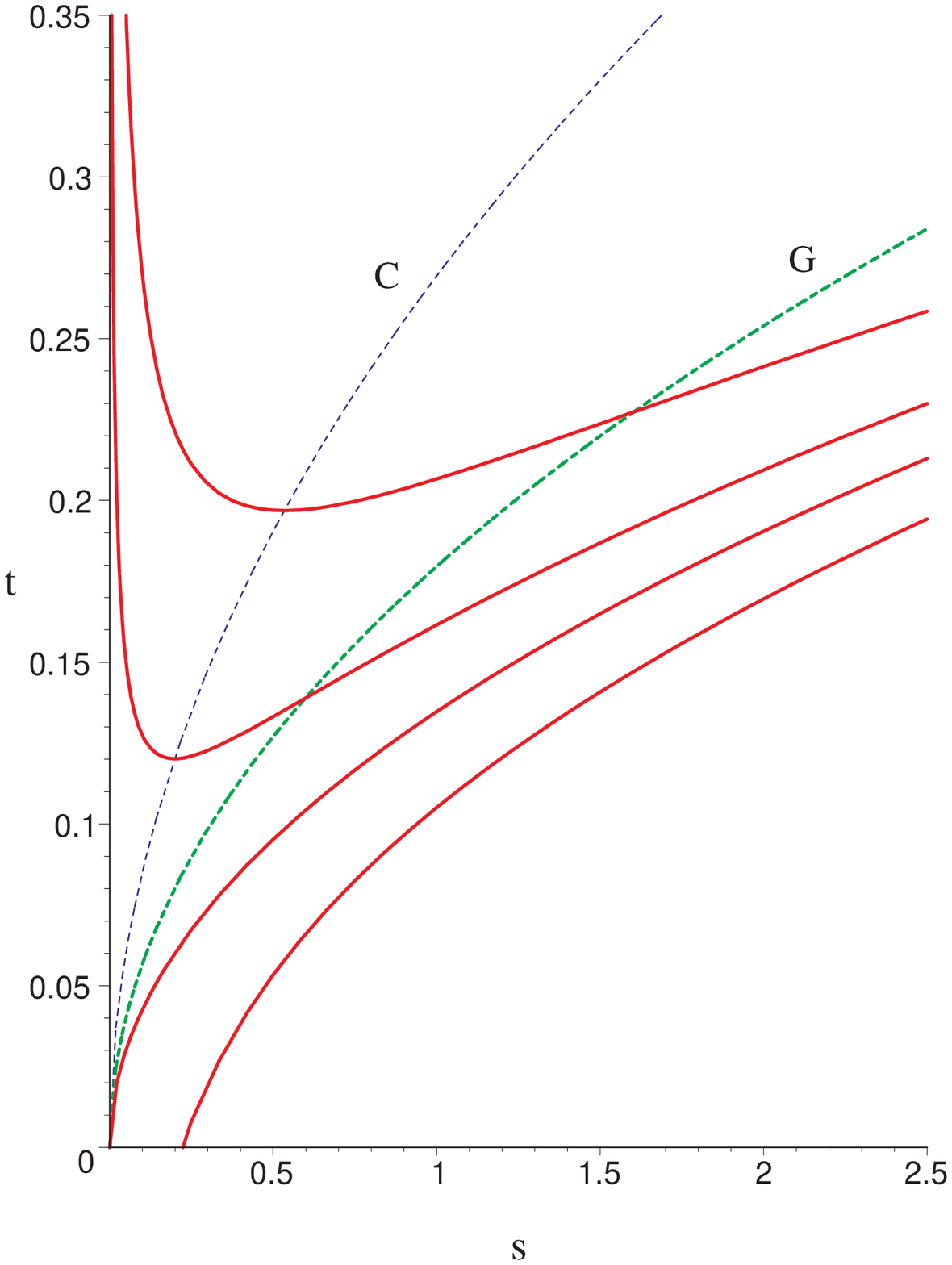}
\caption{Plot showing isopotential curves in the $t-s$ plane of the RN-AdS black holes with $\phi = 0.7$, $0.9$, $1$ and $1.1$ from top to bottom. 
The dotted green curve and dotted blue curve above it are the Gibbs and $c_{\phi}$ curves of fig.(\ref{rnads3}). }
\label{rnads4}
\end{minipage}
\end{figure}
In fig.(\ref{rnads3}) we show the grand canonical stability curve, i.e the $c_{\phi}$ spindodal curve, together with the green coloured  ``Gibbs curve''  representing the 
zeroes of the Gibbs potential. The two curves are labelled as ``C'' and ``G'' respectively. The three grey dotted curves represent isopotential curves, with the topmost 
one at a value of $\phi$ greater than one. In the region bounded by the Gibbs curve, the black hole is unstable against AdS as its free energy is positive. 
Since the $c_{\phi}$-spinodal curve lies fully inside the Gibbs curve, we can say that the locally unstable black hole solutions are also always globally unstable. Thus
 we see that in the region bounded by the $c_{\phi}$ curve the black hole solutions exist but they are unphysical, whereas in the region lying in between 
 the $c_{\phi}$ curve and the Gibbs curve metastable black holes exist. Outside the Gibbs curve , the black holes are both locally as well as globally stable. 
 In fig.(\ref{rnads4}), we show isopotential plots in red of $t$ vs. $s$ for increasing values of $\phi$ from top to bottom, with the green dotted curve and the blue 
 dotted curve being the Gibbs curve and the $c_{\phi}$-spinodal curve respectively of fig.(\ref{rnads3}). For the top two isopotentials, having ${\phi}< 1$, the temperature 
 shows a turning point when the isopotentials cross the $c_{\phi}$ curve. This is exactly the Davies phase transition behaviour mentioned earlier in the
 context of AdS-Schwarzschild black holes. The turning point 
 temperature or the Davies temperature $t_d$, can be obtained in terms of the potential $\phi$ by eliminating $s$ from eq.(\ref{rnphi}) and eq.(\ref{rncphidiv}), 
 and then using eq.(\ref{rntgc}),

\begin{equation}
\label{rntb}
t_{d}=\frac{\sqrt{3}}{2}\,{\frac {\sqrt {1-{\phi}^{2}}}{\pi }}=\frac{\sqrt{3}}{2}\,t_{hp}
\end{equation}

 Thus, for $\phi<1$, only the thermal AdS exists for $0<t<t_d$, while a metastable black hole exists along with thermal AdS for $t_d<t<t_{hp}$. For $t>t_{hp}$ the stable black hole is preferred over the thermal AdS space-time. The lowest isopotential in fig.(\ref{rnads4}) is a typical one with ${\phi}>1$, for which there is only a single black hole branch which is locally as well as globally stable at all temperatures. The extremal entropy for these isopotentials can be obtained from eq.(\ref{rntgc}), as
 \begin{equation}
 s_{ex}(\phi)=\frac{1}{3}\,\pi \, \left( {\phi}^2-1 \right)
 \label{rnsextgc}
 \end{equation}
 For ${\phi}=1$ the black hole has a vanishing entropy at zero temperatures. Having discussed the phase behaviour of RN-AdS black holes in terms of the
 parameters suitable for describing the geometry of their equilibrium state space, we now turn to the discussion of the latter. 

\subsection{State Space Scalar Curvature for RN-AdS Black Holes }
 
 In this subsection, we will investigate the phase behaviour of RN-AdS black holes using thermodynamic geometry. The topic of thermodynamic geometry
 is by now well known, and we refer the reader to \cite{rupp} for a comprehensive review, and to \cite{paper} for the details relevant to black hole thermodynamics.
The thermodynamic line element for the RN-AdS black holes is a dimensionless quantity symbolically expressed as
 \begin{equation}
 \label{rnline}
 dl^2=g_{\mu\nu}dx^{\mu}dx^{\nu}=g_{mm}dm^2+2g_{mq}dmdq+g_{qq}dq^2,
 \end{equation}
 where the indices $\mu, \nu$ in the metric $g_{\mu\nu}$ range over the rescaled variables $m$ and $q$. We note that thermodynamic geometry cannot be 
 meaningfully defined in regions of the equilibrium state space where the thermodynamic line element is negative, as the interpretation of the thermodynamic 
 length between two neighbouring states as the measure of the probability of thermal fluctuation between them would then break down. The constraint of a positive 
 definite line element requires that the deteminant formed by the metric $g_{\mu\nu}$ as well as its prinicipal minor, $g_{mm}$ in this case, be 
 positive definite, \cite{rupp}. This is the familiar Le Chatelier's condition for local thermodynamic stability. It can be verified that this condition is satisfied for all 
 regions lying outside the $c_{\phi}$-spinodal curve in fig.(\ref{rnads3}), i.e, in regions where the black hole is locally stable.

The state space scalar curvature pertaining to the thermodynamic geometry for the RN-AdS black holes, first calculated in \cite{aman}, is re-expressed in the scaled thermodynamic variables as
\begin{equation}
\label{rnR}
R = {\frac {-9s \left( 3\,{s}^{2}+{q}^{2}{\pi }^{2} \right)  \left( {s}^{2}-s\pi +{q}^{2}{\pi }^{2} \right) }
{ \left( -3\,{s}^{2}-s\pi +{q}^{2}{\pi }^{2} \right)  \left( 3\,{s}^{2}-s\pi +{q}^{2}{\pi }^{2} \right) ^{2}\\
\mbox{}}}
\label{curv}
\end{equation}

In what follows, we will elucidate  novel interpretations of the curvature in terms of its zeroes and divergences in relation to the spinodal curves alluded to earlier. The analysis that follows from this perspective is entirely new and provides interesting insights
into the behaviour and significance of the state space scalar curvature. We begin by noting that the curvature can be symbolically written in the following suggestive manner
\begin{equation}
\label{rnRsmb}
R =-9s \left( 3\,{s}^{2}+{q}^{2}{\pi }^{2} \right)\frac{{\mathcal{N}}(g)}{\mathcal{N}(t){\mathcal{D}(c_{\phi})}^2}
\end{equation}
where the symbols $\mathcal{N}(g)$ $\mathcal{N}(t)$ and $\mathcal{D}(c_\phi)$ represent the numerators of  the Gibbs potential and the temperature,
and the denominator of $c_{\phi}$ respectively.  
This clearly shows that $R$ diverges at extremality and along the $c_{\phi}$-spinodal curve, both of which are at the boundary of the thermodynamically 
stable physical region. Interestingly, it goes to zero along the ``Gibbs curve'' of fig.(\ref{rnads3}) and has opposite sign of the Gibbs free energy everywhere. 
Let us see if we can substantiate this. 

Note that the fact that $R$ diverges along the $c_{\phi}$-spinodal curve is
quite reasonable as the scalar curvature pertains to the two dimensional  state space Riemannian geometry which is directly related to the fluctuations in both the thermodynamic variables $m$ and $q$. This would imply 
that $R$ inherits the instabilities of the grand canonical ensemble, as reflected in its divergence along the $c_{\phi}$-spinodal curve. In fact, a similar argument 
can explain the divergence at extremality, where the heat capacity changes sign on crossing to the unphysical naked singularity region. However, we believe that
the issue of extremality in thermodynamic geometry is subtle and we will ignore this divergence in our future discussions.

Since $R$ naturally describes the phase behaviour in the grand canonical ensemble, it would be convenient to re-express it in terms of the potential $\phi$ and $t$. 
To that end, we first express  $R$ in terms of $\phi$ and $s$, using eq.(\ref{rnphi}) and eq.(\ref{rnR}), as follows:
\begin{equation}
\label{rnRphi}
R(\phi,s)=-9\,{\frac {{\pi }^{2}{\phi}^{4}-{\pi }^{2}{\phi}^{2}+4\,\pi \,{\phi}^{2}s-3\,\pi \,s+3\,{s}^{2}}{ \left( \pi \,{\phi}^{2}-\pi -3\,s
 \right)  \left( 3\,s+\pi \,{\phi}^{2}-\pi  \right) ^{2}}}
\end{equation}
Solving eq.(\ref{rntemp}) for the entropy in terms of temperature, we obtain two solutions corresponding to the stable branch and the unstable branch. The 
stable branch solution is
\begin{equation}
\label{rnstbr}
s=\frac{1}{3}\,\pi \,{\phi}^{2}-\frac{1}{3}\,\pi +\frac{4}{3}\,t \left( \frac{2}{3}\,t{\pi }^{3/2}+\frac{1}{3}\,\sqrt {4\,{t}^{2}{\pi }^{3}+3\,\pi \,{\phi}^{2}-3\,\pi } \right) {\pi
}^{3/2}
\end{equation}
On substituting the expression for $s$ from eq.(\ref{rnstbr}) into eq.(\ref{rnRphi}) we obtain an expression for $R$ for the positive-definite region in 
terms of $\phi$ and $t$ as
\begin{eqnarray}
\label{rnRphit}
&&R(\phi,t)=\nonumber\\\nonumber\\
\hspace{-0.1in}&&\frac{9}{4}\,{\frac {\phi^2\left(18{\phi}^{2}+48t^2\pi^2-27 + 18\pi tu\right)-42{\pi }^{2}{t}^{2}-15t{\pi}u+32{
t}^{4}{\pi }^{4}+16{\pi }^{3}{t}^{3}u + 9}{{\pi }^{2}t \left( 2{\pi}t
+u \right)  \left( 3{\phi}^{2}-3+4{\pi }^{2}{t}^{2}+2t{\pi}u
 \right) ^{2}}}\nonumber\\
\end{eqnarray}
where the expression $u$ is given by
\begin{eqnarray*}
u=\sqrt {4\,{\pi }^{2}{t}^{2}+3\,{\phi}^{2}-3}
\end{eqnarray*}
\begin{figure}[t!]
\begin{minipage}[b]{0.3\linewidth}
\centering
\includegraphics[width=1.7in,height=1.7in]{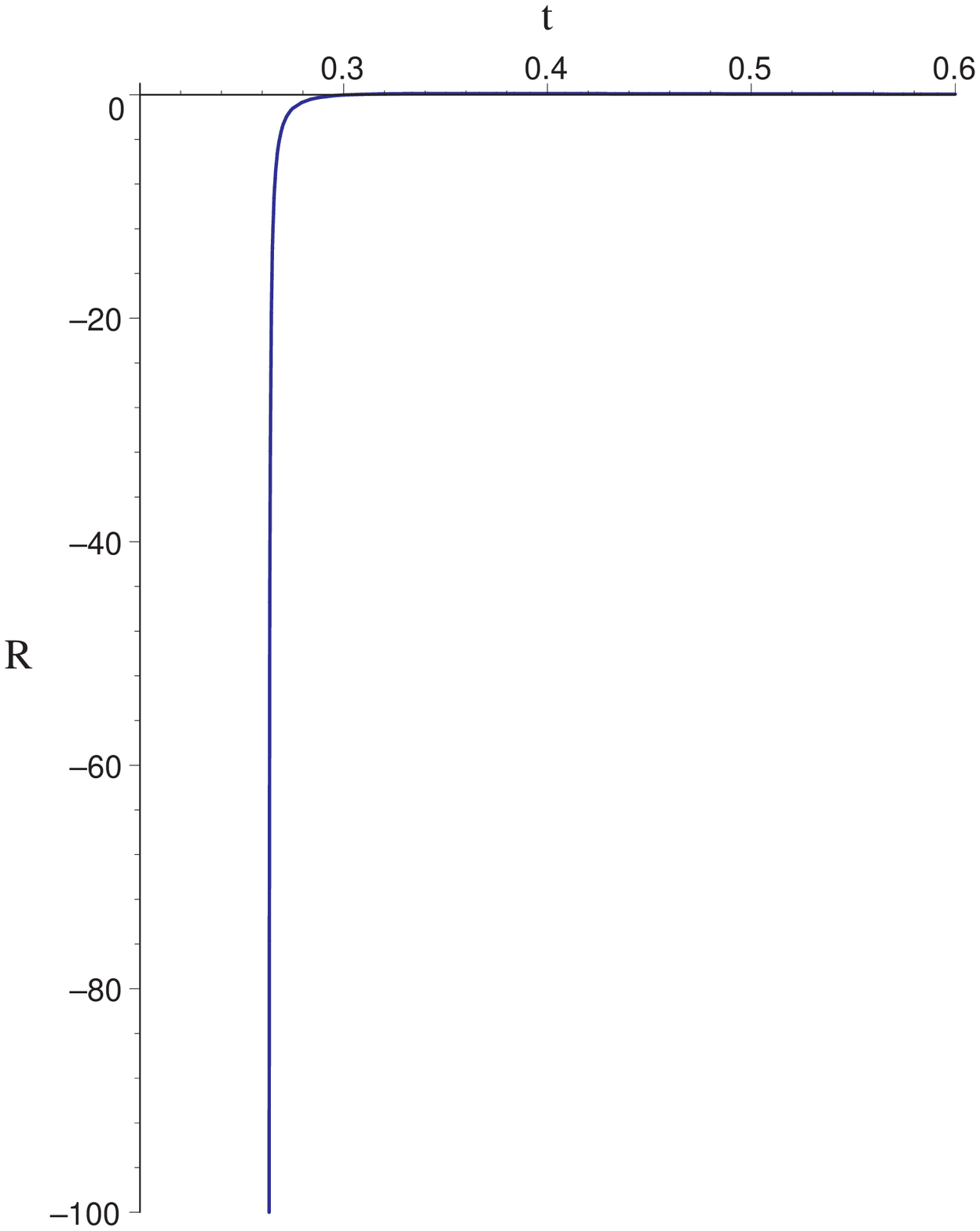}
\caption{Isopotential plot of $R$ vs. $t$ with $\phi$ fixed at $0.03$. $R$ has a negative divergence at $t_b=0.263$}
\label{rnR1}
\end{minipage}
\hspace{0.2cm}
\begin{minipage}[b]{0.3\linewidth}
\centering
\includegraphics[width=1.7in,height=1.7in]{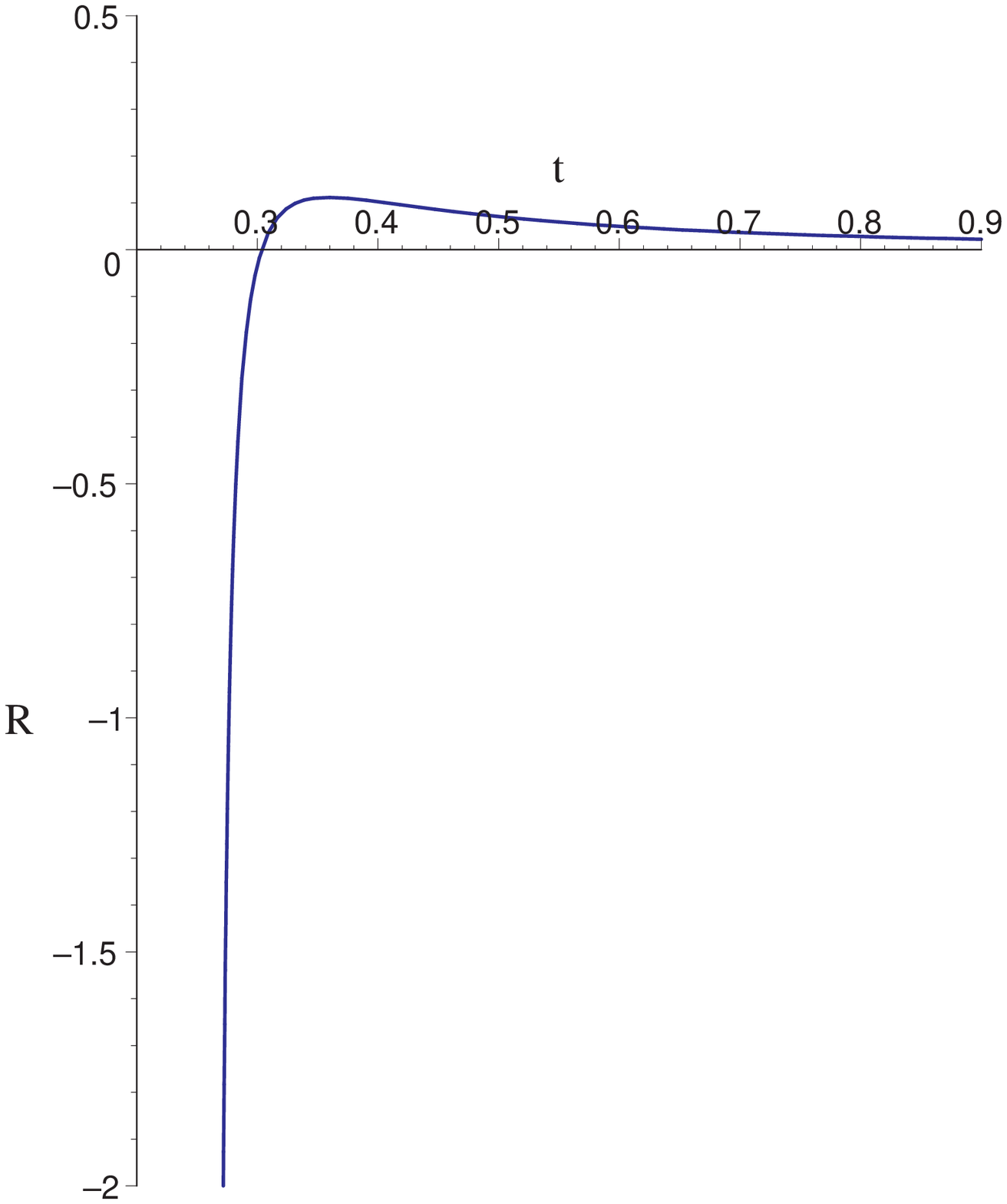}
\caption{Close-up of fig.(\ref{rnR1}). $R$ crosses zero and becomes positive at the Hawking-Page temperature $t_{hp}=0.304$.}
\label{rnR2}
\end{minipage}
\hspace{0.2cm}
\begin{minipage}[b]{0.3\linewidth}
\centering
\includegraphics[width=1.7in,height=1.7in]{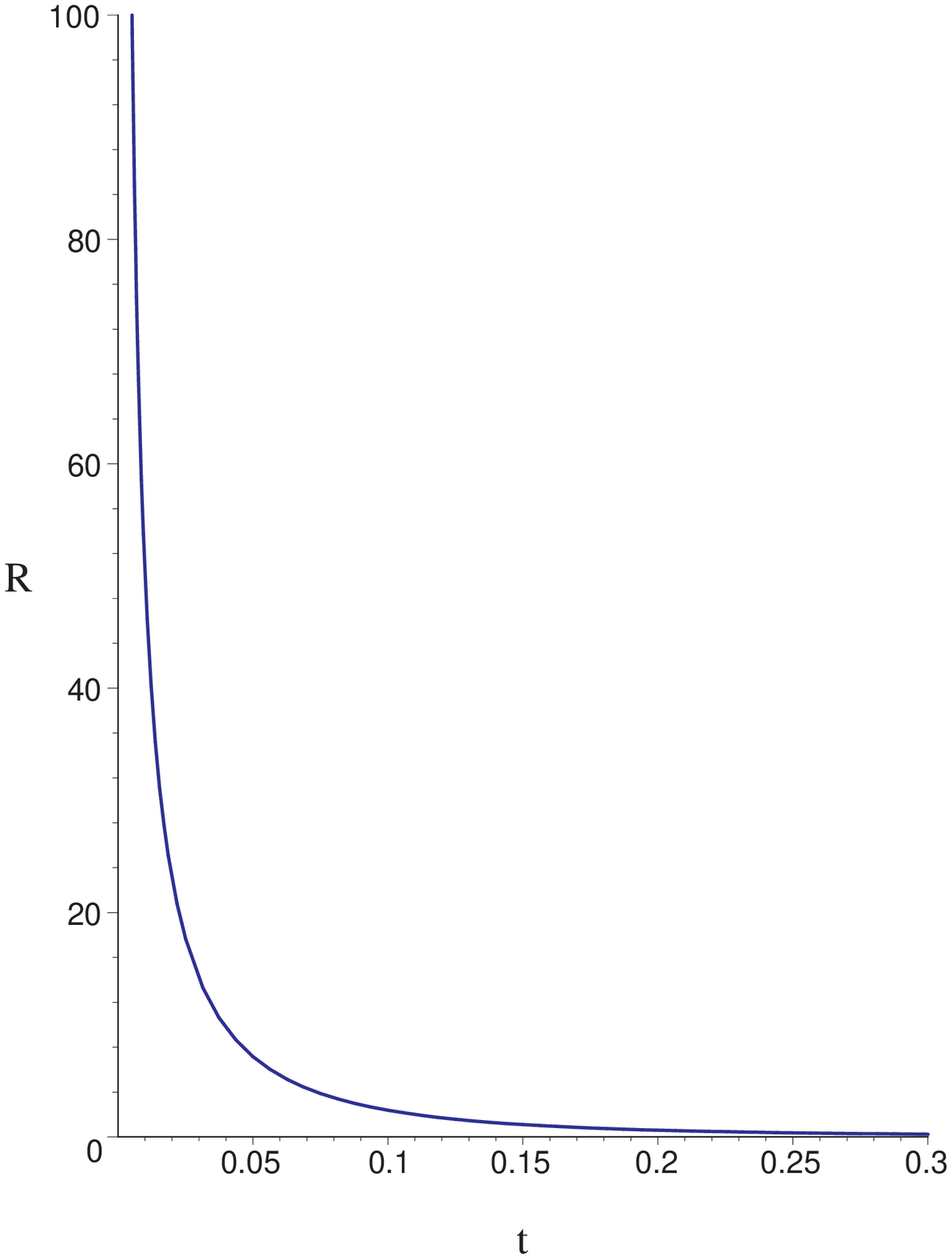}
\caption{Isopotential plot of $R$ vs. $t$ with $\phi$ fixed at $1.3$. $R$ diverges at $t$=$0$ and remains positive at all $t$.}
\label{rnR3}
\end{minipage}
\end{figure}
We now investigate $R$, as obtained in eq.(\ref{rnRphit}), graphically by plotting $R$ vs. $t$ for various potentials. These plots pertain to the RN-AdS
black hole in a grand canonical ensemble. 
Fig.(\ref{rnR1}) is a generic plot of $R$ vs. $t$ at a potential $\phi<1$, and fig.(\ref{rnR2}) is a close-up of the same near the zero crossing of $R$. We clearly see 
that $R$ shows a negative divergence at the Davies temperature $t_d$, eq.(\ref{rntb}), corresponding to the turning point of the isopotential plots like the ones 
shown in fig.(\ref{rnads4}). 

The scalar curvature remains negative for the metastable phase of the black hole, changing sign to positive at the Hawking-Page 
temperature $t_{hp}$, eq.(\ref{rnthp}). Fig.(\ref{rnR3}) shows a generic $R$ vs. $t$ plot for RN-AdS black holes with $\phi\geq 1$. From fig.(\ref{rnads4}) it can be 
seen that these black holes exist at all temperatures and are locally as well as globally stable. The curvature mirrors this by remaining positive at all temperatures. 
Further, it can be seen from eq.(\ref{rnRphit}) that for all values of the potential, in the limit of large $t$, $R$ asymptotes to zero from the positive side as,
\begin{equation}
\label{rnRlimit}
R\,{\sim}\,\frac{1}{t^2}
\end{equation}
Eq. (\ref{rnRphit}) and the analysis presented thereafter are the main results of this subsection. We will have more to say about the issue of sign of the
state space curvature for RN-AdS black holes once we have discussed the phase behaviour and thermodynamic geometry for the 
Kerr-AdS black holes, to which we presently turn.

\section{Kerr-AdS Black Holes}

In this section, we begin by reviewing certain aspects of the thermodynamics of Kerr-AdS black holes, before we discuss the geometry of the equilibrium state
space of the same. As earlier, our analysis will be directly in terms of the thermodynamic parameters which are relevant for the analysis of the state space geometry. 
The Kerr-Ads black holes are thermodynamically characterized by their mass and the angular momentum. On setting $q$=$0$ in eq.(\ref{reducedsmarr}) we obtain the rescaled Smarr relation 
for the Kerr-AdS black hole as,
\begin{equation}
\label{krsmarr}
m=\frac{1}{2}\,{\left({\frac { {{s}^{2}{\pi }^{2}+4\,{\pi }^{4}{j}^{2}+4\,{j}^{2}{\pi }^{3}s+2\,{s}^{3}\pi +{s}^{4}}}{ {s}{\pi }^{3}}}\right)}^{\frac{1}{2}}
\end{equation}
The temperature, $t$, and the angular velocity, $\omega$, are obtained by differentiating the expression for mass or equivalently by setting $q=0$ in the
corresponding formulae for the KN-AdS black holes in section (2.1). 

The condition for extremality obtained by setting the numerator of $t$ to zero leads to,
\begin{equation}
\label{krext}
j_{ex}(s)=\frac{s}{2{\pi }^{2}}\,({\pi }^{2}+4\,\pi \,s+3\,{s}^{2})^{1/2}
\end{equation}
We note that the angular velocity $\omega$ that enters into the thermodynamics through the differentiation of  the Smarr formula is not identical to the angular velocity at the horizon, ${\omega}_h$. Instead, it turns out to be the difference in the angular velocity at horizon and that at the boundary of the spacetime, $\omega={\omega}_h-{\omega}_{\infty}$, \cite{calda},\footnote{i.e, it is measured with respect to a non-rotating frame at infinity, \cite{gibb}. } This also turns out to be the angular velocity of the rotating Einstein universe existing at the boundary of  the asymptotically AdS space-time. This agrees well with the AdS/CFT picture and also shows that consistent thermodynamics up to infinity can be defined only when $\omega <1$. A useful quantity for our calculations in Kerr-AdS case is the expression relating $j$ to $\omega$ and $s$
\begin{equation}
\label{krjom}
j(\omega,s)=\frac{1}{2}\,{\frac {\omega{s}^{3/2}\sqrt {\pi +s}}{{\pi }^{3/2}\sqrt {\pi +s-{\omega}^{2}s}}}
\end{equation}
The expressions for the heat capacities at constant $j$ and constant $\omega$ are lengthy, so we shall just write down the stability conditions corresponding 
to these heat capacities. The canonical ensemble stability constraint is obtained from the locus of  the divergences of  the specific heat $c_j$,
\begin{equation}
\label{krj1}
j_{1}(s)=\frac{s}{2{\pi }^{2}}\,{\frac {({\pi +s})^{1/2} (-3\,{\pi }^{2}-9\,\pi \,s+2\,Y-6\,{s}^{2})^{1/2}}{(3\,\pi +4\,s)^{1/2}}}
\end{equation}
where $Y$ is given by the expression
\begin{equation}
Y=(3\,{\pi }^{3}+10\,s{\pi }^{2}+15\,{s}^{2}\pi +9\,{s}^{3})^{1/2}(\pi+s)^{1/2}
\end{equation}
Similarly, the stability condition for the grand canonical ensemble is obtained form the divergences of the heat capacity $c_{\omega}$,
\begin{equation}
\label{krj2}
j_2(s)=\frac{s}{2{\pi }^{2}}\,({\pi +s})^{1/2}({2\,s^{1/2}\sqrt {\pi +s}-\pi -s})^{1/2}
\end{equation}
Just like in the RN-AdS case, the Kerr-AdS black hole in the grand canonical ensemble also has a well defined reference background in  the rotating thermal 
AdS with a zero Gibbs potential and angular velocity equal to that of the boundary angular velocity of the Kerr-AdS spacetime. For positive values of the free energy,
the rotating thermal AdS space is more stable then the Kerr-AdS black hole, while the stability is reversed via a Hawking-Page phase transition when the free 
energy changes sign and becomes negative. The Gibbs free energy in the grand canonical ensemble may be obtained through the relation
\begin{equation}
\label{krg}
g=m-ts-{\omega}j
\end{equation}
 It can be checked that the locus of the Hawking -Page phase transitions, or the zeroes of $g$ are provided by the implicit relation
 \begin{equation}
 \label{krgzero}
 j_3(s)=\frac{s}{2{\pi }^{2}}\,( {{s}^{2}-{\pi }^{2}})^{1/2}
\end{equation}

\begin{figure}[t!]
\begin{minipage}[b]{0.5\linewidth}
\centering
\includegraphics[width=3in,height=2.5in]{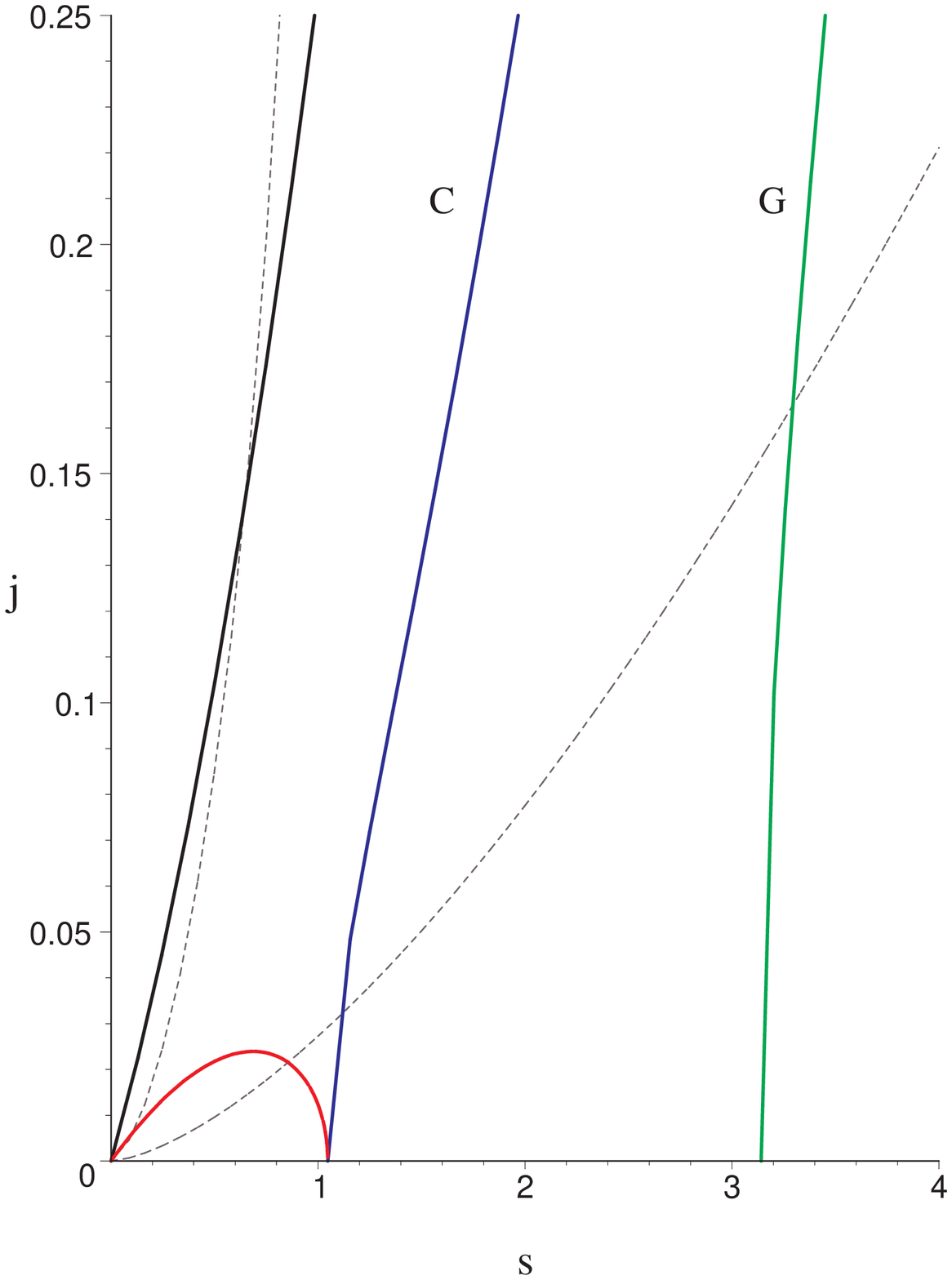}
\caption{Kerr-AdS plots in $s$-$j$ plane. From left to right in order are the black extremal curve, red semi circular $c_j$-spinodal curve, the blue $c_{\omega}$-spinodal curve and the green Gibbs curve. Two grey dottted isopotentials have $\omega$=$0.3$ (lower) and $\omega$=$1.9$ (upper).}
\label{krads1}
\end{minipage}
\hspace{0.6cm}
\begin{minipage}[b]{0.5\linewidth}
\centering
\includegraphics[width=2.7in,height=2.7in]{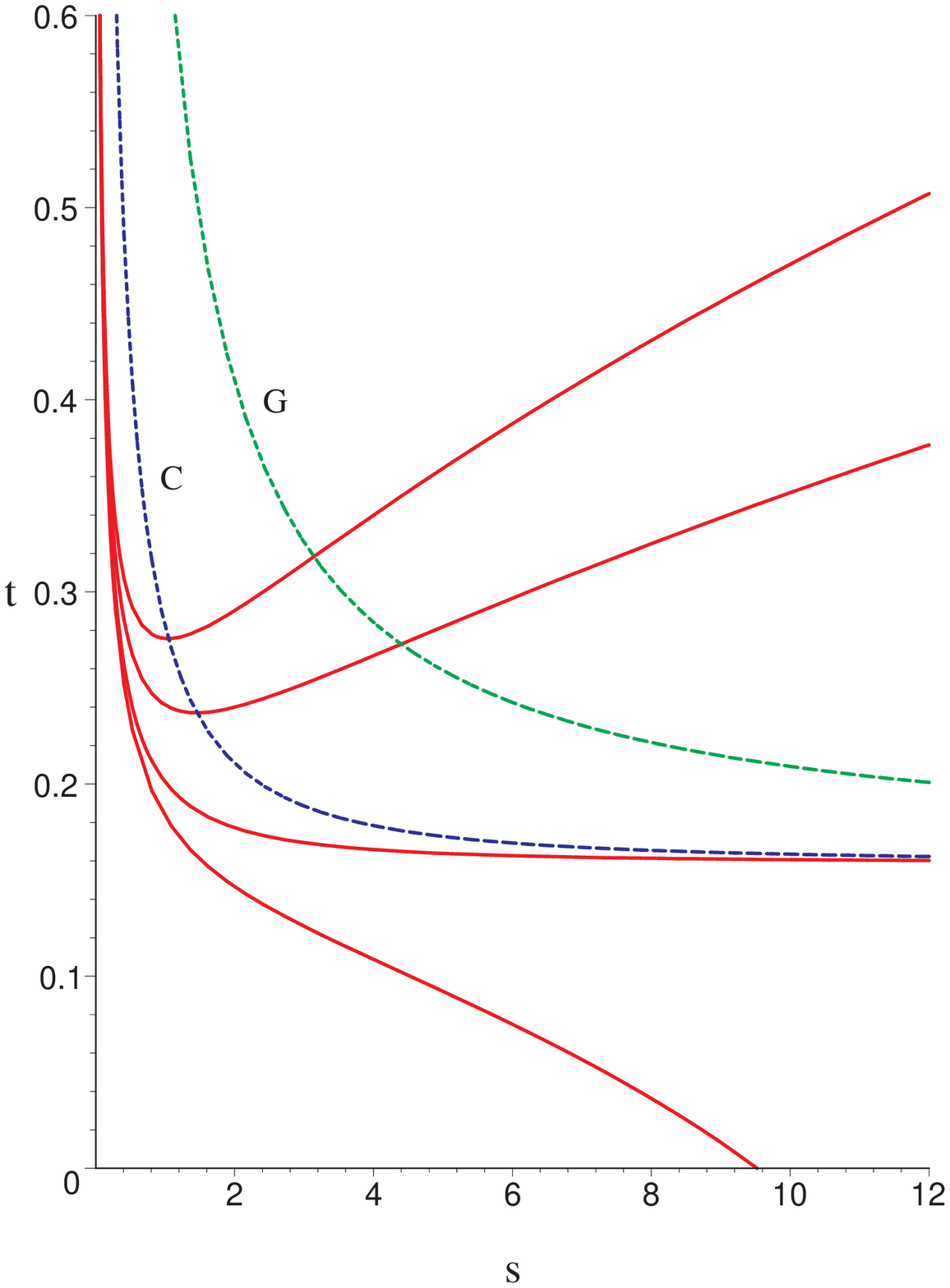}
\caption{$t$ vs. $s$ plots for Kerr-AdS black holes showing isopotential curves in red colour, with the angular velocity $\omega$ ranging from $0$, $0.7$, $1$ and $1.1$ in order from top to bottom. The dotted green curve and dotted blue curve below it are the Gibbs and $c_{\omega}$ curves of fig.(\ref{krads1}). }
\label{krads2}
\end{minipage}
\end{figure}

Now, keeping in mind that the thermodynamic geometry analysis will pertain to the grand canonical ensemble, we express the temperature $t$ and the 
Gibbs free energy $g$ in terms of $\omega$ and $s$ by making use of eq.(\ref{krjom}). These expressions turn out to be 
\begin{equation}
\label{krtom}
t(\omega,s)=\frac{1}{4}\,{\frac {{\pi }^{2}+4\,\pi \,s-2\,\pi \,{\omega}^{2}s+3\,{s}^{2}-3\,{\omega}^{2}{s}^{2}}{s^{1/2}({\pi +s})^{1/2}({\pi +s-{\omega}^{
2}s})^{1/2}{\pi }^{3/2}}} 
\end{equation}
\begin{equation}
\label{krgom}
g(\omega,s)=\frac{1}{4}\,{\frac {\sqrt {s} \left( {\pi }^{2}+{\omega}^{2}{s}^{2}-{s}^{2} \right) }{\sqrt {\pi +s}\sqrt {\pi +s-{\omega}^{2}s}{\pi }^{3/2}}}
\end{equation}
Eliminating $s$ from eq.(\ref{krgom}) and eq.(\ref{krtom}), we obtain the Hawking-Page temperature as a function of $\omega$,
\begin{equation}
t_{hp}=\frac{1}{2\pi}\,\frac {(1+A+2\,\sqrt{A})}{( \sqrt{A}+1)}
\end{equation}
where $A=1-{\omega}^2$.

In fig.(\ref{krads1}) we plot the the extremal curve, the stability curves and two isopotential curves using eq.(\ref{krext}), eq.(\ref{krj1}), eq.(\ref{krj2}), and eq.(\ref{krjom}). 
The black line is the extremal curve, to the left of which lies the naked singularity region. The blue line, labelled as ``C'', is the grand canonical stability curve (or the 
$c_{\omega}$-spinodal curve), to the right of which the heat capacity $c_{\omega}$ is positive. The right most green curve, labelled as ``G'', is the Hawking-Page 
curve, to the right of which the Gibbs energy of the black hole becomes negative. The small semi-circular red curve connecting the extremal curve and the 
$c_{\omega}$-spinodal curve is the canonical stability curve or the $c_j$ curve, below which the heat capacity $c_j$ is negative. The canonical ensemble 
displays a liquid-gas like phase behaviour similar to the RN-AdS case, \cite{calda}, where, for $j<j_c=0.0239$, the Kerr-AdS black hole undergoes a first 
order transition between its small black hole and the large black hole phase. At the critical point $j_c$, which is the maxima of the $c_j$-spinodal curve, the 
constant $j$ line in fig.(\ref{krads1}) becomes tangent to the $c_{j}$-spinodal curve. The two dotted grey isopotential curves in fig.(\ref{krads1}) have been 
obtained using eq.(\ref{krjom}). The lower curve is a generic one for $\omega <1$ while the upper one is a typical curve with $\omega >1$.

In order to investigate the grand canonical ensemble more closely in fig.(\ref{krads2}) we plot some isopotential curves (red in colour) in the $s$-$t$ plane using eq.
(\ref{krtom}). The blue-dotted and the green-dotted curve above it are the $c_{\omega}$-spinodal curve and the Gibbs curve respectively of fig.(\ref{krads1}). 
The black hole is globally stable above the dotted green curve while it is locally unstable below the dotted blue curve. For $\omega <1$ the black holes exhibit a 
typical Davies phase behaviour . The Davies temperature can be obtained by substituting the expression for $j$ from eq.(\ref{krj2}) into the expression for 
temperature 
\begin{equation}
\label{tturnker}
t_d=\frac{1}{2}\,{\frac {2\,\pi \,\sqrt {\pi +s}+3\,\sqrt {\pi +s}s-\pi \,\sqrt {s}}{\sqrt {3\,s+4\,\pi }\sqrt {\sqrt {s}+2\,\sqrt {\pi +s}}{s}^{3/4}
\pi }}
\end{equation}
The black hole remains metastable between the temperatures $t_d$ and $t_{hp}$, with the rotating thermal AdS being the preferred solution as it has a lower 
free energy.  Beyond the Hawking-Page transition temperature the Kerr-AdS black hole becomes globally stable.
Notce that the $\omega $=$1$ 
isopotential curve asymptotes to the line $t=1/2\pi$ from below, whereas the $c_{\omega}$ curve and the Gibbs curve asymptotes to this line 
from above, as $s$ tends to infinity.
This clearly shows that for ${\omega}\geq 1$ the Kerr-AdS black hole is both locally as well as globally unstable at all temperatures. 

\subsection{Scalar curvature of Kerr-AdS Black Holes and its Comparison with RN-AdS.}
We now proceed to discuss the thermodynamic geometry of the Kerr-AdS black holes by first writing down the metric in the equilibrium state space in terms of the relevant thermodynamic variables $j$ and $m$ both of which are allowed to fluctuate.
The line element in the state space with the usual significance is
\begin{equation}
 \label{krline}
 dl^2=g_{\mu\nu}dx^{\mu}dx^{\nu}=g_{mm}dm^2+2g_{mj}dm dj+g_{jj}dj^2,
 \end{equation}
The scalar curvature pertaining to the state space metric is a lengthy expression. We can write it down symbolically as
\begin{equation}
\label{krR}
R =\frac{\mathcal{P}_{Kerr}}{\mathcal{N}(t){\mathcal{D}(c_{\omega})}^2}
\end{equation}
where $\mathcal{P}_{kerr}$ is a polynomial expression of more than $30$ terms and of degree $18$ in $s$ and $8$ in $j$. 
The symbols $\mathcal{N}$ and $\mathcal{D}$ represent the numerator and denominator respectively of their arguments. Just like the in RN-AdS case 
the scalar curvature encodes the divergences in the 
grand canonical ensemble of the Kerr-AdS black hole. A salient difference with the RN-AdS black hole is that the curvature never crosses to the positive side 
and asymptotes to zero from the negative side as $t$ tends to infinity. 
\begin{figure}[t!]
\begin{minipage}[b]{0.5\linewidth}
\centering
\includegraphics[width=3in,height=2.5in]{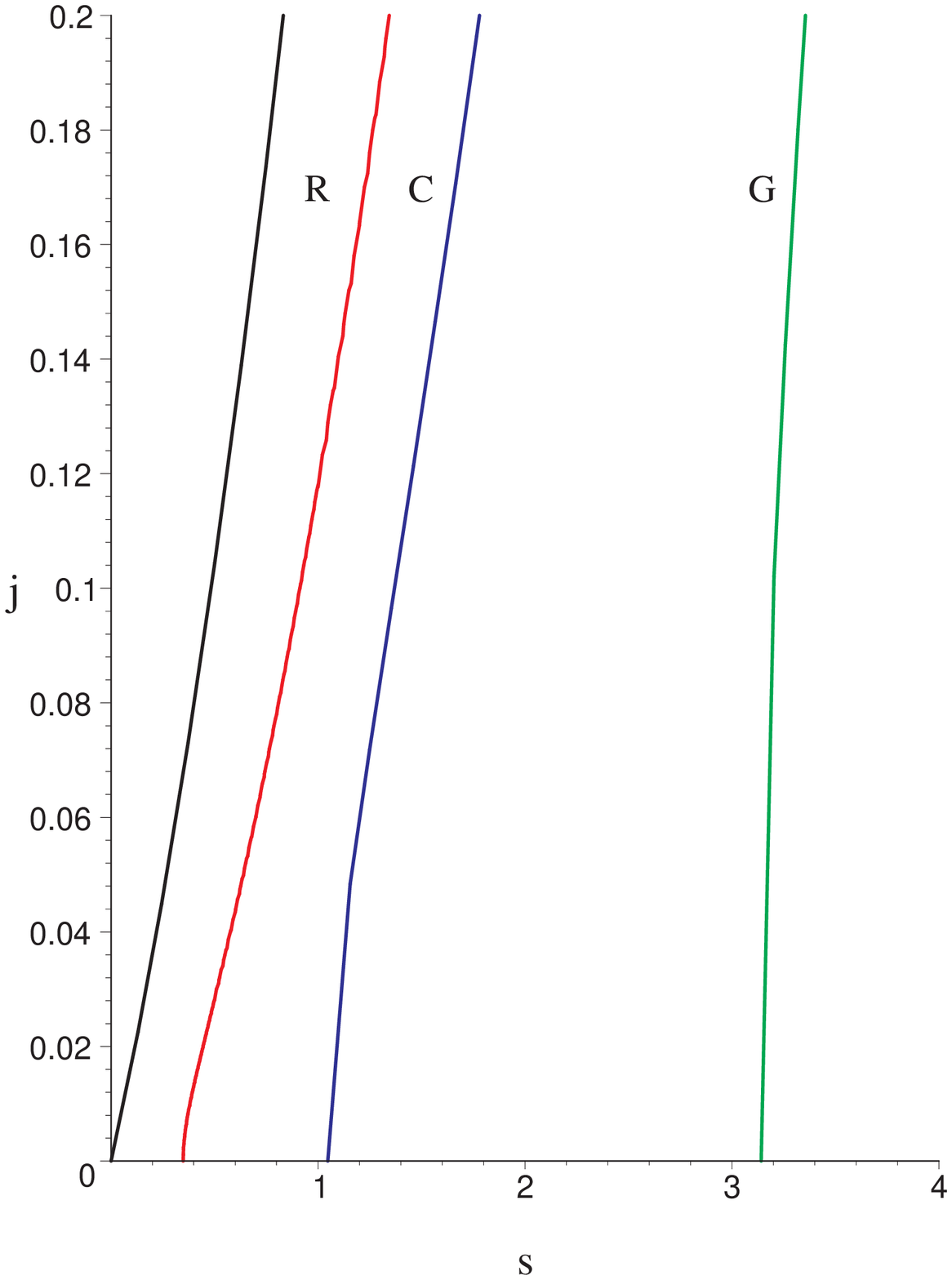}
\caption{Kerr-AdS plots in the $q$-$j$ plane. From left to right in order are the black extremal curve, zeroes of $R$ in red, and the blue and green $c_{\omega}$ and the Gibbs curve respectively.}
\label{krads3}
\end{minipage}
\hspace{0.6cm}
\begin{minipage}[b]{0.5\linewidth}
\centering
\includegraphics[width=2.7in,height=2.7in]{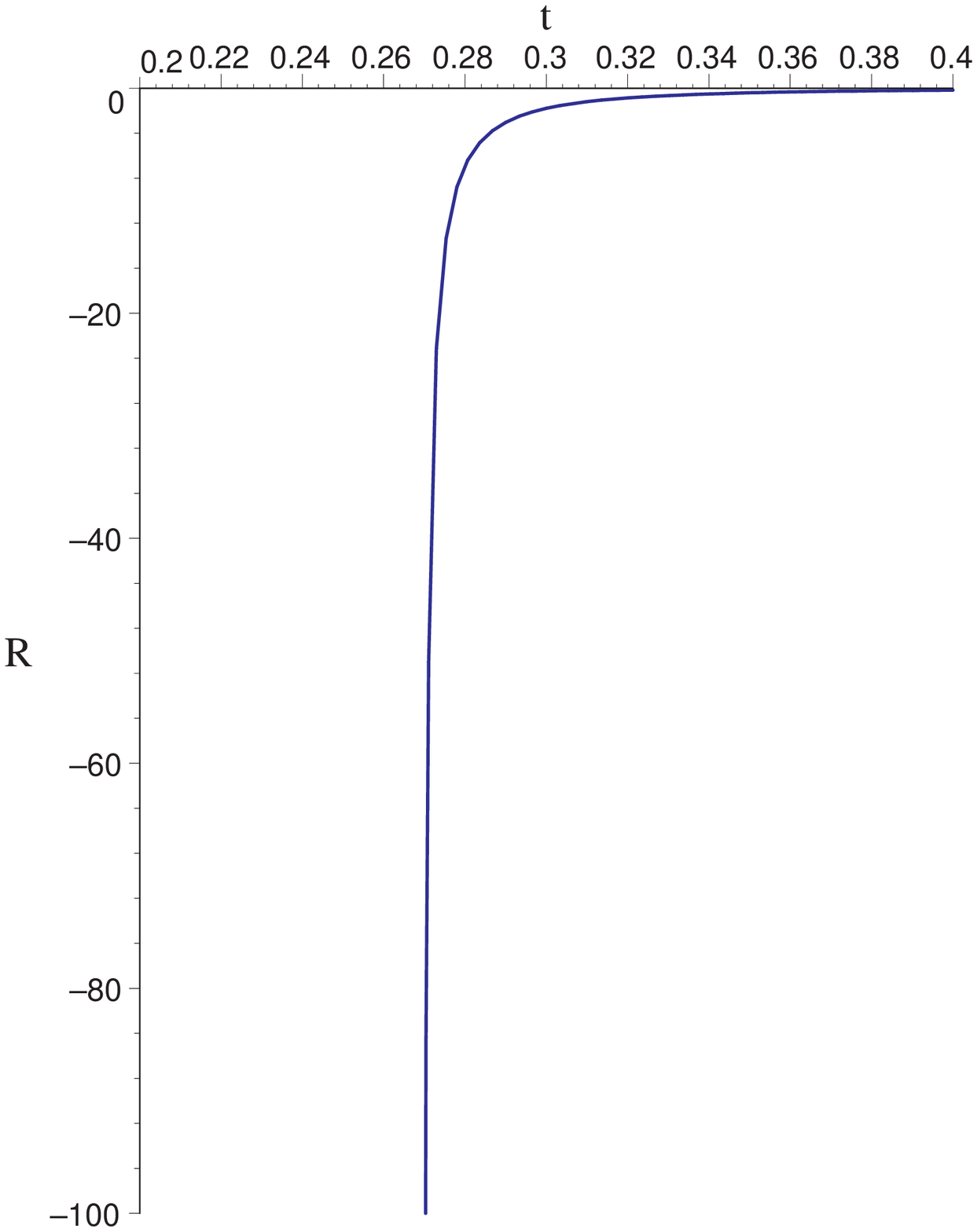}
\caption{Isopotential plot of state space scalar curvature $R$ vs. $t$ for the Kerr-AdS black hole with $\omega$ fixed at a value of 0.3. $R$ diverges at $t_2=0.269$ and remains negative at all $t$. }
\label{krads4}
\end{minipage}
\end{figure}
This is apparent from fig.(\ref{krads3}) where the locus of zeroes of scalar curvature 
(marked as ``R'') is shown to lie to the left of the $c_{\omega}$-spinodal curve and hence in the unstable region of the black hole. Thus, in the case of the 
Kerr-AdS black holes the state space scalar curvature carries no signature of a Hawking-Page phase transition. Substituting the expression for $j$ from eq.(\ref{krjom}) 
into the expression for the curvature we can obtain $R$ as a function of $\omega$ and $s$. With a similar expression for $t$ in terms of $\omega$ and $s$  
from eq.(\ref{krtom}) we can obtain the asymptotic behaviour of $R$ in terms of $t$,
\begin{equation}
R\,{\sim}\,-\frac{1}{t^4}
\end{equation}
We should point out here that the exact significance of the sign of $R$ is still an unsettled issue in the context of black holes.

Further exploring the behaviour of the scalar curvature in the RN-AdS and the Kerr-AdS black holes we now go to the AdS-Schwarzschild limit of both the black 
holes \emph{via} their respective grand canonical ensembles, i.e, by setting their respective potentials, $\phi$ or $\omega$, to zero. This ensures that the 
fluctuations in the respective charges, $q$ or $j$, are still non-zero and hence a thermodynamic geometry analysis may be undertaken. Besides, it appears 
more natural to recover the AdS-Schwarzschild limit through the grand canonical ensemble of these black holes as there is no difference in the nature of the 
isopotential curves of fig.(\ref{rnads4}) or fig.(\ref{krads2}), on setting $\phi$ or $\omega$ to zero. The AdS-Schwarzschild black holes continue to show a stable 
and an unstable branch starting at a turning point temperature and on further increasing the temperature undergo a Hawking-Page transition to a globally 
stable black hole   On the other hand there is certainly a change in the isocharge curves when the charges $q$ or $j$ are set to zero in the 
canonical ensemble of RN-AdS (fig.(\ref{rnads2})) or Kerr-AdS black holes. This is manifest in the disappearance of the small black hole branch.

The AdS-Schwarzschild limit of the scalar curvature is different in the two cases. For the RN-AdS black holes the curvature becomes
\begin{equation}
\label{Rrnsc}
R_1=-27\,{\frac {s \left( \pi -s \right) }{ \left( 3\,s+\pi  \right)  \left( \pi -3\,s \right) ^{2}}},\,\,\,\,\,q=0
\end{equation}
For the Kerr-AdS black holes this limit becomes
\begin{equation}
\label{Rkrsc}
R_2={\frac { \left( {\pi }^{2}-3\,\pi \,s-54\,{s}^{2} \right) \pi }{s \left( 3\,s+\pi  \right)  \left( \pi -3\,s \right) ^{2}}},\,\,\,\,\,j=0
\end{equation}

\begin{figure}[t!hbp]
\centering
\includegraphics[width=3in,height=2.5in]{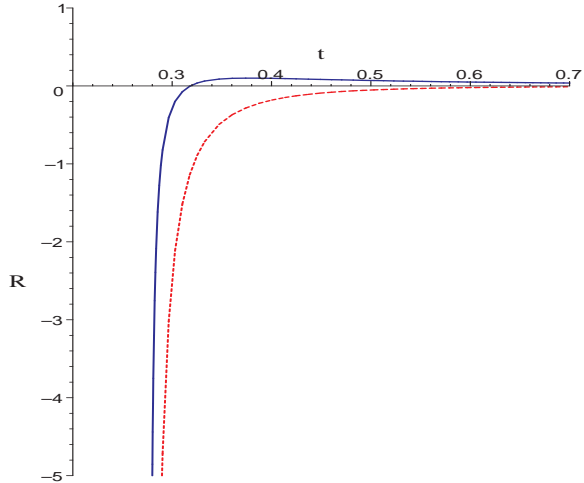}
\caption{Zero potential plot of RN-AdS black hole in blue and the Kerr-AdS black hole in dotted red}
\label{krads5}
\end{figure}

In fig.(\ref{krads5}) we plot the scalar curvature for the RN-AdS black hole in the grand canonical ensemble at zero potential vis a vis that of the Kerr-AdS black hole in the grand canonical ensemble at zero angular velocity. The zero potential RN-AdS curvature clearly encodes the Hawking-Page transition of the AdS-Schwarzschild black hole by crossing from negative to positive value at $t_{hp}=1/{\pi}$. This is the main result of this subsection.

\section{KN-AdS Black Holes in the Grand Canonical Ensemble}

The thermodynamic behaviour of Kerr-Newman AdS black holes has been extensively discussed in \cite{calda} with reference to both the canonical and the grand canonical ensembles, whereas the thermodynamic geometry of these black holes has been discussed in details in our previous work \cite{paper}. Therefore we shall be brief in our discussions here and use the section mainly to compare the thermodynamic curvture of KN-AdS black holes with the RN-AdS and Kerr-AdS case.

Thermodynamic geometry is not applicable in the canonical ensemble for these black holes as both the charge and the angular momentum remain constrained 
with the mass as the only fluctuating thermodynamic variable, thereby rendering the thermodynamic metric trivial. However, we add that these black holes
show interesting thermodynamic behaviour in the canonical ensemble, where, for a range of fixed charges $q$ and $j$, they display a liquid gas 
like phase coexistence behaviour culminating in criticality much like in the canonical ensembles for RN-AdS and Kerr-AdS black holes. In the grand canonical 
ensemble all the three ``charges'' $m$ , $q$, and $j$ are unconstrained and the black hole is in a thermal, mechanical and electrical equilibrium with its 
reservoir held at constant temperature, electric potential and angular momentum respectively. The quantities $t$, $\omega$ and $\phi$, which serve as 
control parameters for the grand canonical ensemble, have been obtained previously in terms of $s$, $q$ and $j$ by differentiating the scaled Smarr relation, 
eq.(\ref{reducedsmarr}). Using the expression for these conjugate variables, the heat capacity at constant potential and angular velocity, $c_{\phi\omega}$, 
and other susceptibilities may then be obtained. Referring the reader to the works mentioned at the beginning of this section for details, we now briefly 
describe the thermodynamic behaviour in the grand canonical ensemble. It will be convenient to express $q$ and $j$ in terms of the control 
parameters $\phi$ and $\omega$ by inverting eqs. (\ref{knphiom1}) and (\ref{knphiom}).
\begin{eqnarray}
q &=& \frac{\phi\left[\pi s\left(\pi + s - \omega^2s\right)\left(\pi + s\right)\right]^{1 \over 2}}{\omega^2s\pi - \pi s - \pi^2}\nonumber\\
j &=& \frac{\omega s^{\frac{3}{2}}\left(\pi\phi^2 + s - \omega^2s + \pi\right)\left[\left(s + \pi\right)\left(s + \pi - \omega^2s\right)\right]^{1 \over 2}}
{2\pi^{\frac{3}{2}}\left(\omega^2s - \pi - s\right)^2}
\label{knqjomphi}
\end{eqnarray}
Indeed, these expressions turn out to be useful since, using these, every other function like the temperature, Gibbs free energy, heat capacity or the state space scalar curvature 
may be re-expressed in terms of $\phi$,$\omega$ and $s$, a form particularly suited to the grand canonical ensemble.

The thermal AdS space-time rotating  with a fixed value of $\omega$ and at a constant pure gauge potential $\phi$ can serve as a reference background for the black holes in 
this ensemble, \cite{calda}. Therefore, when the Gibbs free energy of the black hole, defined as
\begin{equation}
\label{knGibbs}
g=m-ts-{\phi}q-{\omega}j
\end{equation}
is positive, the rotating thermal AdS space-time  is globally preferred while for negative $g$ the black hole spacetime becomes more stable than AdS space. Setting $g(\omega,\phi,s)$ 
to zero we can obtain $\phi$ as an implicit function of $\omega$ and $s$. We call this function $\phi_1$, where
\begin{equation}
\label{kngzero}                                                                        
{\phi}_1(\omega,s)=-{\frac {\sqrt {-{\pi }^{2}s-{\pi }^{3}+{\omega}^{4}{s}^{3}-\pi\,{\omega}^{2}{s}^{2}+{s}^{2}\pi +{\pi }^{2}{\omega}^{2}s
-2\,{\omega}^{2} {s}^{3}+{s}^{3}}}{\sqrt {\pi }\sqrt {2\,{\omega}^{2}\pi \,s+{\omega}^{2}{s}^{2}-{s}^{2}-2\,\pi \,s-{\pi }^{2}}}}
\end{equation}                                                                             

Let us briefly recapitulate the phase behaviour for these black holes in the grand canonical ensemble. 
For $\omega,\phi<1$ KN-AdS black holes exhibit a Davies phase behaviour similar to RN-AdS and Kerr-AdS in the grand canonical ensemble with their 
respective potentials less than unity. There are no black hole solutions up to the Davies temperature $t_d$, beyond which a locally stable and an unstable branch 
exist at all temperatures. For $t_d<t<t_{hp}$, the black hole remains in a metastable state. On further increasing the temperature to $t_{hp}$ and above, the 
locally stable branch changes from metastable to globally stable via a Hawking-Page phase transition, as its Gibbs potential changes sign from positive to 
negative. For $\phi\geq 1$, $\omega<1$ a globally as well as locally stable black hole branch exists at all temperatures, with an extremal solution at 
zero temperature. Along the divergence in $c_{\phi\omega}$ (i.e, setting the denominator of $c_{\phi\omega}$ to zero), we can express $\phi$ as follows             
\begin{eqnarray}
\label{kncdiv}
&&{\phi}_2(\omega,s)= \sqrt {\pi +s-{\omega}^{2}s}\times\nonumber\\\nonumber\\                                                                                         
&&\hspace{-0.5in}{\frac {\sqrt
{-3\,{s}^{4}{\omega}^{4}+6\,{\omega}^{2}{s}^{4}-3\,{s}^{4}-4\,\pi                     
\,{\omega}^{4}{s}^{3}+12\,\pi \,                                                      
{\omega}^{2}{s}^{3}-8\,{s}^{3}\pi +6\,{\pi }^{2}{\omega}^{2}{s}^{2}-6                 
\,{s}^{2}{\pi }^{2}+{\pi }^{4}}}{\sqrt {\pi }\sqrt {{s}^{4}+{\pi }^{4}                
+4\,{s}^{3}\pi +4\,{\pi }^{3}s+6\,{s}^{2}{\pi }^{2}-4\,{\pi }^{3}{                    
\omega}^{2}s-4\,\pi \,{\omega}^{2}{s}^{3}-6\,{\pi }^{2}{\omega}^{2}{s}                
^{2}-2\,{\omega}^{2}{s}^{4}+{s}^{4}{\omega}^{4}}}}\nonumber\\                         
\end{eqnarray}                                                                        

For the case of $\omega\geq 1$ for which a globally as well as locally unstable branch exists at all temperatures
\footnote{for ${\omega}=1$ the unstable branch exists for all $t$ above a minimum value that depends on ${\phi}$}
the thermodynamics is not consistently defined as the rotating Einstein universe at the boundary moves faster than light.

With the mass, charge and angular momentum all unconstrained the thermodynamic state space becomes three dimensional and the state space metric may be written as
\begin{equation}
\label{knmetric}
g_{\mu\nu}(m,q,j)\equiv\, \left(
  \begin{array}{ccc}
g_{mm} & g_{mq} & g_{mj} \\
g_{qm} & g_{qq} & g_{qj} \\
g_{jm} & g_{jq} & g_{jj} \\
\end{array}
\right)
\end{equation}

It can be verified that the line element corresponding to this metric is positive definite in regions where the heat capacity $c_{\omega\phi}$ is 
positive, \cite{paper}. Evidently, scalar curvature is not the only independent measure of curvature in three dimensions. However, following \cite{rupp1}, 
where, for conventional thermodynamic systems, it was shown that in a higher dimensional state space $R$ can still be associated with the correlation 
volume, we shall restrict our investigation to the state space scalar curvature only. The state space scalar curvature may be written symbolically as

\begin{equation}
\label{knR}
R=\frac{\mathcal{P}_{KN}}{\mathcal{N}(t)\mathcal{D}(c_{\phi\omega})^2}
\end{equation}
 where $\mathcal{P}_{KN}$ is a lengthy polynomial expression of more than $300$ terms and of degree $26,24,$ and $10$ in $s$, $q$ and $j$ respectively. 
 The divergence of $R$ along the $c_{\phi\omega}$-spinodal ``surface'' in the $s$-$q$-$j$ parameter space reflects the fact that the thermodynamic 
 metric pertains to a grand canonical ensemble with all the thermodynamic variables set to fluctuate.
 Using the substitutions in eq.(\ref{knphiom}), the scalar curvature can be written in terms of $\phi,\omega$ and $s$, which is in a form suited for analysis in the 
 grand canonical ensemble.
 
 The curvature has been studied in our previous work, \cite{paper}, so we shall not elaborate on it in any detail. The general form of the $R$ vs. $t$ 
 plots for $\phi <1,\omega <1$ and $\phi \geq 1,\omega <1$ is the same as the corresponding plots for RN-AdS black holes, as shown in 
 fig.(\ref{rnR1}), fig.(\ref{rnR2}) and fig.(\ref{rnR3}). However, now the zero crossing of $R$ for the case $\phi <1$ is no more at the same temperature as the 
 zero crossing of the free energy. The presence of $j$ fluctuations separates the zeroes of $R$ (``the R curve'') and the zeroes of Gibbs free energy, $g$.
\begin{figure}[t!]
\begin{minipage}[b]{0.3\linewidth}
\centering
\includegraphics[width=1.7in,height=1.7in]{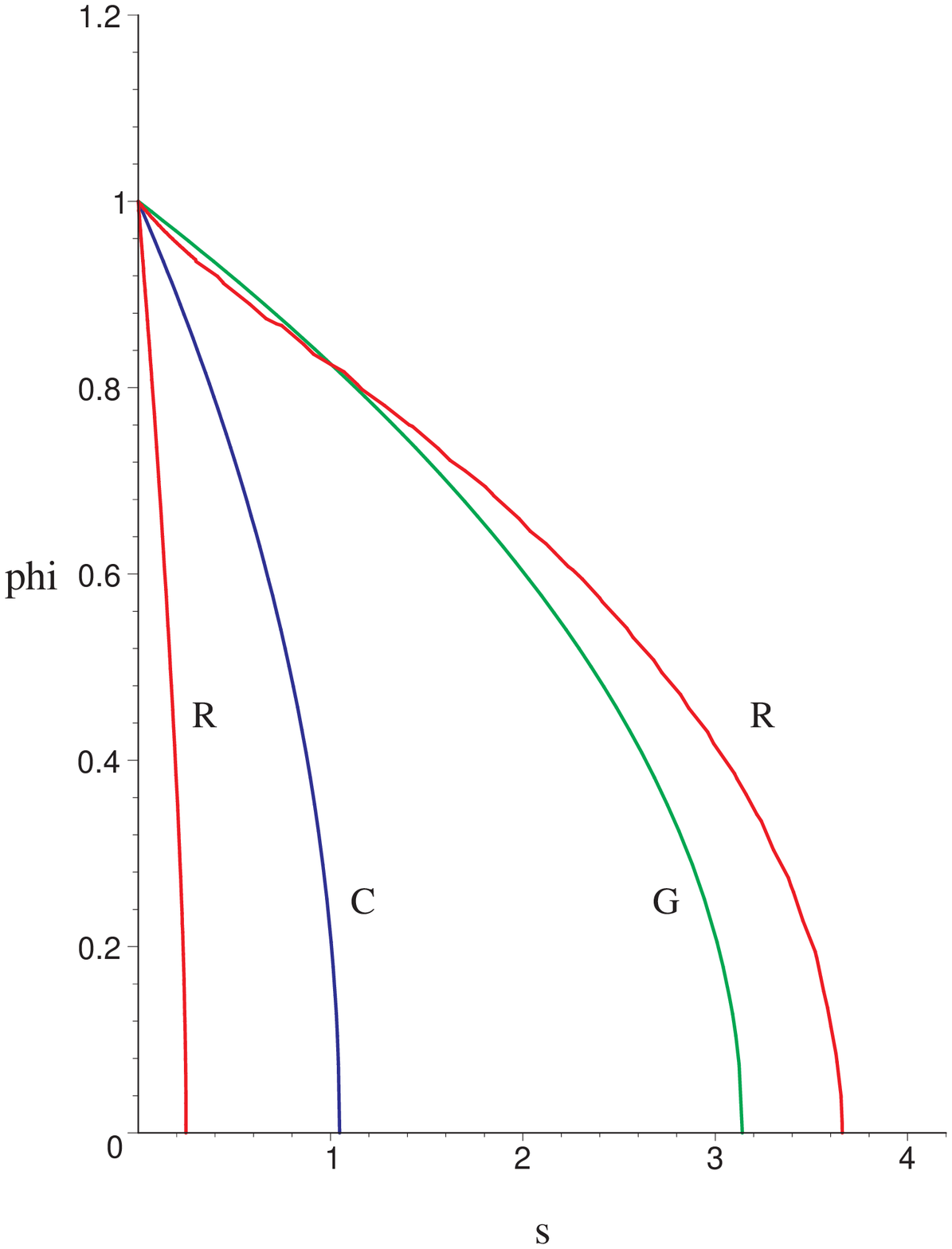}
\caption{$\phi$-$s$ plot of infinities of $c_{\phi\omega}$ and zeroes of R and g,
at $\omega=0.$}
\label{knads1}
\end{minipage}
\hspace{0.2cm}
\begin{minipage}[b]{0.3\linewidth}
\centering
\includegraphics[width=1.7in,height=1.7in]{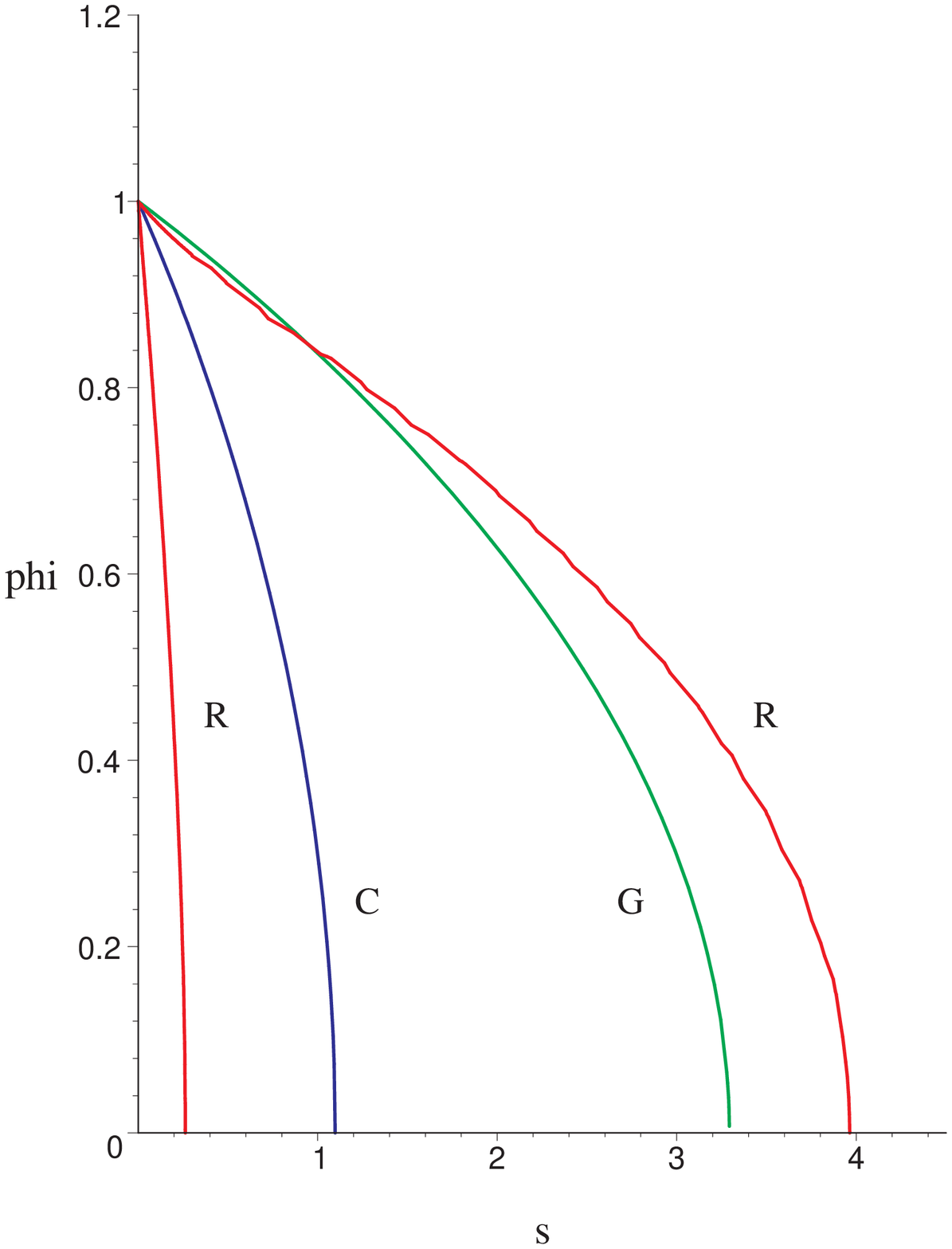}
\caption{$\phi$-$s$ plot of infinities of $c_{\phi\omega}$ and zeroes of R and g,
at $\omega=0.3$}
\label{knads2}
\end{minipage}
\hspace{0.2cm}
\begin{minipage}[b]{0.3\linewidth}
\centering
\includegraphics[width=1.7in,height=1.7in]{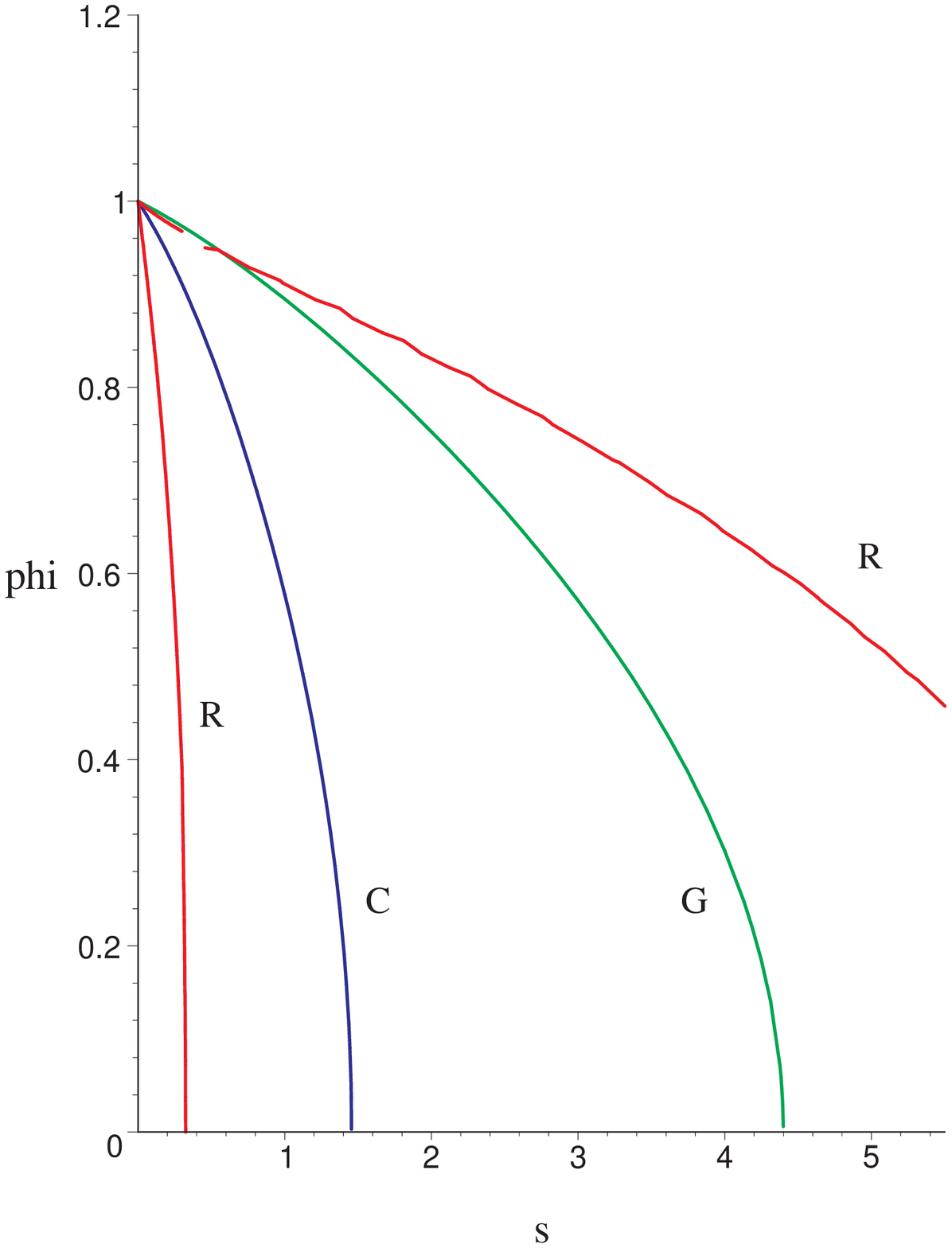}
\caption{$\phi$-$s$ plot of infinities of $c_{\phi\omega}$ and zeroes of R and g,
at $\omega=0.7$.}
\label{knads3}
\end{minipage}
\end{figure}

Let us investigate this effect in more detail. In figs(\ref{knads1}), fig.(\ref{knads2})and fig.(\ref{knads3}) we plot the zeroes of $R$, the Gibbs curve 
and the $c_{\phi\omega}$-spinodal curve in the $\phi$-$s$ plane for different fixed values of $\omega$ in an increasing order from left to right. The 
first figure in the series has $\omega$=$0$ and is an ``RN-AdS'' black hole in the grand canonical ensemble with a non zero fluctuation in $j$ about a zero 
mean value. Whereas without the $j$ fluctuations the zeroes of $R$ and $g$ would have coincided, now the two curves intersect each other 
at ${\phi}_0\approx0.82$ with an increasing separation for smaller values of ${\phi}$. It can be verified that while for $\phi<{\phi}_0$ the scalar curvature 
crosses zero and becomes positive at a temperature, $t_R$ above the Hawking-Page temperature $t_{hp}$, for $\phi>{\phi}_0$ it crosses zero at 
a temperature $t_R$ which is slightly below the Hawking Page phase transition temperature $t_{hp}$. The qualitative pattern remains the same for higher 
values of $\omega$, even though the separation between the $R$ curve and 
the Gibbs curve at lower $\phi$ progressively increases. 
To get an idea of the order of difference in $t_R$ and $t_{hp}$ let us find them for two different ranges of $\phi$ and $\omega$ values. At the potentials $\omega=0.9, \phi=0.1$, for which the R-curve and the Gibbs curve are well separated according to the previous discussion, the two temperatures turn out to be $t_R=0.309$ and $t_{hp}=0.226$. Whereas, with $\phi=0.9$, $\omega=0.1$, for which the curves are nearly coincident, the temperatures turn out to be $t_R=0.132$, $t_{hp}=0.136$. 
Interestingly, for all $\omega$ there exists a $\phi={\phi}_0$  for which the R curve and the Gibbs 
curve intersect, i.e, $t_R=t_{hp}$.
\begin{figure}[t!hbp]
\centering
\includegraphics[width=3in,height=2.5in]{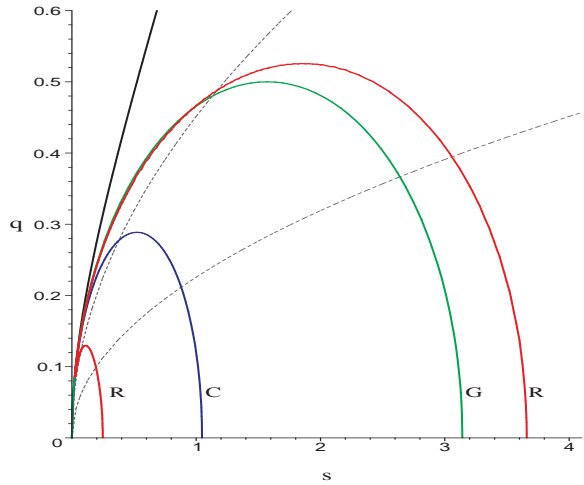}
\caption{Plot of infinties of $c_{\phi\omega}$ and zeroes of $R$ and $g$ in the
$q$-$s$
plane at $j=0$. Grey dotted isopotential curves are at $\phi=0.4$ below
and $\phi=0.8$ above.}
\label{knads4}
\end{figure}

\begin{figure}[t!]
\begin{minipage}[b]{0.5\linewidth}
\centering
\includegraphics[width=3in,height=2.5in]{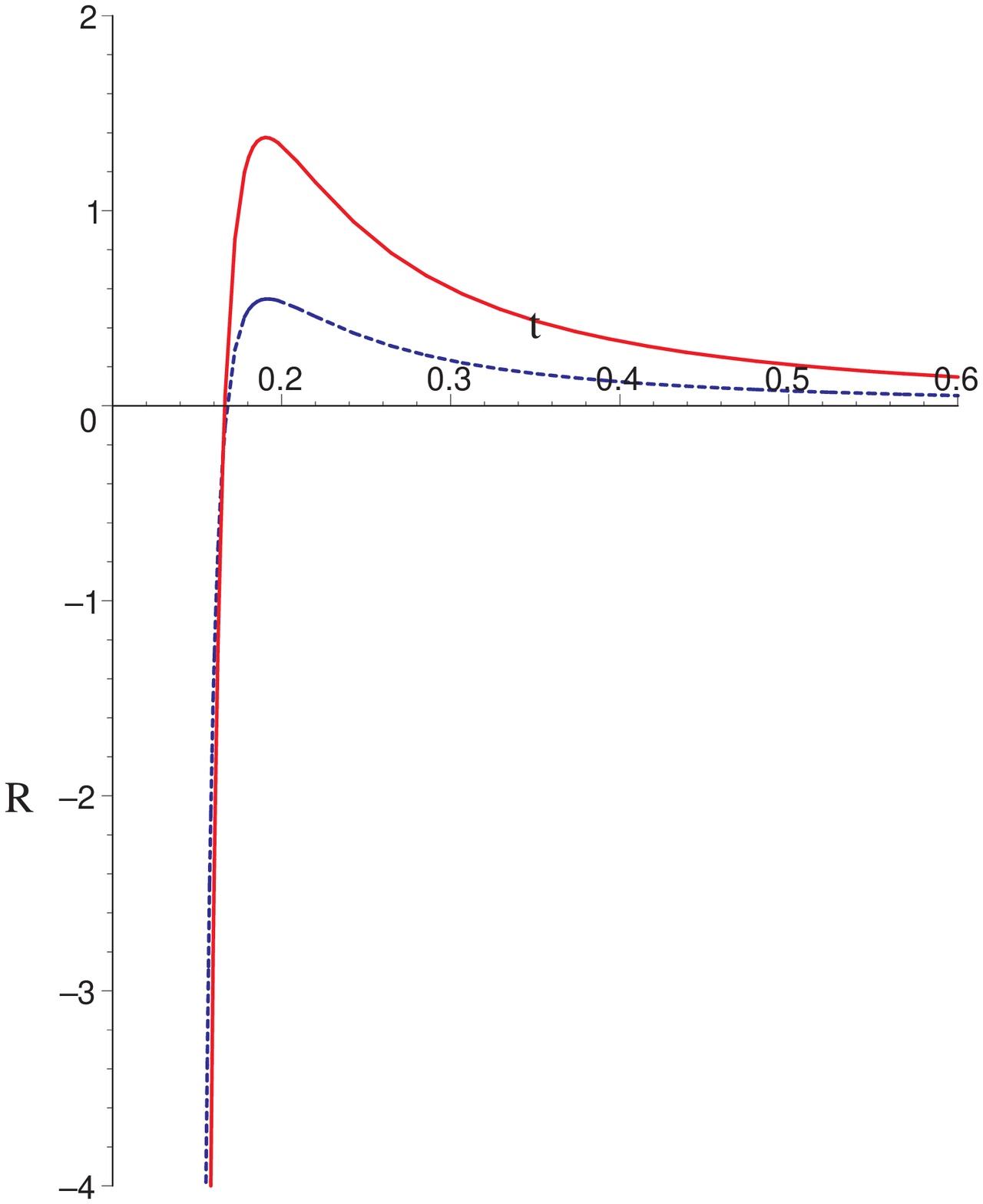}
\caption{$R$ vs $t$ plot showing KN-AdS curvature in red at $\omega=0,\phi=0.85$ and RN-AdS curvature in dotted blue at $\phi=0.85$  }
\label{knads5}
\end{minipage}
\hspace{0.6cm}
\begin{minipage}[b]{0.5\linewidth}
\centering
\includegraphics[width=2.7in,height=2.7in]{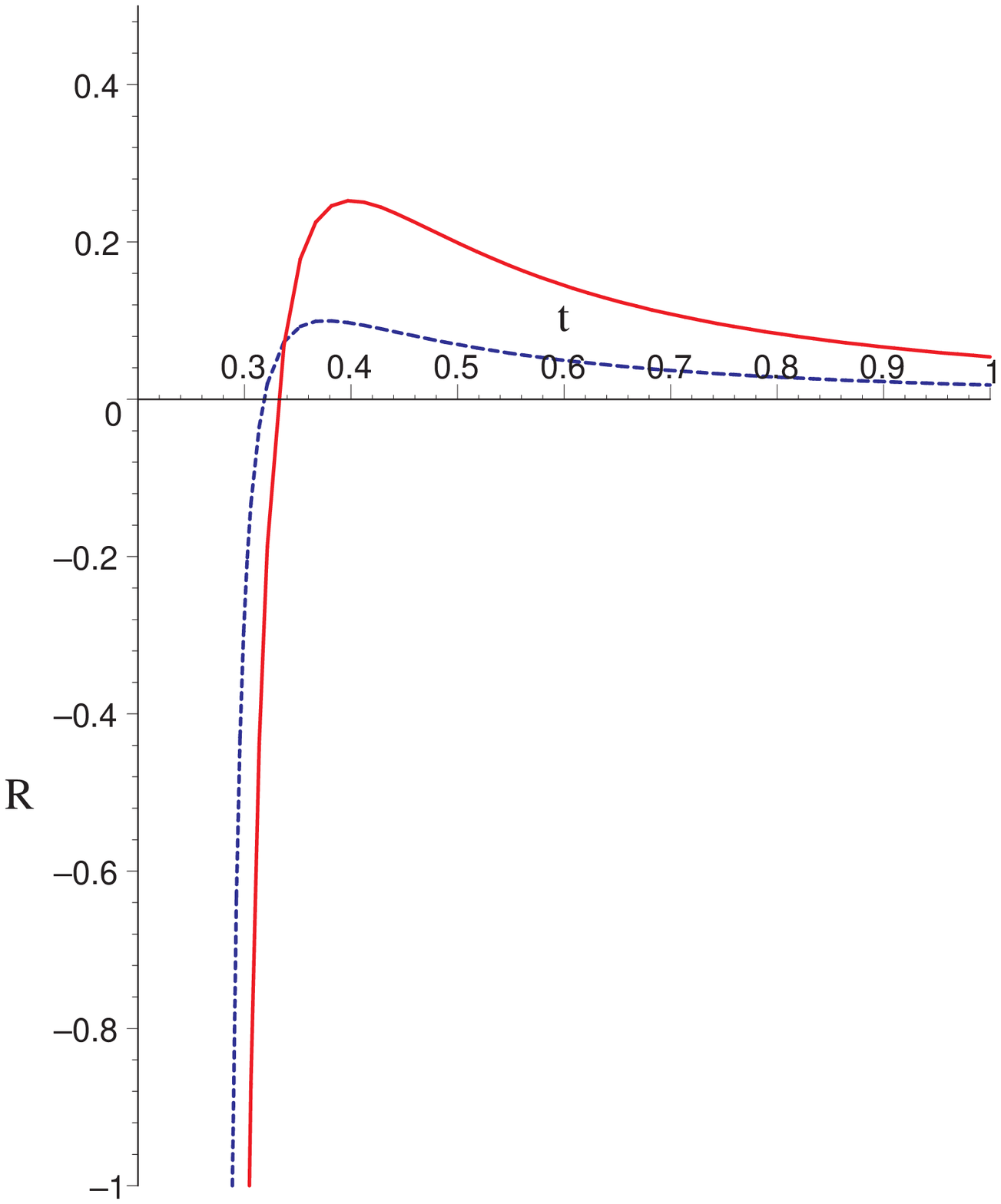}
\caption{$R$ vs $t$ plot showing KN-AdS curvature in red at $\omega=0,\phi=0$ and RN-AdS curvature in dotted blue at $\phi=0$ .}
\label{knads6}
\end{minipage}
\end{figure}

For a comparison with the RN-AdS black hole we redraw fig.(\ref{knads1}) in the $j=0$ section of the $s$-$q$-$j$ parameter space to obtain fig.(\ref{knads4}). The green Gibbs curve marked ``G'' and the blue $c_{\phi\omega}$-spinodal curve marked ``C'' are same as the curves in fig.(\ref{rnads3}) while the R-curve is marked in red. The small R-curve lies in the unstable region and so is irrelevant. Comparing fig.(\ref{rnads3}), where the R-curve coincides with the  Gibbs curve, and fig.(\ref{knads4}), we can make out the difference between the absence of $j$ and finite fluctuations about a zero mean value of $j$. Two isopotentials in dotted grey have been shown for a correspondence with fig.(\ref{knads1}).  In fig.(\ref{knads5}) we compare the $R$ vs. $t$ plot of the KN-AdS scalar curvature in its ``RN-AdS'' limit with the plot 
of the RN-AdS curvature, both of them at $\phi=0.85$. Expectedly, given the large value of the potential, the two curvatures are very similar, with the  zero of the 
KN-AdS curvature slightly different from the zero of RN-AdS curvature owing to $j$ fluctuations.

We now take the ``AdS-Schwarzschild'' limit of the Kerr-Newman thermodynamic curvature by setting  both $q$ and $j$ to zero in its expression. The corresponding black hole system is a KN-AdS black hole held at $\phi=\omega=0$, because of which the mean values of $j$ and $q$ remain at zero, with non-zero fluctuations around the mean. The curvature turns out to be
\begin{equation}
\label{knsc}
R_3={\frac {{\pi }^{3}-6\,s{\pi }^{2}-90\,{s}^{2}\pi +81\,{s}^{3}}{s \left( 3\,s+\pi  \right)  \left( \pi -3\,s \right) ^{2}}}
\end{equation}

The behaviour of $R_3$ is very similar to the RN-AdS curvature in the AdS-Schwarzschild limit, $R_1$, obtained in eq.(\ref{Rrnsc}) and shown in fig.(\ref{krads5}). The zero crossing of $R_3$, however, is different owing to the $j$ fluctuations, and takes place at $t_R=0.331$ which is greater than the zero crossing of $R_1$ at $t_{hp}=1/\pi=0.318$. We show this in fig.(\ref{knads6}), where the curvatures $R_3$ and $R_1$ have been plotted together for comparison.

For large temperatures $R$ decays to zero in the same way as the RN-AdS black hole curvature, 
i.e, $R\,{\sim}\,{1/t^2}$ as $t\to\infty$.

\section{KN-AdS Black Holes in the Mixed Ensembles}

The Kerr-Newman AdS black hole in the grand canonical ensemble has all its thermodynamic charges, $m$, $j$ and $q$ unconstrained. However, note that the electric charge, $q$, and the angular momentum, $j$, are quite unlike each other physically. It is therefore reasonable to consider ensembles with restricted fluctuations, in which one of the two thermodynamic charges, $j$ or $q$, is held fixed, while the other fluctuates. We may term such ensembles as ``mixed'' ensembles, since the black hole can be said to be in a canonical ensemble with respect to the constrained charge, while it is in a grand canonical ensemble with respect to its fluctuating charge, which it exchanges with its surrounding at a constant conjugate potential. This would lead one to consider ``mixed'' susceptibilities like $c_{j\phi}$, the heat capacity at constant angular momentum and electric potential, and, similarly, $c_{q\omega}$ etc. For the case of the asymptotically flat Kerr-Newman black holes such mixed heat capacities were considered in \cite{lousto}, where their critical exponents were calculated, while \cite{rupp3} and \cite{rupp4} considered all possible thermodynamic fluctuations in these black holes with extensive discussions on their stability and thermodynamic geometry. KN-AdS black holes with restricted fluctuations were investigated extensively by us in a previous work, \cite{paper}, where we had discussed their phase behaviour and certain aspects of thermodynamic curvature. It was observed that in the mixed ensemble the KN-AdS black holes posessed a rich phase structure, exhibiting liquid gas like phase coexistence regions and critical points. In the following we shall briefly outline the phase structure of these black holes from a somewhat new perspective, before proceeding to discuss their critical behaviour in the next section.

\begin{figure}[t!]
\begin{minipage}[b]{0.5\linewidth}
\centering
\includegraphics[width=3in,height=2.5in]{12ab.bmp}
\caption{Picture showing the shifting of the $c_{\phi}$ spinodal curve along the $c_j$
spinodal curve. Critical behaviour exists only up to the maxima of $c_j$ curve at
$j_c$=$0.0239$.  }
\label{mxads1}
\end{minipage}
\hspace{0.6cm}
\begin{minipage}[b]{0.5\linewidth}
\centering
\includegraphics[width=2.7in,height=2.7in]{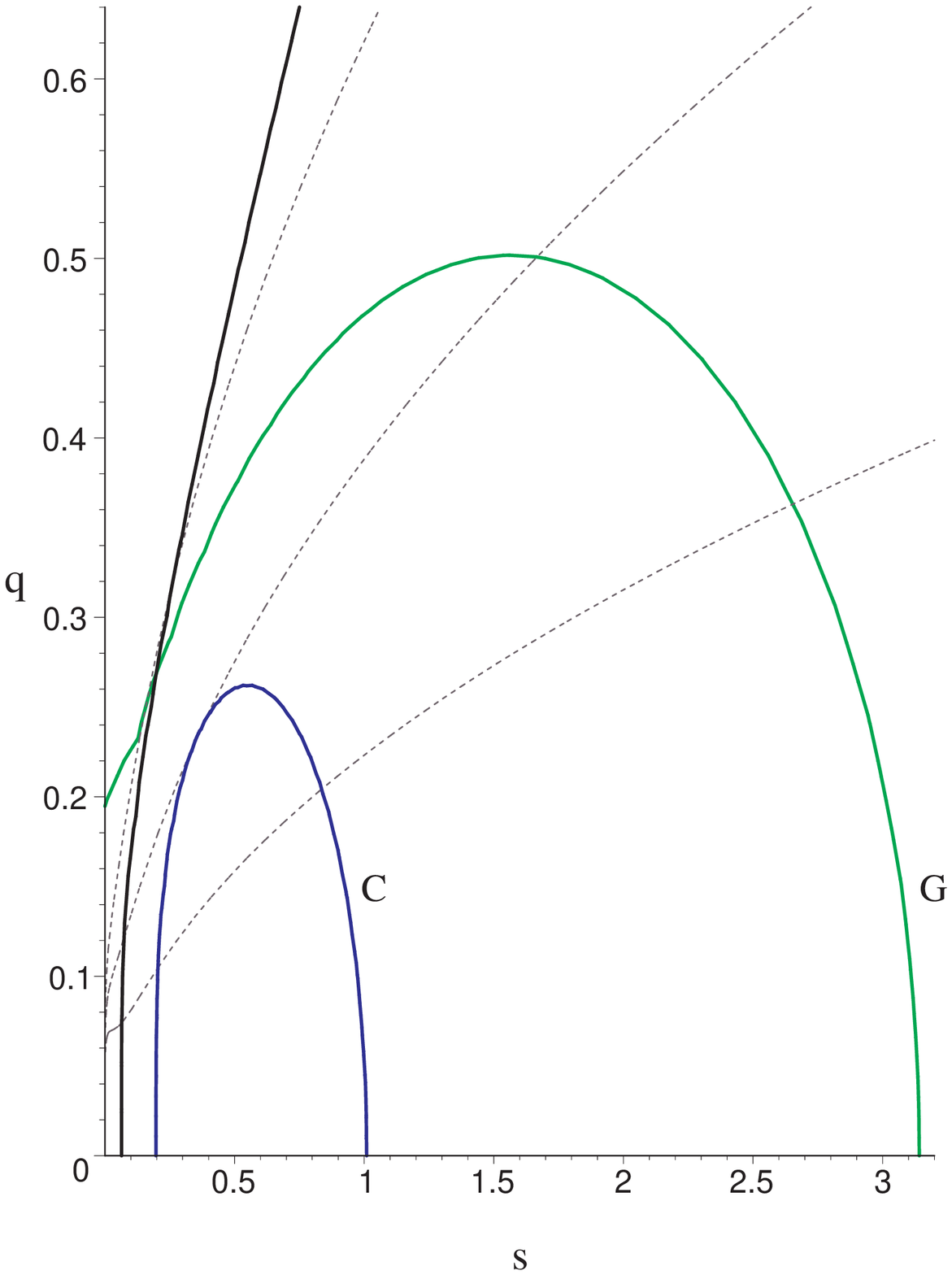}
\caption{Plot of extremal, Gibbs and $c_{\phi}$ curves in the $q$-$s$ plane for fixed $j$
ensemble, with $j=0.011$. Isopotential curves in dotted grey are at
$\phi=0.395,0.687({\phi}_c)$ and $1.1$ from bottom to top.}
\label{mxads2}
\end{minipage}
\end{figure}

The fixed $j$-ensemble is formed by adding a constant angular momentum, $j$, to the RN-AdS black hole in a grand canonical ensemble. Since this black hole does not exchange its angular momentum
$j$ with the reservoir it is ``mechanically'' isolated, even though it continues to maintain an electrical equilibrium with its surroundings. The free energy for this ensemble is given by
\begin{equation}
\label{A}
A=m-ts-\phi q
\end{equation}
where the conjugate quantities are now given by their full KN-AdS thermodynamic relations, eqs. (\ref{knphiom1}), (\ref{knphiom}) and eq.(\ref{kntemp}). This free energy, ``A'', remains constant during a first order phase transition, exhibiting a multivalued branched structure or a swallow tail shape in a free energy vs. temperature plot. However, now the zeroes of the free energy do not correspond to a Hawking-Page phase transition. This is because thermal AdS with a fixed angular momentum, $j$, is no more a solution to the Einstein equations, and therefore cannot serve as a reference background in action calculation with fixed $j$ and fixed $\phi$ boundary conditions. For more details we refer the reader to our previous paper \cite{paper} and the references therein.

In fig.(\ref{mxads1}) we pictorially depict the effect of adding constant $j$ by drawing the stability curves in the three dimensional $s$-$q$-$j$ parameter space. As $j$ becomes non zero, the $c_{\phi}$ spinodal curve in the $q$-$s$ plane of RN-AdS black hole begins to ``slide'' along the $j$ axis in such a way that its base always lies on the $c_j$ spinodal curve in the $j$-$s$ plane of Kerr-AdS black hole. As a consequence of this, some amount of parameter space becomes available to the left of the (shifted) $c_{\phi}$ curve\footnote{More precisely, it is the constant $j$ section of the $c_{j\phi}$ spinodal surface but we shall continue to name it as $c_{\phi}$ curve.}. Therefore, now the isopotential plots can cut the $c_{\phi}$ curve \emph{twice}, thus effecting an abrupt change from the Davies phase behaviour for $j=0$ to a liquid-gas like phase behaviour with a small black hole branch and a large black hole branch. In fig.(\ref{mxads2}) we plot the $c_{\phi}$ stability curve in the $q$-$s$ plane for a fixed $j$. We also plot the zeroes of the free energy, A, labelling it as the ``G curve''.
 A comparison of fig.(\ref{mxads2}) with fig.(\ref{rnads1}) for RN-AdS black hole clearly brings out the shifting of the ``C'' curve to the right as discussed above.
Among the representative isopotentials shown in the figure, those that intersect the $c_{\phi}$ curve twice will have their $t$-$s$ curves typically like those in fig.(\ref{rnads2}), with a first order transition between the small and the large black hole branches. The isopotential curve tangent to the $c_{\phi}$ curve becomes the critical curve, along which the black hole undergoes a second order phase transition at the tangent, which is therefore the critical point for a given value of $j$. For $\phi$ values 
beyond the critical potential the $j$-ensemble black hole changes continuously from the small black hole phase to the large black hole phase without undergoing 
any phase transition. Referring back to the pictorial depiction in fig.(\ref{mxads1}) we can infer that the $c_{\phi}$ curve eventually shrinks and disappears as it 
reaches the maxima of the $c_j$ spinodal at $j=j_c=0.0239$, which is the critical point of the Kerr-AdS black hole in the canonical ensemble. For $j>j_c$ the black holes in 
the fixed $j$-ensemble has a unique stable branch at all temperatures. 

    \begin{figure}[t!]
    \begin{minipage}[b]{0.5\linewidth}
    \centering
    \includegraphics[width=3in,height=2.5in]{abcd.bmp}
    \caption{Picture showing the shifting of the $c_{\phi}$ spinodal curve along the $c_j$
    spinodal curve. Critical behaviour exists only up to the maxima of $c_j$ curve at
    $j_c$=$0.0239$.  }
    \label{mxads3}
    \end{minipage}
    \hspace{0.6cm}
    \begin{minipage}[b]{0.5\linewidth}
    \centering
    \includegraphics[width=2.7in,height=2.7in]{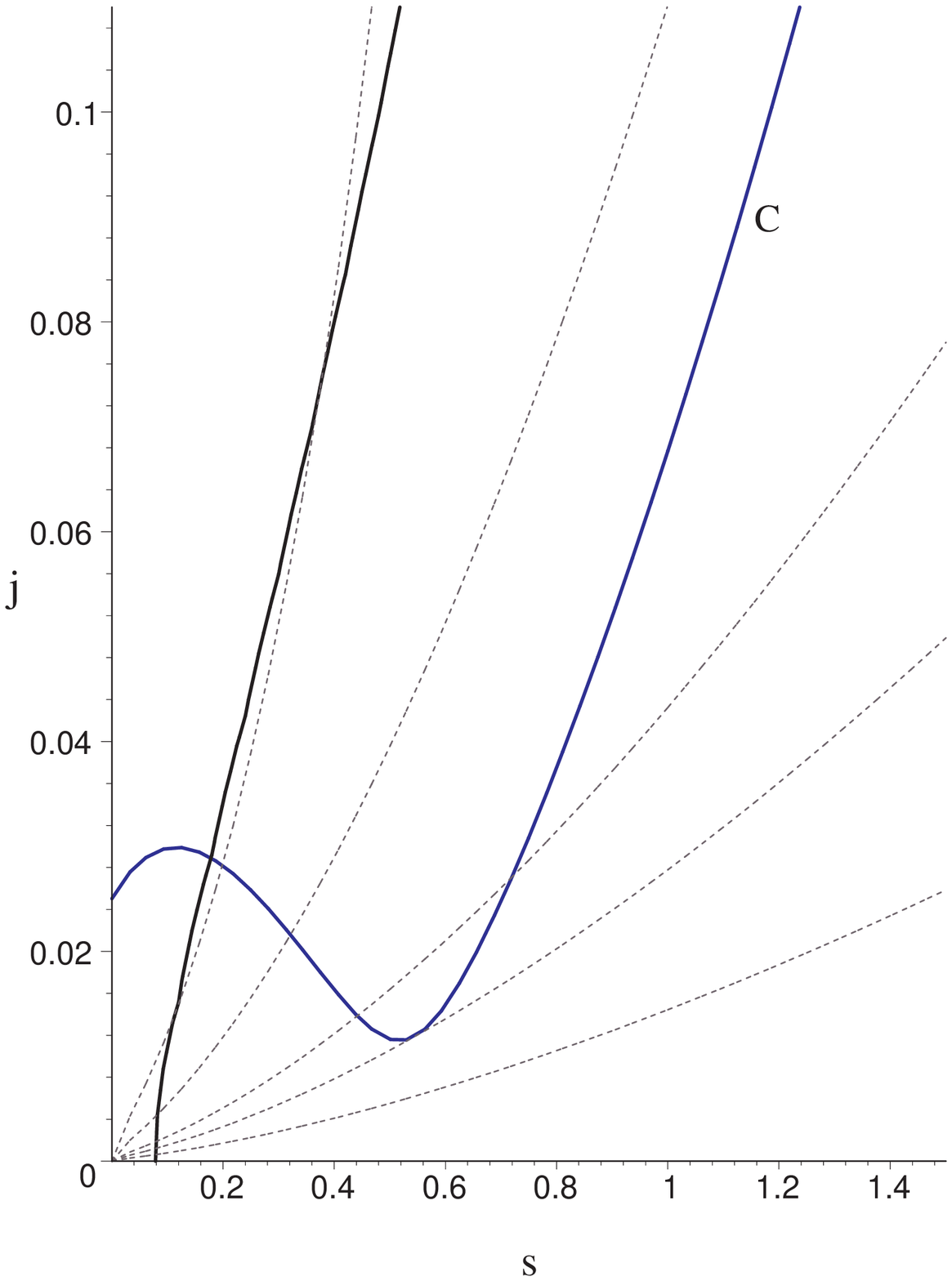}
    \caption{Plot showing the extremal curve and the $c_{\omega}$ curve in the $q$-ensemble, with $q=0.17>q_c$. The isopotentials are at $\omega=0.15,0.286 ({\omega}_c), 0.74,1,2.12$ respectively from bottom to top}
    \label{mxads4}
    \end{minipage}
    \end{figure}

The fixed $q$ ensemble is similarly formed by adding a constant charge $q$ to the Kerr-AdS black hole in a grand canonical ensemble, so that the black hole
remains ``electrically `` isolated while continuing to maintain its mechanical equilibrium with the reservoir. The free energy for the $q$- ensemble is given as
\begin{equation}
\label{B}
B=m-ts-\omega j
\end{equation}
where the conjugate potentials are once again given by eqs. (\ref{knphiom1}), (\ref{knphiom}) and eq.(\ref{kntemp}). Similar to the $j$-ensemble this free energy exhibits a swallow tail shape in a first order transition. However, the $q$-ensemble has a richer phase behaviour
than the fixed $j$ ensemble, as we will show below.

In the three dimensional pictorial depiction of fig.(\ref{mxads3}) we can observe the effect of adding a constant $q$ term. The $c_{\omega}$ curve in the $j$-$s$ plane corresponding to the Kerr-AdS black hole translates along the $c_q$ spinodal curve belonging to the RN-AdS black hole in the $s-q$ plane at the base. 
As a result of this an additional branch of $c_{\omega}$ curve appears to the left of the $c_q$ curve, giving rise to a low temperature stable black hole branch 
separated from the stable large black hole branch by an unstable region. On further increasing the charge the two $c_{\omega}$ branches join together at the 
maxima of the $c_q$ curve at $q=q_c=1/6$,  which is the critical point of the RN-AdS black hole in the canonical ensemble. Interestingly, up to $q=q_c$, there are 
no critical points, with a discontinuous phase transition between the small and the large black hole taking place for all values of angular velocity  less than 
unity\footnote{For $\omega>1$ there are no stable large black hole branches for any fixed $q$.}
\begin{figure}[t!]
\begin{minipage}[b]{0.5\linewidth}
\centering
\includegraphics[width=3in,height=2.5in]{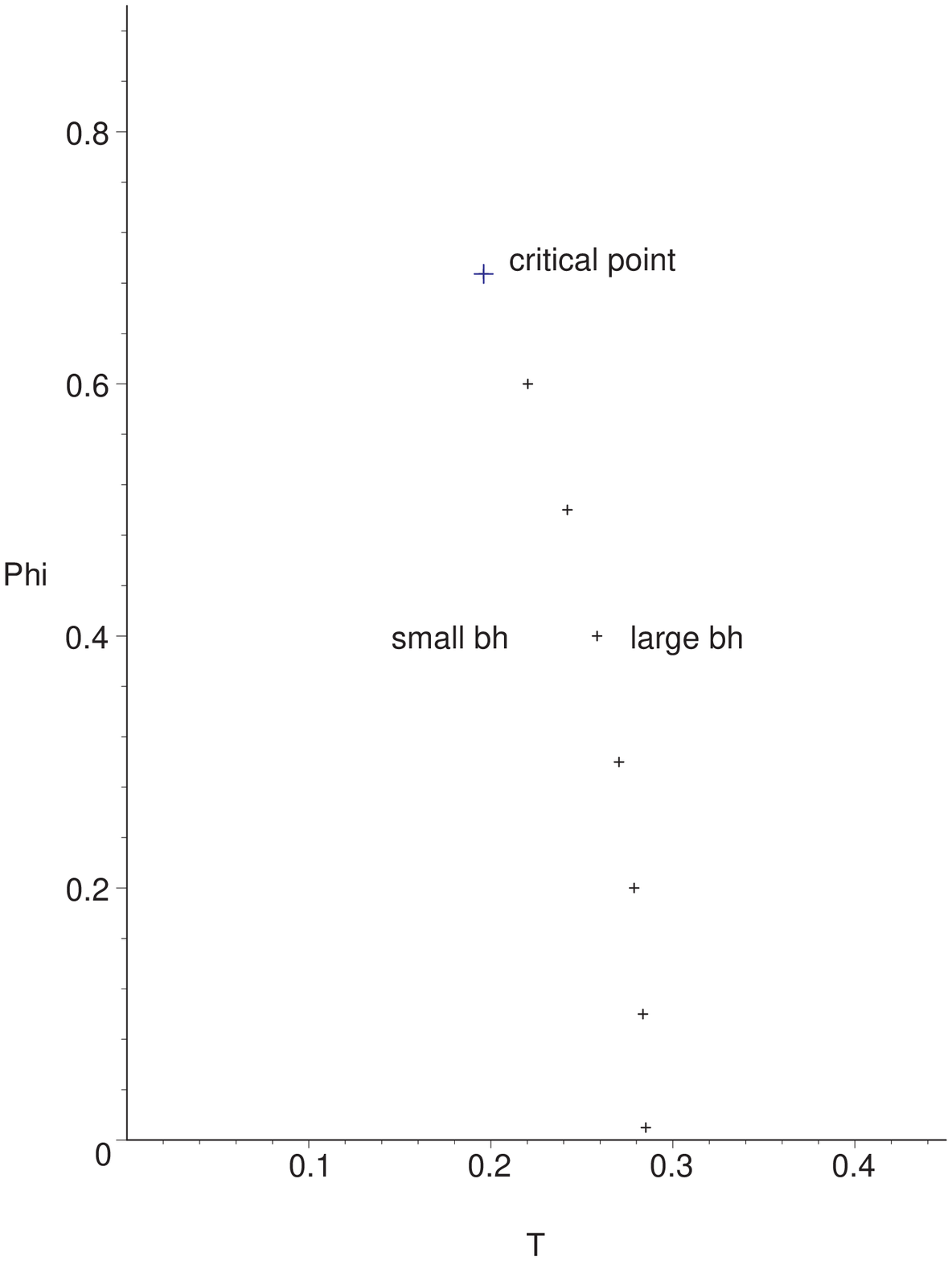}
\caption{Phase coexistence curve in the $\phi$-$t$ plane of the $j=0.011$
ensemble.}
\label{mxads5}
\end{minipage}
\hspace{0.6cm}
\begin{minipage}[b]{0.5\linewidth}
\centering
\includegraphics[width=2.7in,height=2.7in]{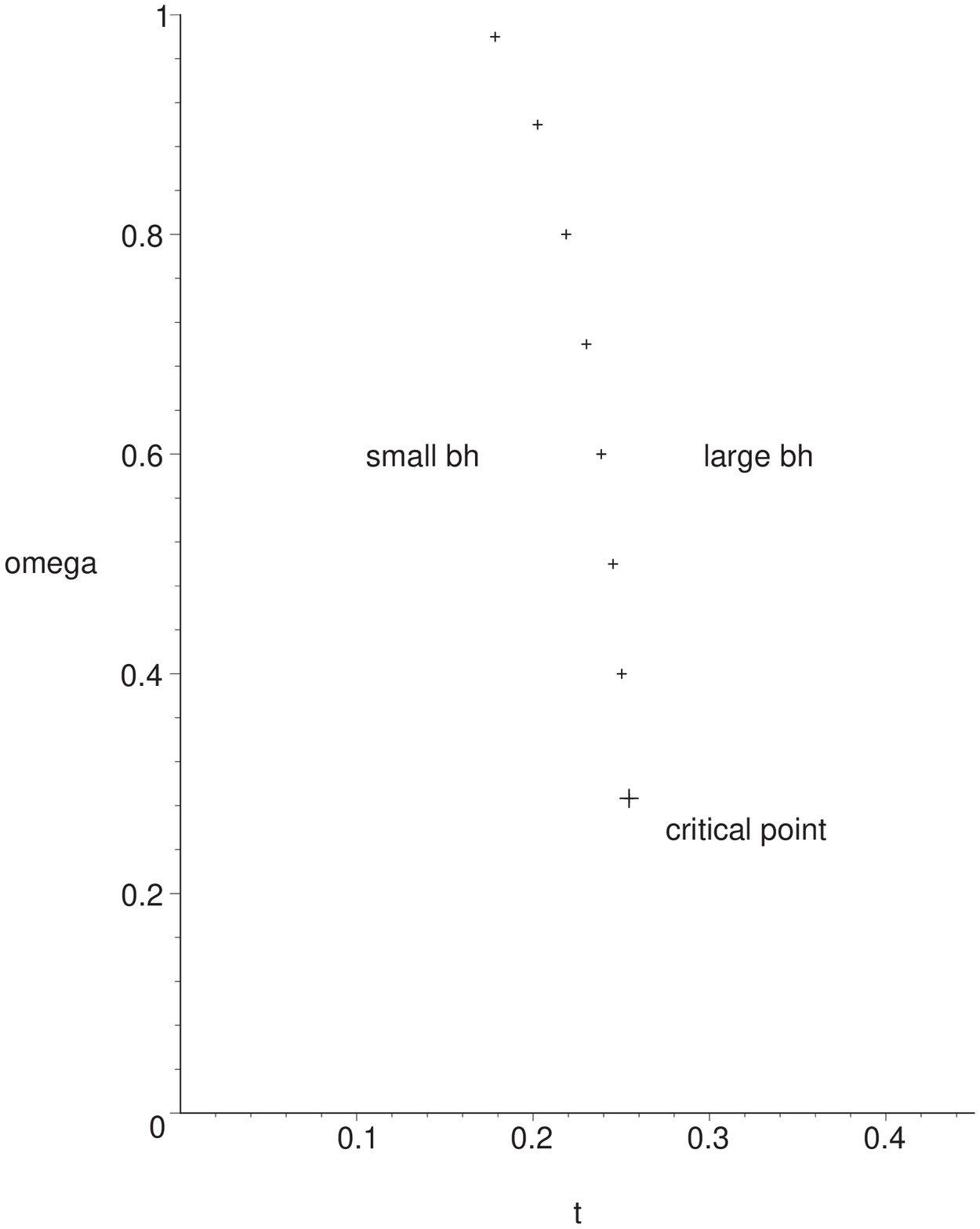}
\caption{Phase coexistence curve in the $\omega$-$t$ plane of the $q=0.17$
ensemble.}
\label{phasecoexfixedq}
\end{minipage}
\end{figure}
At the same time, for $q>q_c$, critical point exists for all values of $q$. In fig.(\ref{mxads4}), we look at the critical behaviour of the $q$-ensemble in a greater detail by plotting the $c_{\omega}$ curve in the $j$-$s$ plane for a fixed $q$ value greater than $q_c$. The heat capacity is positive below the $c_{\omega}$ curve. 
The critical isopotential curve, having angular velocity ${\omega}_c$, is the one which is tangent to the $c_{\omega}$ stability curve. For $1>\omega>{\omega}_c$ 
the black hole shows a liquid-gas like phase coexistence behaviour with a first order transition between the small black hole and the large black hole.

To end this subsection, we plot, in fig.(\ref{mxads5}), the coexistence curve
in the $\phi$-$t$ plane for the fixed $j$ ensemble. Notice the analogy with the liquid-vapour coexistence curve in the pressure-temperature plane ($p-t$) for fluids, with the correspondence $\phi\leftrightarrow p$.
The slope of the coexistence curve can be understood through the Clausius-Claperyon
equation. At any point on the coexistence curve having co-ordinates $(t,\phi)$, the Gibbs free energies of the small black hole, $A_1$, and the large black hole, 
$A_2$, have the same value, \emph{i.e} $A_1=A_2$. On moving along the coexistence curve to a nearby point, $(t+{\Delta}t,\phi+{\Delta}\phi)$, the new free 
energies of the two branches will still be equal to each other, thus implying ${\Delta}A_1={\Delta}A_2$. Given that the variation in Gibbs free energy is given 
by ${\Delta}A=-s{\Delta}t-q{\Delta}\phi$ we obtain
\begin{equation}
-s_1{\Delta}t-q_1{\Delta}\phi=-s_2{\Delta}t-q_2{\Delta}\phi
\end{equation}

where the subscripts $1$ and $2$ indicate the small black hole and the
large black hole respectively. From this equation we obtain the slope of the coexistence curve as
\begin{equation}
\frac{{\Delta}\phi}{{\Delta}t}=-\frac{s_2-s_1}{q_2-q_1}
\end{equation}
The slope of the coexistence curve is clearly negative here, while it is positive in the
case of fluids because of the sign convention for pressure which treats the pressure applied \emph{on} the fluid
as positive. 

For the fixed $q$-ensemble, the phase coexistence plot is shown in fig.(\ref{phasecoexfixedq}). The slope is again negative, as can be
verified by a calculation analogous to the one presented above. However, this has qualitatively
different features from that of the fixed $j$-ensemble curve of fig.(\ref{mxads5}).

\subsection{ Critical exponents in the mixed ensembles}

In systems exhibiting second order phase transitions the order parameter and various
susceptibilities display a power law behaviour near the critical point. This is essentially
because of the dominance near criticality of the long range correlated fluctuations, as a
result of which the system description becomes scale invariant. The scaling exponents or
the critical exponents for the  power law dependence of these quantities are seen
to be the same for a wide variety of systems. This feature of universality again follows
from the divergence of the correlation length near the critical point, which renders the
details of microscopic interactions irrelevant. In fact, the various systems can be
classified under separate universality classes, based on the spatial dimension of the
system and that of the order parameter.
We will first list the critical exponents corresponding to various thermodynamic functions
for the liquid gas system, as an example. The divergence of the specific heat at
constant volume above and below the critical temperature goes as follows:
\begin{eqnarray}
c_v&{\sim}&(t_r)^{-\alpha}\nonumber\\
c_v&{\sim}&(-t_r)^{-\alpha'}
\end{eqnarray}
where $t_r=\frac{t_c-t}{t_c}$.
The isothermal susceptibility
${\kappa}_t=({\frac{1}{\rho}}{\frac{{\partial}{\rho}}{{\partial}p}})|_t$ diverges as
\begin{eqnarray}
{\kappa}_t&~{\sim}~&(t_r)^{-\gamma}\nonumber\\
{\kappa}_t&~{\sim}~&(-t_r)^{-\gamma'}
\end{eqnarray}
The order parameter, ${\Delta}{\rho}={\rho}-{\rho}_{cr}$, varies along the coexistence
curve as
\begin{eqnarray}
{\Delta}{\rho}~{\sim}~(t_r)^{\beta}
\end{eqnarray}
where   $(t<t_c)$. Finally, along the critical isotherm, the variation of the order
parameter with its intensive variable is given as
\begin{eqnarray}
{\Delta}{\rho}~{\sim}~(p_r)^{1/{\delta}}.
\end{eqnarray}
where $p_r=\frac{p_c-p}{p_c}$.
The critical behaviour can be neatly summarized in the ``static scaling hypothesis'',
which is one of the central results of renormalization group theory, \cite{hes},
\cite{fish}. According to this, in the vicinity of the critical point, the singular part
of the free energy is a generalized homogeneous function (GHF) of its variables. Thus, for
a liquid gas system for example, we have
\begin{equation}
\label{scaling}
G_{s}({\lambda}^{a_t}{\Delta}t,{\lambda}^{a_p}{\Delta}p)={\lambda}G_{s}({\Delta}t,{\Delta}p),
\end{equation}
where ${\Delta}t$ and ${\Delta}p$ are deviations from their critical values and $a_t$ and
$a_p$ are called the scaling powers of temperature and pressure. It then follows from the
properties of GHFs that all the derivatives and Legendere transforms of the free energy
will also be GHFs. The critical exponents ${\alpha},{\beta},{\gamma},{\lambda}$, which can
now be obtained by differentiating the fundamental equation (\ref{scaling}), are seen to
depend on \emph{two} independent quantities $a_t$ and $a_p$ . This leads to two
independent scaling relations between the exponents.
\begin{eqnarray}
\label{wid}
 {\alpha} &+& 2{\beta} + {\gamma}=2~~~~~~~\mbox{(Widom equality).}\\
 {\gamma} &=& {\beta}({\delta}-1)~~~~~~~~~~\mbox{(Rushbrooke equality).}{\nonumber}
\end{eqnarray}

Another consequence of eq. (\ref{scaling}) is that the primed and unprimed exponents like
$\alpha$, $\alpha'$ and $\gamma$, $\gamma'$ are the same.
The critical exponents and scaling relations were first discussed in the context of
Kerr-Newman black holes in \cite{lousto}, and were worked out in detail for RN-AdS black
holes in \cite{wu}. As discussed previously, in both the `$q$' and the `$j$' ensembles the
KN-AdS black holes exhibit a liquid gas like phase behaviour, with a typical
liquid-vapour like coexistence curve culminating in a critical point (fig.(\ref{mxads3})). We
now discuss the critical exponents and scaling relations in these ensembles.

Since the expressions for thermodynamic functions are too lengthy and algebraically
complicated, they cannot be inverted to obtain the equation of state which, in turn, may
be used to calculate the exponents, as is the standard procedure.

These can, however, be obtained iteratively using an expansion of the
thermodynamic quantities in the neighbourhood of the critical point in powers of the
(small) path variable along which the critical point is approached. We shall describe this
method using the constant $j$ ensemble as a concrete example.
Let us look at the scaling behaviour near critical point along the equipotential curves.
For any two points $({q},{s})$ and $({q}+{\Delta}q,{s}+{\Delta}s)$ lying on a curve of
constant potential in the $q$-$s$ parameter space, see fig.(\ref{mxads2}), the following relation will hold
\begin{equation}
\label{crisopot}
{\phi}({q}+{\Delta}q,{s}+{\Delta}s)={\phi}({q},{s})=\mbox{constant}
\end{equation}
Since the motion is constrained on an equipotential curve, the variations of $s$ and $q$
are not independent of each other. With $q$ chosen as the independent path variable, $s$
now becomes an implicit function $s(q)$, whose form has to be determined iteratively.
To this end, we write down a symbolic expansion of $s({q})$ as a power series in
${\Delta}q$ around its value, $s(\tilde{q})=\tilde{s}$, at an arbitrary point,
$q=\tilde{q}$, on the equipotential curve.
\begin{equation}
\label{sex}
s(\tilde{q}+{\Delta}q)-s(\tilde{q})={\Delta}s=s_1(\tilde{q}){\Delta}q+\frac{1}{2!}s_2(\tilde{q})({\Delta}q)^2
+\frac{1}{3!}s_3(\tilde{q})({\Delta}q)^3+...
\end{equation}
The coefficients $s_1$, $s_2$, etc can now be obtained recursively from  eq.
(\ref{crisopot}) by first expanding ${\phi}(\tilde{q}+{\Delta}q,s(\tilde{q}+{\Delta}q))$
in a Taylor series in ${\Delta}q$ and ${\Delta}s$ about the point ($\tilde{q},\tilde{s}$),
then plugging in the expansion of ${\Delta}s$ from eq. (\ref{sex}) and requiring that the
final expression vanish at all orders ${{\Delta}q}$, ${{\Delta}q}^2$, etc. Thus, it can be
verified that

\begin{eqnarray}
s_1&=&-\frac{{\phi}_q}{{\phi}_s} \nonumber\\
s_2&=&-\frac{1}{\phi_s^3}\left[\phi_{qq}\phi_s^2 - 2\phi_{qs}\phi_q\phi_s + \phi_{ss}\phi_q^2\right] \nonumber \\
s_3 &=& -\frac{1}{\phi_s^5}\left[\phi_{qqq}\phi_s^5 - 3\phi_{qqs}\phi_q\phi_s^3 + 3\phi_{qss}\phi_q^2\phi_s^2 
-\phi_{sss}\phi_q^3\phi_s -3\phi_{qs}\phi_{qq}\phi_s^3 \right. \nonumber\\
& & \mbox{} \left. + 6\phi_{qs}^2\phi_s^2\phi_q  - 9\phi_{qs}\phi_{ss}\phi_s\phi_q^2 
+ 3\phi_{ss}\phi_q\phi_{qq}\phi_s^2 + 3\phi_{ss}\phi_q^3\right]
\label{exponents}
\end{eqnarray}
where the symbol ${\phi}_s$ represents $\partial{\phi}/\partial{s}$ and ${\phi}_{sq}$
represents ${\partial}^2{\phi}/{\partial{s}\partial{q}}$, etc. Having obtained these
coefficients we can now expand the temperature $t(q,s)$ near the critical point as a power
series in the \emph{order parameter} ${\Delta}q=q-q_{cr}$, by approaching the critical
point along the critical curve ${\phi}(q,s)={\phi}_{cr}$. 
 \footnote{The critical points are obtained numerically owing to the inability to invert
many of the algebraic expressions involved}
Using the coefficients from eq. (\ref{exponents}), it can be verified that
\begin{equation}
\label{crtemp}
t(q_{cr}+{\Delta}q,s_{cr}+{\Delta}s)=t_{cr}+ \frac{1}{3!}t_{3}({\Delta}q)^3+~\mbox{higher
order terms,}
\end{equation}
where $t_3$ is a positive constant of order unity whose exact value depends on the
critical point. Similarly, it can be verified that the inverse of specific heat at
constant potential, $c_{\phi}${\footnote{Here, $c_{\phi}$ is the restriction of
$c_{j\phi}$ on constant $j$ sections of the parameter space}}, goes as

\begin{equation}
\label{crspecific}
{c_{\phi}^{-1}(q_{cr}+{\Delta}q,s_{cr}+{\Delta}s)}=\frac{1}{2!}c_2({\Delta}q)^2+~\mbox{higher
order terms,}
\end{equation}
where $c_2$ is a positive constant. Further, one can check that the capacitance
${\chi}_t=({\partial}q/{\partial}{\phi})|_{t}$ also diverges in the same way as the
specific heat.
\begin{equation}
\label{crcap}
{\chi}_t^{-1}~{\sim}~({\Delta}q)^2
\end{equation}
Thus, in order to find the scaling exponents for $c_{\phi}$, for example, we can readily
infer from eq. (\ref{crtemp}) and eq. (\ref{crspecific}) that
\begin{equation}
\label{crcphi}
c_{\phi} ~{\sim}~ (t_r)^{-3/2}.
\end{equation}

A completely analogous calculation gives the scaling behaviour of various thermodynamic
quantities along the critical isotherm, ${t}(q,s)={t}_{cr}$. For this, we first rewrite
eq.(\ref{crisopot}) along isotherms.
\begin{equation}
{t}({q}+{\Delta}q,{s}+{\Delta}s)={t}({q},{s})=\mbox{constant}
\end{equation}
Proceeding in a similar manner, we obtain the coefficients corresponding to eq. (\ref{exponents})
\begin{eqnarray}
s_1^{'}&=&-\frac{{t}_q}{{t}_s}\nonumber\\
s_2^{'}&=&-\frac{1}{t_s^3}\left[t_{qq}t_s^2 - 2t_{sq}t_st_q + t_{ss}t_q^2\right]
\end{eqnarray}
with $s_3^{'}$ having a similar expression as $s_3$ of eq. (\ref{exponents}), and as before, the symbol ${t}_s$, $t_{sq}$  
stand for $\partial{t}/\partial{s}$ and ${\partial}^2{t}/{\partial{s}\partial{q}}$, etc. On plugging these
coefficients into the expansion of ${\phi}(q,s)$ around the critical point, we can
numerically verify that
\begin{equation}
\label{crop}
{\phi}(q_{cr}+{\Delta}q,s_{cr}+{\Delta}s)={\phi}_{cr}+
{\frac{1}{3!}}{\phi}_{3}({\Delta}q)^3+~\mbox{higher order terms,}
\end{equation}
where ${\phi}_3$ is a positive constant whose value depends on the critical point.
Similarly, it can be verified that along the critical isotherm, specific heat $c_{\phi}$
goes as
\begin{equation}
{c_{\phi}^{-1}(q_{cr}+{\Delta}q,s_{cr}+{\Delta}s)}=\frac{1}{2!}c_2^{'}({\Delta}q)^2+~\mbox{higher
order terms,}
\end{equation}
where $c_2^{'}$ is positive.
Having elucidated our general approach, we now list out the critical exponents
corresponding to both the ensembles.

For the fixed $j$ ensemble we obtain
\begin{equation}
\label{crtj}
{\alpha}_j=2/3, ~~{\beta}_j=1/3,~~ {\gamma}_j= 2/3,~~ {\delta}_j=3 ~.
\end{equation}
where the subscript $j$ indicates the ensemble.
Note that these exponents follow both the scaling relations eq. (\ref{wid}). Moreover, these exponents are exactly the same as the critical exponents for 
RN-AdS black holes in the canonical ensemble, \cite{johnson1} and \cite{wu}. \footnote{the critical exponents of Kerr-AdS black holes in the canonical 
ensemble, which we have checked independently, are the same as RN-AdS exponents}

The critical exponents for the $q$-ensemble can be obtained by a similar calculation. We
first consider entropy as an implicit function $s(j)$ along the critical isopotential
curve, ${\omega}(j,s)={\omega}_{cr}$, or the critical isotherm, $t(j,s)=t_{cr}$, and then
write down a symbolic expansion for it around the critical point in powers of the order
parameter ${\Delta}j=j-j_{cr}$. The coefficients of expansion, obtained recursively just
as in the $j$-ensemble, are then used to calculate various critical exponents. We list
them here:
\begin{equation}
\label{crtq}
{\alpha}_q=2/3, ~~{\beta}_q=1/3,~~ {\gamma}_q= 2/3,~~ {\delta}_q=3 ~.
\end{equation}
Which is exactly the same as in the fixed $j$ ensemble, and again, these exponents follow both the scaling laws of eq. (\ref{wid}). 

We are now in a position to make a very interesting observation. The fact that the critical exponents calculated in this subsection 
matches exactly with those for the RN-AdS black holes and the Kerr-AdS black holes (with appropriate identification of the thermodynamic parameters) indicates 
a kind of universality in the critical behaviour. Black holes being non extensive thermodynamic systems, thermodynamic behaviour
of these are ensemble dependent. However, for critical points in different ensembles, the KN-AdS black hole seems to 
have an universal set of critical exponents, irrespective of the ensemble chosen. 

The calculation of the critical exponents for the KN-AdS black holes, together with the observation made in the last paragraph
constitute the main results of this subsection. In the last subsection, we briefly comment on scaling relations of the Ruppeiner curvature
for the mixed ensembles. 

\subsection{Scalar curvature for KN-AdS Black Holes in the Mixed Ensembles }

The state space scalar curvature $R$ for the constant $j$ and constant $q$ ensembles can
be symbolically expressed as
\begin{equation}
\label{scalar}
R_j=\frac{{\mathcal{P}}_j}{s{\mathcal{N}}(t){\mathcal{D}}(c_{\phi})^2}~~~~;~~~~
R_q=\frac{{\mathcal{P}}_q}{s{\mathcal{N}}(t){\mathcal{D}}(c_{\omega})^2}
\end{equation}
where the symbols $\mathcal{N}$ and $\mathcal{D}$ represent the numerator and denominator
respectively of their argument, and $\mathcal{P}_j$ is a polynomial of about $250$ terms
while $\mathcal{P}_q$ is a polynomial of about 200 terms. Here the subscripts $j$ and $q$
are used to label the ensembles.
We observe that $R$ diverges along the spinodal curve (which includes the critical point) and
the extremal curve. As discussed in \cite{paper}, inside the coexistence region bounded by
the spinodal curve the line element of the corresponding state space scalar curvature is negative, and
hence thermodynamic geometry is not applicable in that region. Similarly, the extremal
curve separates the physical region from the naked singularity region where thermodynamic
geometry is once again not applicable owing to a negative line element in the
corresponding state space.\footnote{The thermodynamics at extremality is more subtle as it
is governed by quantum fluctuations and not thermal fluctuations. Like the previous cases we shall ignore the divergences near extremality.} 
We will mainly be
interested in the divergence of $R$ at criticality.
Before we pass on to discuss the behaviour of scalar curvature near the critical point, we
note the following observation at extremality. For both the ``mixed'' ensemble black holes
considered, the product of the scalar curvature and the specific heat at fixed charge has a
limiting value of unity at $t=0$. Thus, for the $j$ ensemble, we have
\begin{eqnarray}
\lim_{t \to 0}{R_{j}c_{q}}=1~~~~,
\end{eqnarray}
where $R_j$ is the scalar curvature at fixed $j$ and $c_q$ is the restriction of the
specific heat $c_{jq}$ on the constant $j$ section. Similarly, for the fixed $q$ ensemble
we obtain
\begin{eqnarray}
\lim_{t \to 0}{R_{q}c_{j}}=1~~~~.
\end{eqnarray}
where $R_q$ is the scalar curvature at fixed $q$ and $c_j$ is the restriction of the
specific heat $c_{jq}$ on the constant $q$ section. This agrees with the result obtained
in \cite{rupp3} for asymptotically flat black holes.
As was observed in \cite{paper}, $R$ always shows a negative divergence at the critical
points of both the $q$ and the $j$ ensembles. This is similar to its behaviour in
conventional systems displaying criticality. Moreover, the denominator of $R_j$ contains
the square of the denominator of $c_{\phi}$ or $({\partial}q/{\partial}{\phi})|_{t})$ and
that of $R_q$ contains the square of the denominator of $c_{\omega}$ or
$({\partial}j/{\partial}{\omega})|_{t})$. That $R$ shares its divergence with  specific
heat at constant potential or with the susceptibilities is not unexpected as its
calculation envisions the black hole in a grand canonical ensemble, with the relevant
charges fluctuating. 

In conventional thermodynamic systems the state space scalar
curvature is interpreted as a correlation volume  ${\xi}^d$, where $\xi$ is the
correlation length and $d$ is the spatial dimension of the system, \cite{rupp}. We may
define a critical exponent ${\vartheta}$ corresponding to the scaling of $R$ as
\begin{equation}
R~{\sim}~t_{r}^{-\vartheta}
\end{equation}
Using the critical exponent for the correlation length,
\begin{equation}
\label{corr}
{\xi}~{\sim}~(t_r)^{-\nu}
\end{equation}
and the hyperscaling relation, (\cite{rupp}, \cite{hes}, \cite{fish}),
\begin{equation}
\label{hyper}
\hspace{1.2in}d\nu=2-\alpha~~~~~\mbox{Josephson equality}
\end{equation}
we can write down the scaling relation for $\vartheta$ as
\begin{equation}
\label{rscale}
\hspace{1.26in}\vartheta=2-\alpha~~~~~\mbox{Ruppeiner equality}
\end{equation}
where we have termed the scaling relation for curvature as ``Ruppeiner equality''.
Using eq. (\ref{crcphi}) and eq. (\ref{rscale}), it can be further verified that
\begin{equation}
\label{rct}
\hspace{-0.5in}R\hspace{0.02in}c\hspace{0.02in}t_{r}^2=K~~,
\end{equation}
where $c$ is the specific heat capacity and $K$ is a constant having dimensions of
entropy. Using ``two scale factor universality'' (see references in \cite{rupp}) the
constant term $K$ can be seen to be a combination of critical exponents,
 \begin{equation}
 \label{2scale}
 K=-{\beta}({\delta}-1)({\beta}{\delta}-1){\kappa}_B
 \end{equation}
 Thus, the r.h.s of eq. (\ref{rct}) is a universal constant in a given equivalence class
of critical exponents. On putting in the values of critical exponents for different
universality classes, it can be verified that $K$ is in general negative, which in turn
implies that $R$ is negative near the critical point.

However, in the case of black holes, as observed in \cite{rupp3}, the state space scalar
curvature  is a dimensionless number as opposed to its dimension of volume in conventional
thermodynamic systems. Besides, there is no interpretation of a correlation length as such for
black holes. Thus, $R$ can no longer be interpreted as a correlation volume.
Nevertheless, with the understanding that it is a measure of ``correlations'' at the
horizon, \cite{rupp3}, we can still continue to investigate its scaling behaviour near
criticality.
\begin{figure}[t!]
\begin{minipage}[b]{0.5\linewidth}
\centering
\includegraphics[width=2.7in,height=2.7in]{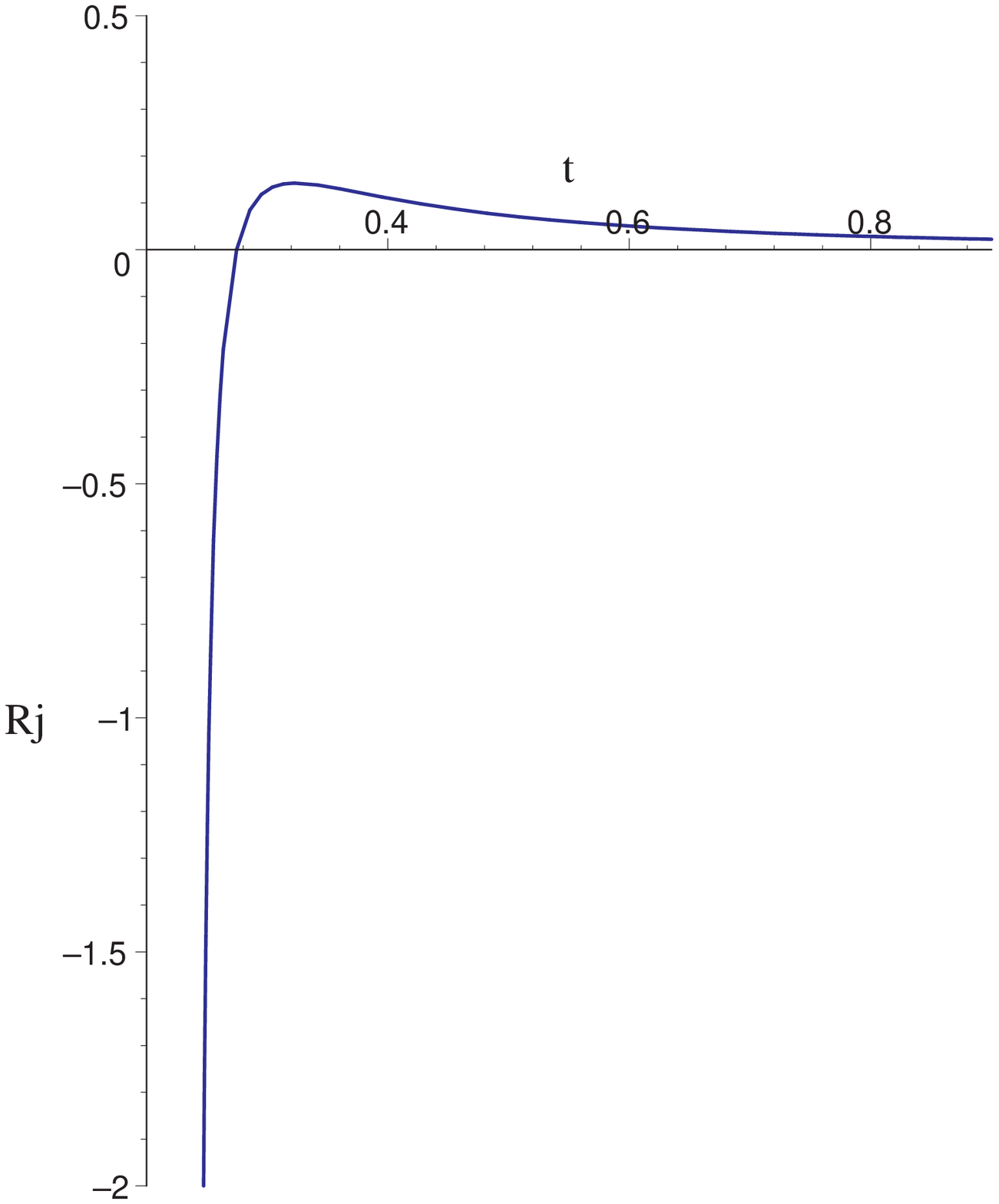}
\caption{Plot of $R_j$ vs. $t$ for the large black hole phase along an isopotential curve,
$\phi=0.5$. $j$ is fixed at a value of $0.011$.}
\label{r1}
\end{minipage}
\hspace{0.6cm}
\begin{minipage}[b]{0.5\linewidth}
\centering
\includegraphics[width=2.7in,height=2.7in]{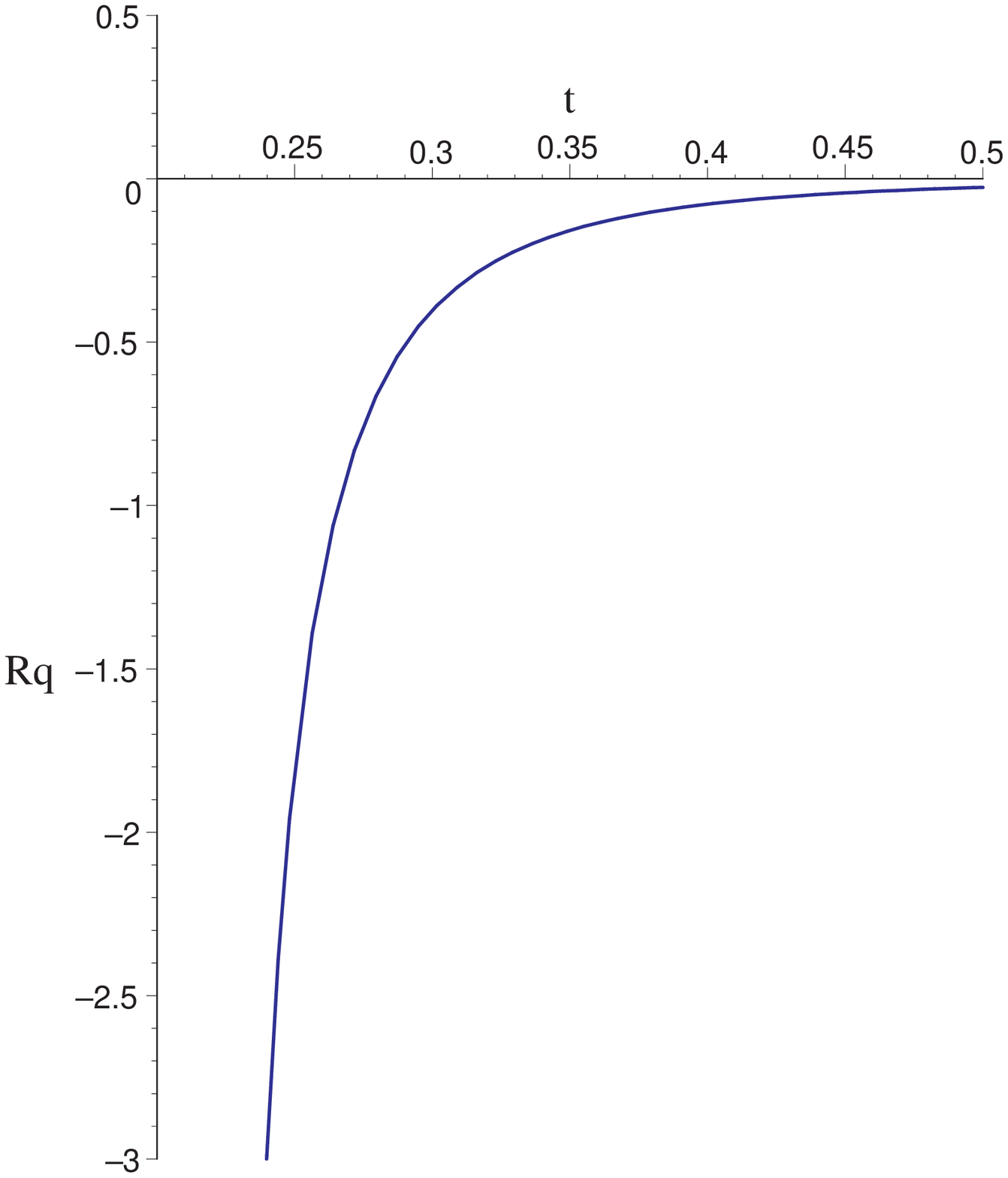}
\caption{Plot of $R_q$ vs. $t$ for the large black hole phase along an isopotential curve,
$\omega=0.8$. $q$ is fixed at a value of $0.16$.}
\label{r2}
\end{minipage}
\end{figure}
 Recalling the observation that $R$ contains the square of the denominator of specific heat at constant potential we can write down its scaling beahviour as
\begin{equation}
R_j~{\sim}~t_r^{-2{\alpha}_j}~~~;~~~R_q~{\sim}~t_r^{-2{\alpha}_q}
\end{equation}
where the subscripts indicate constant $q$ or constant $j$ ensemble. Recalling from eqns
(\ref{crtj}) and (\ref{crtq}) that ${\alpha}_j={\alpha}_q=2/3$, it can be verified that
the scaling relation for $\vartheta$, eq.(\ref{rscale}), still holds in spite of the
Josephson equality not being a meaningful one in the context of black holes. Similarly,
the validity of eq.(\ref{rscale}) also ensures that eq.(\ref{rct}) too holds in the case
of the mixed ensemble black holes. Thus, close to the critical point
\begin{equation}
 R_j\hspace{0.02in}c_{\phi}\hspace{0.02in}t_{r}^2=K_j~~~,~~~
R_q\hspace{0.02in}c_{\omega}\hspace{0.02in}t_{r}^2=K_q
\end{equation}
However, the constants $K_j$ and $K_q$ do not follow eq.(\ref{2scale}) as ``two scale
factor universality'' is not valid for black holes. We add that, as observed in
\cite{paper}, $R$ is always negative in the critical region.

\begin{figure}[t!]
\begin{minipage}[b]{0.5\linewidth}
\centering
\includegraphics[width=2.7in,height=2.7in]{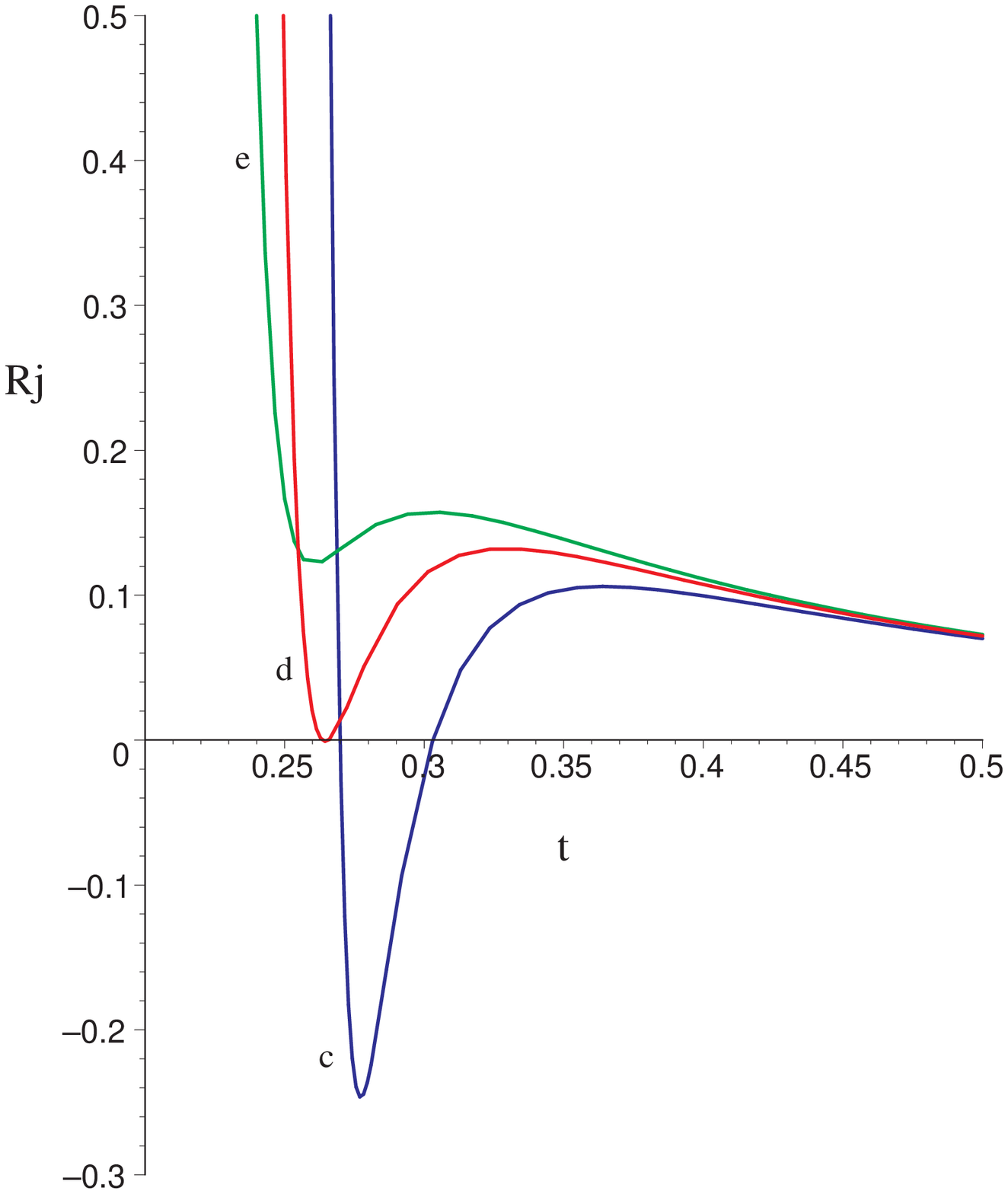}
\caption{Plot of $R_j$ vs. $t$ along isopotential curves with ${\phi}_c=0.12$,
${\phi}_d=0.4$ , ${\phi}_e=0.5$ . $j$ is fixed at a value of $0.043$.}
\label{jr1}
\end{minipage}
\hspace{0.6cm}
\begin{minipage}[b]{0.5\linewidth}
\centering
\includegraphics[width=2.7in,height=2.7in]{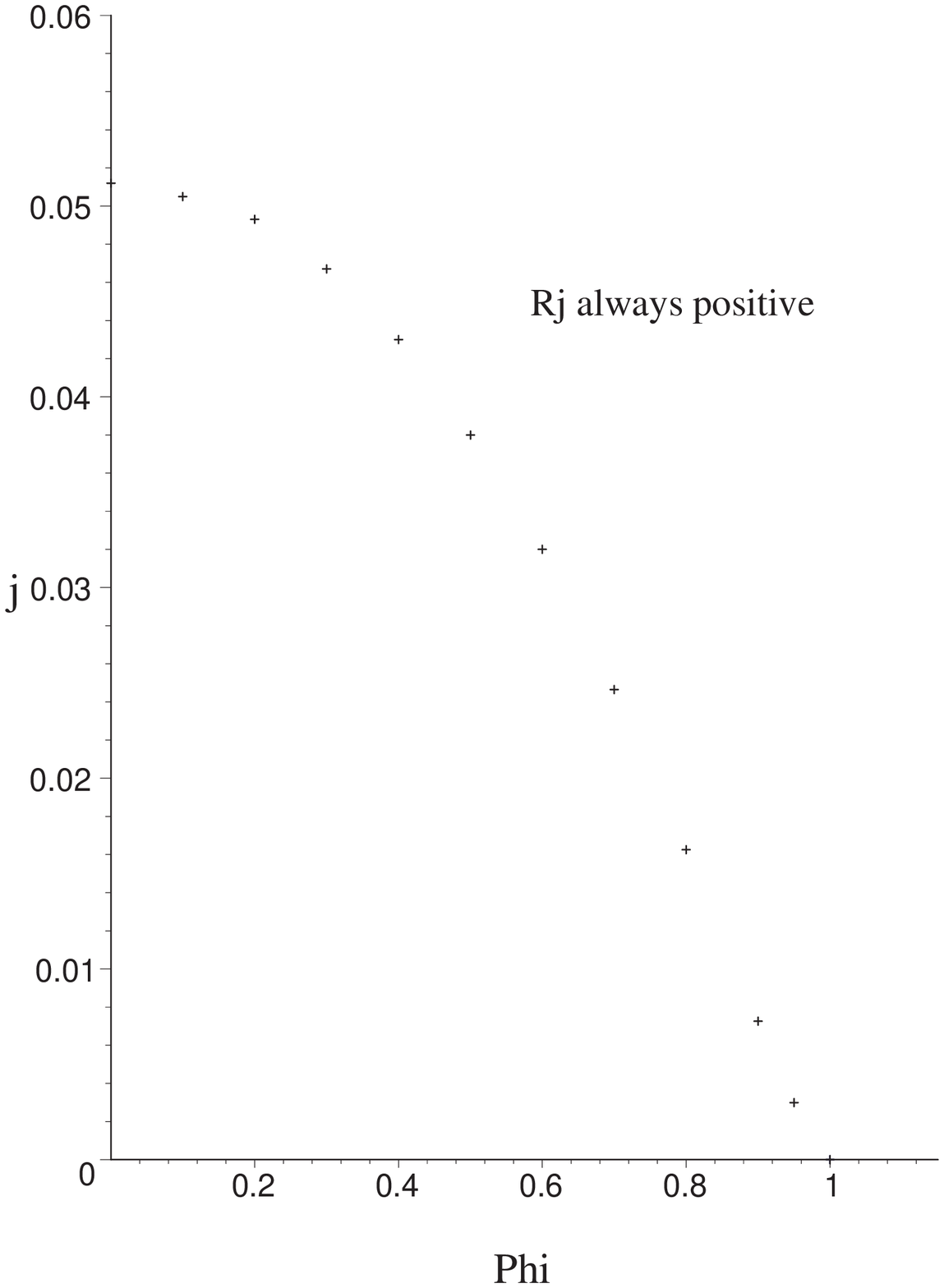}
\caption{Locus of points in the $j$-$\phi$ plane above which $R>0$ for all $t$. The curve
intersects the $j$ axis at $j=0.0512$.}
\label{jr2}
\end{minipage}
\end{figure}

Let us now discuss the behaviour of $R$ for large black holes in the two mixed ensembles.
In fig.(\ref{r1}) and fig.(\ref{r2}) we show generic plots of $R_j$ vs $t$ and $R_q$ vs
$t$ respectively in the large black hole phase. It can be seen from the figures that while
in the $j$-ensemble the state space scalar curvature always changes sign from negative to
positive on increasing the temperature, in the $q$-ensemble the curvature corresponding to
the large black hole phase remains negative at all temperatures. These figures are similar
to figs. (1a) and (2a) for the RN-AdS and Kerr-AdS black holes respectively. In both cases we can see from eq. (\ref{scalar}) that the state space scalar curvature
asymptotes to zero as $s$, and therefore $t$, tends to infinity. However, it can be
verified that, they do so from opposite sides of the $t$-axis. For large $t$, their asymptotic behaviour becomes
\begin{equation}
R_j\sim\frac{1}{t^2}~~,~~R_q\sim-\frac{1}{t^4}
\end{equation}
Notice that the asymptotic behaviour remains the same as the RN-AdS and Kerr-AdS black hole, \emph{i.e}, the $j=0$ or the $q=0$ case respectively. These observations lead us
to conclude that while the constrained, non fluctuating charges have a decisive role to
play in the thermodynamics of small black holes at low temperatures, they do not play a
significant role in the thermodynamics of high temperature large black holes. It also serves to re-emphasize the different nature of $q$ and $j$ fluctuations, as was noticed previously.

We can make a further observation for the constant $j$ black holes in relation to the sign
of $R$. Fig.(\ref{jr1}) shows a plot of $R_j$ vs. $t$ for three isopotential curves, with
its angular momentum $j$ fixed at a value of $0.043$. The black hole does not exhibit
criticality since its angular momentum is above the critical value $j_c=0.0239$, fig.(2a).
However, on following the behaviour of $R$ along the isopotential curves ${\phi}_c$,
${\phi}_d$, and ${\phi}_e$  we notice that there is a ``critical'' value of
${\phi}={\phi}_d$ beyond which $R$ remains positive at \emph{all} temperatures. This is similar to the case of $\phi>1$ for RN-AdS black holes, shown in fig.(\ref{rnR3}). In
fig.
(\ref{jr2}) we plot a locus of such ``critical'' points in the $j$-$\phi$ plane. From the
figure it can also be inferred that there is a minimum value of the fixed angular momentum,
$j=j_p=0.0512$, above which the scalar curvature remains positive at all temperatures for any
value of the potential ${\phi}$.

\begin{figure}[t!]
\begin{minipage}[b]{0.3\linewidth}
\centering
\includegraphics[width=1.7in,height=1.7in]{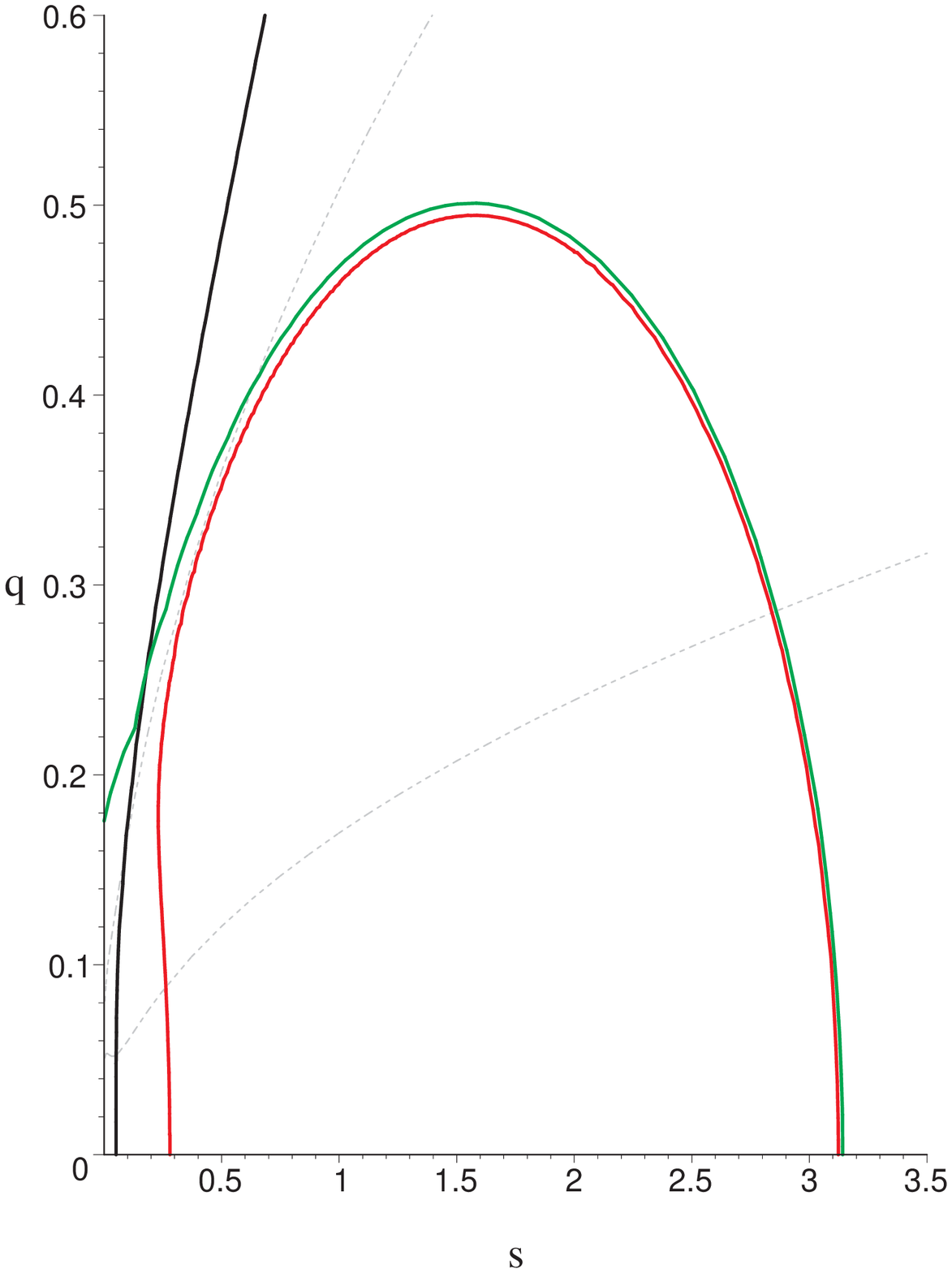}
\caption{Locus of zeroes of $R$, $t$ and the free energy for $j=0.009$. Grey dotted isopotentials are at $\phi=0.3$ below and $\phi=0.9$ above.}
\label{jr1}
\end{minipage}
\hspace{0.2cm}
\begin{minipage}[b]{0.3\linewidth}
\centering
\includegraphics[width=1.7in,height=1.7in]{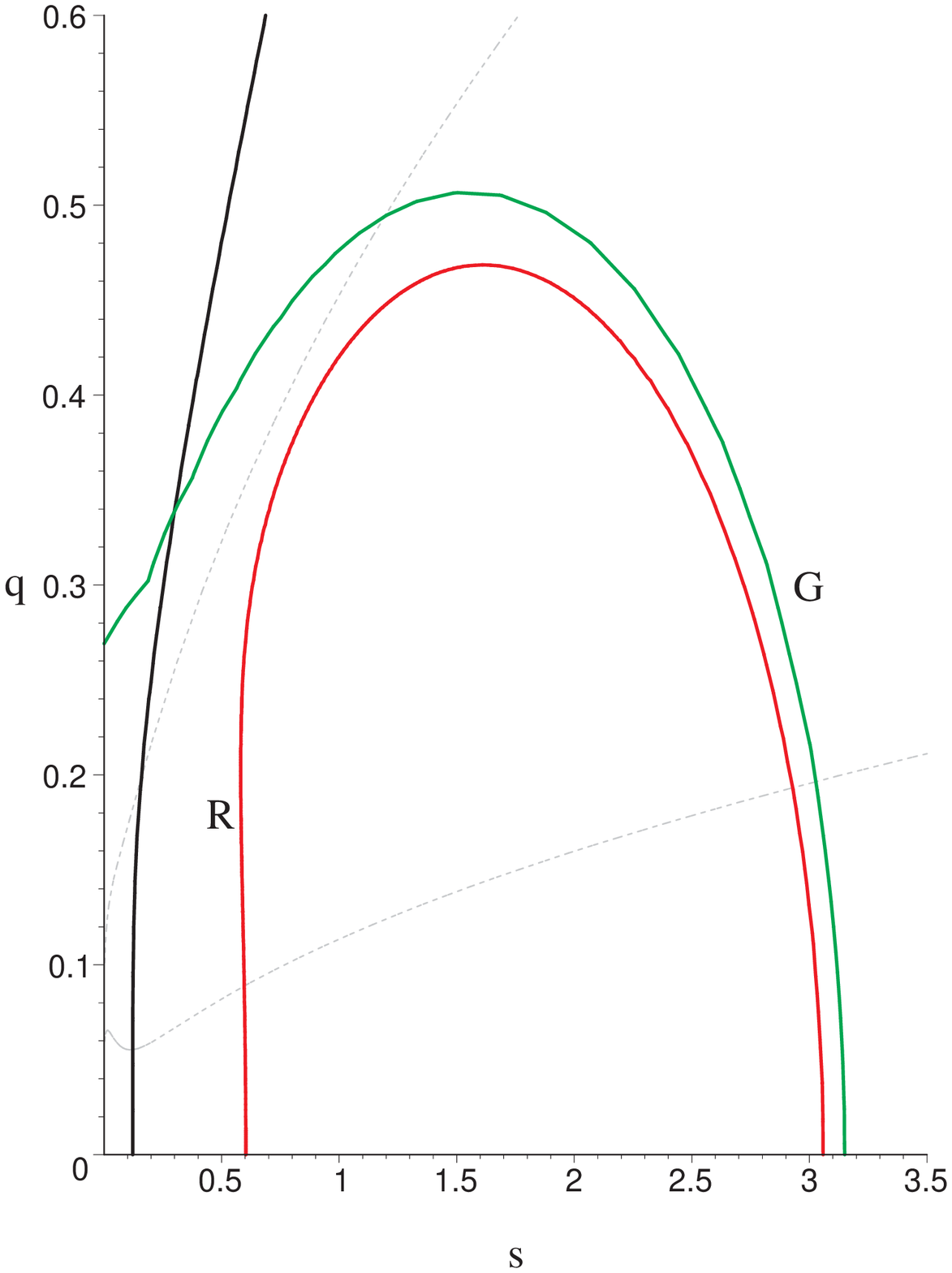}
\caption{Locus of zeroes of $R$, $t$ and the free energy for $j=0.021$. Grey dotted isopotentials are at $\phi=0.2$ below and $\phi=0.8$ above}
\label{jr2}
\end{minipage}
\hspace{0.2cm}
\begin{minipage}[b]{0.3\linewidth}
\centering
\includegraphics[width=1.7in,height=1.7in]{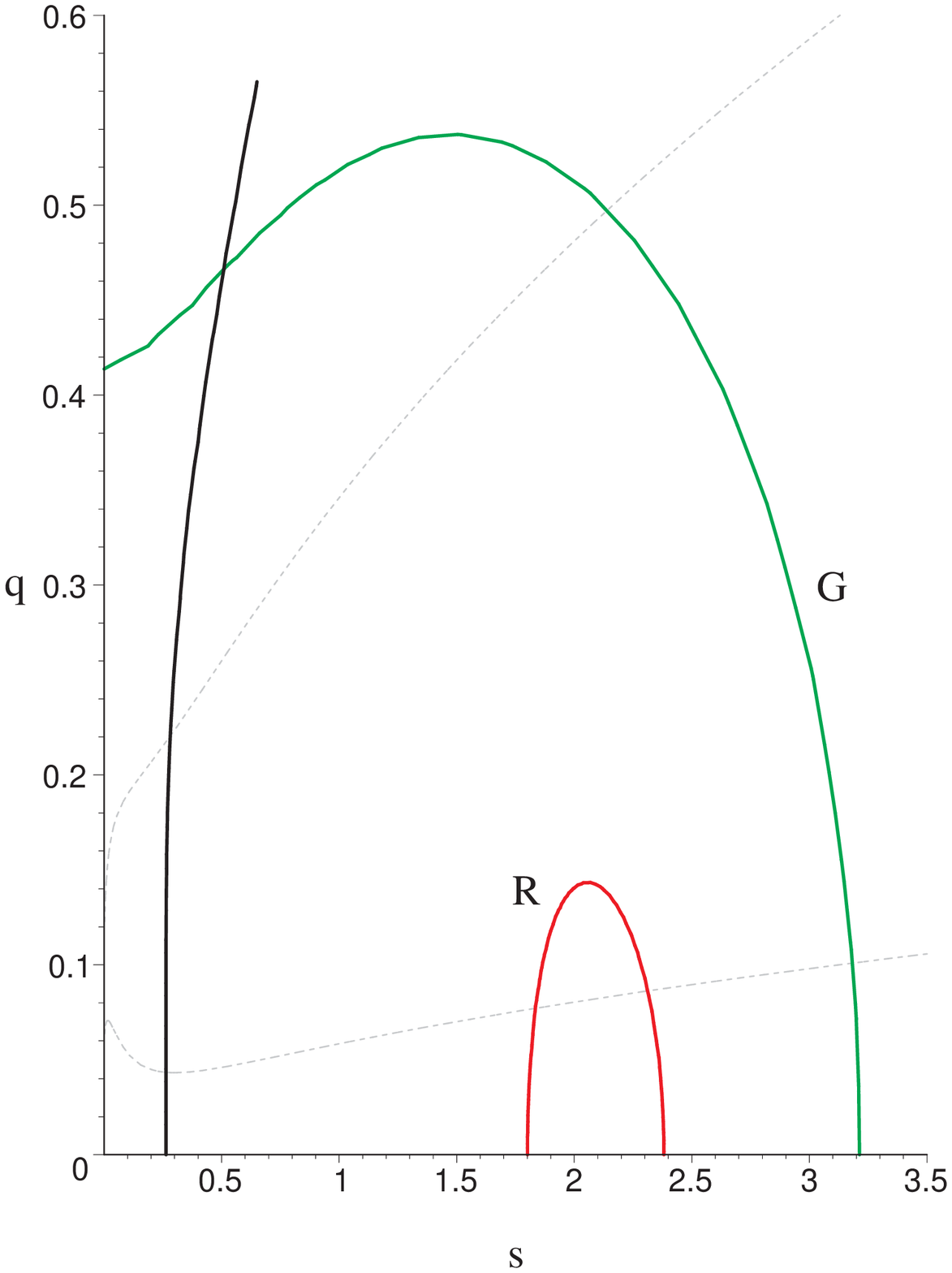}
\caption{Locus of zeroes of $R$, $t$ and the free energy for $j=0.0497$. Grey dotted isopotentials are at $\phi=0.1$ below and $\phi=0.6$ above}
\label{jr22}
\end{minipage}
\end{figure}

Finally, continuing our investigation of the sign of $R$ in the $j$ ensemble we now focus our
attention on the comparison between the zeroes of the scalar curvature and the zeroes of
the free energy. We recall that in the case of RN-AdS black holes (\emph{i.e}, ``$j=0$" ensemble) the 
zeroes of $R$ and the free energy coincide. Let us see how a non-zero $j$
effects a change in this. In each of the figs. (6), (7) and (8), which are labeled by
increasing values of $j$ from left to right, we represent the zeroes of curvature by the
red curve and those of the free energy by the green curve while zeroes of the temperature are represented by the black extremal curve. The scalar curvature is negative under the curve of its zeroes and positive outside while the free energy is positive under its curve and negative
outside. The naked singularity region lies to the left of the extremal curve. On comparing these figures with fig.(\ref{rnads3}) for RN-AdS black holes we can infer that introduction of a non-zero \emph{constant} $j$ causes a separation of the zeroes of $R$ and the free energy. On increasing the value of $j$, the $R$ curve shrinks, thus widening the gap with the free energy curve, until it would finally disappear at $j=0.0512$, fig.(\ref{jr2}). Notice that the pattern of separation of the $R$ curve and the $G$ curve is different from the case of \emph{fluctuating} $j$, as shown in fig.(\ref{knads4}) and in fig.(\ref{knads1}) to fig.(\ref{knads3}) of section $5$.  In the former, constant case, the $R$ curve is always inside the $G$ curve, while in the latter, fluctuating case, the $R$ curve lies on both sides of the $G$ curve. Again, in the constant  $j$ case, for each fixed $j$ below $j_p=0.0512$ there always exists a minimum $\phi$ above which $R$ never crosses zero and is always positive, as mentioned in the preceding paragraph. Beyond $j_p$, $R$ is positive for all $\phi$. In comparison, in the fluctuating $j$ case, $R$ always crosses zero for all $\phi<1$. Moreover, in the latter case, for higher values of $\phi$ the $R$ curve and the $G$ curve are almost coincident. It should be borne in mind, however, that in the mixed ensemble, the zeroes of Gibbs potential do not have the interpretation of a ``Hawking-Page'' curve.

\section{Discussions and Conclusions}

In this paper, we have studied various aspects of criticality, scaling behaviour, and thermodynamic geometry of AdS black holes, following our earlier 
work in \cite{paper}. This completes the comprehensive analysis of the structure of the equilibrium thermodynamic state space geometry of AdS black holes. 
Several observations leading to important new insights regarding the phase structure of these black holes have been obtained in this context. 
The methods used in this analysis are significantly different from the ones conventionally used in the literature.  
The critical exponents for second order phase transitions in the 
mixed ensembles introduced and elucidated in \cite{paper} for the KN-AdS black holes have also been calculated. These have been compared to
the critical exponents for RN-AdS and Kerr-AdS black holes in the canonical ensemble and are found to be identical. This suggests an universality 
in critical phenomena for AdS black holes. 

The state space scalar curvature involving fluctuations in all the thermodynamic variables have been studied in the grand 
canonical ensemble for KN-AdS black holes. Our results establish the significant fact that the fluctuations in the electric 
charge and those in the angular momentum are distinctly different from each other in their respective effects on the geometry of the
thermodynamic state space. One of the main conclusions of this analysis is that for
black holes whose state space geometry involves electric charge fluctuations exhibits a change in sign  of the scalar curvature, which 
closely parallels the corresponding change in sign of the Gibbs free energy. In fact, remarkably 
for the RN-AdS black holes, we establish an exact correspondence between the zeroes of the state space scalar curvature and that of the Gibbs free energy.
It turns out that for the range of parameters for which the black hole is globally
stable against thermal AdS, the scalar curvature is positive and is negative when the black hole is globally unstable. However interestingly, for the Kerr-AdS 
black holes, this phenomenon is notably absent and the scalar curvature remains negative at all temperatures. 
Also, for the case of the fixed $j$ mixed ensemble, we have established a minimum value of the potential $\phi$ above which the scalar 
curvature is positive at all temperatures. 

In order to further explore the effect of  fluctuations on the sign of the scalar curvature, we have considered the full set of fluctuating thermodynamic 
charges in the case of the KN-AdS black hole in the grand canonical ensemble. The effect of fluctuations in the angular velocity and the electric potential
on the sign of the curvature have been studied in details in this context. The significant outcome of this analysis is that a close correspondence 
between the zeroes of the curvature and the Gibbs free energy is observed only at large values of the electric potential. 
Further, the RN-AdS limit of the KN-AdS scalar curvature was also studied, leading to the observation that the fluctuations in the angular momentum 
$j$ causes a shift in the zeroes of the state space scalar curvature. The asymptotic behaviour of the thermodynamic scalar curvature was also studied, 
illustrating significant differences between the RN-AdS and the Kerr-AdS black holes. 

Finally, we have studied the behaviour of the thermodynamic curvature of KN-AdS black holes in the mixed ensembles near criticality, and obtained a hyperscaling
relation. Interestingly, this behaviour closely resembles that of conventional thermodynamic systems. This provides further justification for using 
thermodynamic geometry as an effective tool for analysing phase transitions and critical phenomena in black hole thermodynamics. 

It would be interesting to 
apply our formulation to analyse black hole thermodynamics in higher derivative gravity theories. This may lead to interesting phase structures 
and corresponding critical phenomena for these systems. The same is true for higher dimensional AdS black holes.
Further, the significance of the novel phase behaviour of KN-AdS black holes in the mixed ensembles
established and elucidated by us in this paper and our earlier work \cite{paper}, vis a vis the AdS/CFT correspondence, is an interesting open problem 
for the future.


\begin{thebibliography}{99}
\bibitem{td1}
  R.~M.~Wald,
  ``The thermodynamics of black holes,''
  Living Rev.\ Rel.\  {\bf 4}, 6 (2001)
  [arXiv:gr-qc/9912119].
\bibitem{td2}
D.~N.~Page,
  ``Hawking radiation and black hole thermodynamics,''
  New J.\ Phys.\  {\bf 7}, 203 (2005)
  [arXiv:hep-th/0409024].
  \bibitem{td3}
  R.~Brout, S.~Massar, R.~Parentani and Ph.~Spindel,
  ``A Primer for Black Hole Quantum Physics,''
  Phys.\ Rept.\  {\bf 260}, 329 (1995)
  [arXiv:0710.4345 [gr-qc]].
\bibitem{maldacena}
  O.~Aharony, S.~S.~Gubser, J.~M.~Maldacena, H.~Ooguri and Y.~Oz,
  ``Large N field theories, string theory and gravity,''
  Phys.\ Rept.\  {\bf 323}, 183 (2000)
  [arXiv:hep-th/9905111].
\bibitem{tisza}
L. Tisza, ``Generalized Thermodynamics,'' Pub. MIT Press, Cambridge, MA (1966)
\bibitem{callen}
H. B. Callen, ``Thermodynamics and an Introcution to Thermostatitics,'' Pub. Wiley, New York (1985)
\bibitem{weinhold}
F. Weinhold, J. Chem Phys. {\bf 63} (1075) 2479, {\it ibid} J. Chem Phys. {\bf 63} (1975) 2484.
\bibitem{rupp}
G. Ruppeiner, Rev. Mod. Phys. {\bf 67} (1995) 605, erratum {\it ibid} {\bf 68} (1996) 313.
\bibitem{ferrara}
S.~Ferrara, G.~W.~Gibbons and R.~Kallosh,
  ``Black holes and critical points in moduli space,''
  Nucl.\ Phys.\  B {\bf 500}, 75 (1997)
  [arXiv:hep-th/9702103].
\bibitem{paper}
A.~Sahay, T.~Sarkar and G.~Sengupta,
  ``Thermodynamic Geometry and Phase Transitions in Kerr-Newman-AdS Black Holes,''
  [arXiv:1002.2538 [hep-th]].
\bibitem{johnson1}
A.~Chamblin, R.~Emparan, C.~V.~Johnson and R.~C.~Myers,
``Charged AdS black holes and catastrophic holography,''
Phys.\ Rev.\  D {\bf 60}, 064018 (1999)
\bibitem{wu}
X.~N.~Wu,
``Multicritical Phenomena Of Reissner-Nordstrom Anti-De Sitter Black Holes,''
Phys.\ Rev.\  D {\bf 62}, 124023 (2000).
\bibitem{jan1}
H. Janyszek, R. Mrugala, ``Geometrical Structure of the State Space in Classical Statistical and Phenomenological Thermodynamics,''
Rep. Math. Phys. {\bf 27} (1989) 145.
\bibitem{mirza1}
  B.~Mirza and H.~Mohammadzadeh,
  ``Ruppeiner Geometry of Anyon Gas,''
  Phys.\ Rev.\  E {\bf 78}, 021127 (2008)
  [arXiv:0808.0241 [cond-mat.stat-mech]].
\bibitem{rupp1}
G. Ruppeiner
``Riemannian geometric approach to critical points: General theory'', Phys. Rev. {\bf E 57} (1997) 5135.
\bibitem{calda}
M.~M.~Caldarelli, G.~Cognola and D.~Klemm,
``Thermodynamics of Kerr-Newman-AdS black holes and conformal field
theories,''
Class.\ Quant.\ Grav.\  {\bf 17}, 399 (2000)
[arXiv:hep-th/9908022].
\bibitem{johnson2}
A.~Chamblin, R.~Emparan, C.~V.~Johnson and R.~C.~Myers,
  ``Holography, thermodynamics and fluctuations of charged AdS black holes,''
  Phys.\ Rev.\  D {\bf 60}, 104026 (1999)
  [arXiv:hep-th/9904197].
\bibitem{aman}
J.~E.~Aman, I.~Bengtsson and N.~Pidokrajt,
  ``Geometry of black hole thermodynamics,''
  Gen.\ Rel.\ Grav.\  {\bf 35}, 1733 (2003)
  [arXiv:gr-qc/0304015].
\bibitem{hes}
H.~E.~Stanley,
``Scaling, universality, and renormalization: Three pillars of modern critical phenomena,''
Rev.\ Mod.\ Phys.\  {\bf 71}, S358 (1999).
\bibitem{fish}
Scaling, Universality and Renormalization Group Theory, M.E Fisher.
\bibitem{lousto}
C.~O.~Lousto,
``The Fourth law of black hole thermodynamics,''
Nucl.\ Phys.\  B {\bf 410}, 155 (1993)
[Erratum-ibid.\  B {\bf 449}, 433 (1995)]
  [arXiv:gr-qc/9306014].
\bibitem{rupp3}
G.~Ruppeiner,
  ``Stability And Fluctuations In Black Hole Thermodynamics,''
  Phys.\ Rev.\  D {\bf 75}, 024037 (2007).
\bibitem{rupp4}
  G.~Ruppeiner,
  ``Thermodynamic curvature and phase transitions in Kerr-Newman black holes,''
  Phys.\ Rev.\  D {\bf 78}, 024016 (2008)
  [arXiv:0802.1326 [gr-qc]].
[arXiv:hep-th/9902170].
\bibitem{gibb}
G.~W.~Gibbons, M.~J.~Perry and C.~N.~Pope,
  ``The first law of thermodynamics for Kerr - anti-de Sitter black holes,''
  Class.\ Quant.\ Grav.\  {\bf 22}, 1503 (2005)
  [arXiv:hep-th/0408217].

\end{thebibliography}
\end{document}